\documentstyle{article}
\catcode`\@=11
 
\@addtoreset{equation}{section}
\@addtoreset{equation}{subsection}
\def\theequation{\ifnum\value{subsection}>0\relax
\thesubsection.\arabic{equation}\relax
\else\ifnum\value{section}>0\relax
\thesection.\arabic{equation}\relax
\else\arabic{equation}\fi\fi}
 
\input{epic.sty}
\input{eepic.sty}

\begin{document}

\begin{titlepage}
\begin{flushright}
YCTP-P44-92\\
December 1992\\
USC-93-013
\end{flushright}
\vspace{1cm}
\begin{center}
\LARGE{Fused Potts Models} \\
\vspace{1 cm}
\large{W.M.Koo} \footnote{Work supported in part by 
DOE grant DE-AC02-76ERO3075} 
                                       \normalsize{ and }
\large{H.Saleur} \footnote{Work supported in part by DOE contract 
DE-AC02-76ERO3075 and by
the Packard foundation. Address after january 1993: Dept. of Physics 
and Dept. of Mathematics, USC, University Park, Los Angeles CA 90089}\\
\vspace {1 cm} 
\normalsize{Department of Physics, Yale University\\
P O Box 6666 \\ New Haven, CT 06511, USA\\}
\vspace { 1 cm }
\begin{abstract}
Generalizing the mapping between the Potts model with
nearest neighbour interaction and the six vertex model, we build a 
family of "fused Potts models" related to the   spin $k/2$
 ${\rm U}_{q}{\rm su}(2)$ invariant vertex model and quantum spin chain. 
These Potts models have still variables taking
values $1,\ldots,Q$ ($\sqrt{Q}=q+q^{-1}$)
but they have a set of complicated multi spin interactions. The general
technique to compute these interactions, the resulting lattice geometry,
symmetries, and
the detailed examples of $k=2,3$   are given.

For $Q>4$ spontaneous magnetizations are
computed on the integrable first order phase transition line, 
generalizing Baxter's results for $k=1$. 

For $Q\leq 4$, we discuss 
the full phase diagram of the spin one ($k=2$) anisotropic
and ${\rm U}_{q}{\rm su}(2)$ invariant quantum spin chain
 (it reduces in the limit $Q=4$ ($q=1$) to the much studied
phase diagram of the isotropic spin one quantum spin chain). 
Several critical lines and massless phases are exhibited. 
The appropriate generalization of the 
Valence Bond State method of Affleck et al. is worked out.

\end{abstract}
\vspace {2 cm} 
%PACS Numbers: 04.40.+c, 97.60.Lf
\end{center}
\end{titlepage}

\section{Introduction}
 
\hspace{5 mm}
                       	
Numerous families of integrable lattice models have recently been 
exhibited. In general, these families obey the following pattern. 
One starts with an algebra like $sl(n)$ and one of its representation 
say $\rho$. Using quantum group technology a solution of the Yang 
Baxter equation acting in $\rho\otimes\rho$ can be found, which 
encodes the Boltzmann weights of a {\bf vertex model}. These weights 
are trigonometric and they depend on the quantum group deformation 
parameter $q$ and the spectral parameter. The degrees of freedom are 
the weights of the representation $\rho$, and they sit 
on edges of usually the square lattice.  
Using the quantum group symmetry the model can also be reformulated 
as a {\bf solid on solid model} whose degrees of freedom are highest 
weights and sit now
on the vertices of usually the square lattice. 
For $q$ a root of unity, the model truncates and a restricted 
sos model can be defined.

In the particular case of ${\rm su(2)}$ spin $1/2$, there
exists, besides the six vertex and sos model, a third "equivalent" 
model, the {\bf Potts model} \cite{bax}. Its existence and  relations
with the first two  have a precise  meaning in terms of 
Temperley Lieb algebra representation theory. The purpose of this 
paper is to show that for higher spin, there is also  a Potts model 
naturally associated with the vertex and sos models. This model still 
uses the same set of variables $\sigma=1, \ldots, Q$ but involves  
interactions that  are more complicated than the nearest neighbour 
coupling of the  spin $1/2$ case. That such models exist has been 
known in principle for a long time \cite{syo} but their precise 
definition, and the general algebraic formalism to build them, are 
new to the best of our knowledge. While we were working on that
construction, we became aware of the work of Nienhuis \cite{nih}
where the first member of the hierarchy, a Potts model with nearest
and next to nearest neighbors interactions related to spin 1, has 
already been exhibited. However the techniques used in \cite{nih} and 
in our work are very different.

The construction of what we shall call {\bf $\Gamma_{k}$ Potts model} 
is discussed in details in section 3, after elementary reminders 
about $k=1$ are given in section 2. From another  perspective, 
it gives an interesting light to the differences between
integer and half integer vertex models or quantum spin chains (we can
in particular reformulate the spin $k/2$ ${\rm su(2)}$
 spin chains problems in terms
of $Q=4$ states Potts models). For $k$ greater than one, only 
submanifolds of the full parameter space are integrable. In section 4 
we discuss the simplest integrable line for $Q>4$ where a first 
order phase transition takes place. We compute spontaneous 
magnetizations by generalizing Baxter's calculation  for spin $1/2$. 

We discuss in  section 5 the full phase diagram of the 
$k=2$ model, mainly  in terms of the 
corresponding one dimensional quantum spin chain. This phase diagram 
restricts in the case $Q=4$ to the widely studied one for the spin 1
${\rm su(2)}$ hamiltonian. Critical properties are
studied with particular emphasis on the integrable lines. We also 
construct the valence bond states for the $P_{2}$ projector for 
arbitrary $Q$ thereby extending the construction in \cite{aklt} to 
anisotropic systems (with quantum group symmetry). In opposition to 
\cite{aklt}, phase transitions are encountered as $Q$ varies.
 
Two appendices are included. Appendix A contains the explicit 
expression for the Boltzmann weights of the $k=3$ Potts model. 
Various loop model interpretations  for the family of Potts model, in 
particular for $k=2$, are reviewed briefly in appendix B. 

\pagebreak 
  
\section{The $Q$-state Potts model and staggered six-vertex model}
 
\hspace{5 mm}
 
   We begin by reviewing the $Q$-state Potts model on a square lattice. 
We consider a cylinder made of an $l \times t$ rectangular strip 
$\cal L$ with free boundary condition on the top and bottom rows, 
and periodic in the time direction. The partition function of the 
model with horizontal and vertical couplings $K_{1}$ and $K_{2}$ is 
given by\cite{bax}
\begin{eqnarray}
  Z_{\rm Potts} = \sum_{\{\sigma_{i}\}} \prod_{\langle ij \rangle} 
e^{K_{1} \delta_{\sigma_{i}\sigma_{j}} + K_{2} \delta_{\sigma_{i}
\sigma_{j}}}\;,
\end{eqnarray} 
where the product is over all neighboring horizontal and vertical
links $<ij>$ and $\sigma_{i}$ assumes values $1$ to $Q$.
 
     The column to column transfer matrix $\tau_{{\rm Potts}}$ can be 
expressed as a product of local transfer matrices $X_{2i-1}$ and 
$X_{2i}$ which add, respectively, a horizontal and a vertical link to 
the lattice.  We have
\begin{equation}
      \tau_{{\rm Potts}}=Q^{l/2} \prod_{i=1}^{l}X_{2i-1}
\prod_{i=1}^{l}X_{2i}
\end{equation}
and
\begin{equation}
      Z_{{\rm Potts}}={\rm tr}(\tau)^{t}
\end{equation}      
with   
\begin{equation}
\begin{array}{lll} 
      X_{2i-1}&=&x_{1}+e_{2i-1}\;,\\                          
      X_{2i}&=&1+x_{2}e_{2i}\;,         
\end{array}               \label{eq:X}        
\end{equation}                                
where we define
       \[x_{k}=\frac{e^{K_{k}-1}}{Q^{1/2}}\;\;\; {\rm for}\;
\; k=1,2 \]
and  the operators $e_{i}$'s which propagate $\{\sigma_{i}\}$ to
 $\{\sigma'_{i}\}$ in the time direction,  have matrix elements 
given by\cite{tl}
\begin{equation}
\left \{ \begin{array}{rcl} 
       (e_{2i})_{\sigma,\sigma^{'}}&=&Q^{1/2}
\delta_{\sigma_{i}\sigma_{i+1}}\! \prod_{j=1}^{l} 
\delta_{\sigma_{j}\sigma'_{j}}\;,\\            
              & &                   \\
       (e_{2i-1})_{\sigma,\sigma^{'}}&=&Q^{-1/2} \prod_{j\neq i}
 \delta_{\sigma_{j}\sigma'_{j}}\;,                 
\end{array} \right.                                \label{eq:e}       
\end{equation}
They satisfy the following relations (dropping the subscripts 
from now on)
\begin{equation}
\left \{ \begin{array}{rcl}
     e_{i}^{2}&=&Q^{1/2}e_{i}\;,\\          
     e_{i}e_{i\pm 1}e_{i}&=&e_{i}\;,\\            
     \,[e_{i},e_{j}]&=&0\;\;\;\; ; |i-j|\geq 2 \;.    
\end{array} \right.              \label{eq:TL} 
\end{equation}
The algebra generated by them is known as the Temperley Lieb algebra. 
In the Potts representation the following trace properties 
hold\cite{jon}
\begin{equation}
\begin{array}{rcl}
{\rm tr}\left[w(1,e_{1},\ldots,e_{i-1})e_{i}\right] & = & Q^{-1/2}
{\rm tr}\,w(1,e_{1},\ldots,e_{i-1})\;,\\                         
{\rm tr}\left(e_{i}\right) & = & Q^{-1/2}{\rm tr}{\bf 1}\;,
\;\;\; {\rm and}\\
           {\rm tr}{\bf 1} & = & Q^{l}\;, 
\end{array}         \label{eq:tr}                
\end{equation}         
where $w$ is any word in $1, e_{1},\ldots,e_{i-1}$.
   The Potts model exhibits a duality transformation implemented 
by rewriting the local transfer matrices as follows
\begin{equation}
\begin{array}{lll} 
      X_{2i-1}&=&x_{1}(1+x_{1}^{-1}e_{2i-1})\;,\\                          
      X_{2i}&=&x_{2}(x_{2}^{-1}+e_{2i})\;,         
\end{array}               \label{eq:Xp}        
\end{equation}   
and interchanging the roles of $e_{2i-1}$ and $e_{2i}$;
\begin{equation}
\left \{ \begin{array}{rcl}        
      \left(e_{2i}\right)_{\sigma,\sigma'}&=&Q^{-1/2} 
\prod_{j\neq i} \delta_{\sigma_{j}\sigma'_{j}}\;,\\                             
                                 \\
      \left(e_{2i-1}\right)_{\sigma,\sigma'}
&=&Q^{1/2}\delta_{\sigma_{i}\sigma_{i+1}}\! \prod_{j=1}^{l}
 \delta_{\sigma_{j}\sigma'_{j}}\;,                                             
\end{array} \right.                                \label{eq:ep}       
\end{equation}
which amounts to redefining $X_{2i-1}$ and $X_{2i}$ to be the local 
operators that add a vertical and a horizontal link to the Potts 
lattice respectively. This alternative interpretation of the roles 
of $e_{2i-1}$ and $e_{2i}$ does not alter the algebraic relations 
satisfied by them . We thus have
\begin{equation}
Z_{{\rm Potts}}(x_{1},x_{2})=(x_{1}x_{2})^{lt}Z_{\rm Potts}^{'}
(x_{2}^{-1},x_{1}^{-1})
\end{equation}
where the prime denotes the dual lattice ${\cal L}^{'}$ which differs 
from ${\cal L}$ by boundary effects only. (see fig.(1)) 

\begin{center}  
\setlength{\unitlength}{0.006in}
\begin{picture}(345,218)(0,-10)
\put(126.000,180.000){\arc{10.000}{4.7124}{7.8540}}
\put(6.000,180.000){\arc{10.000}{1.5708}{4.7124}}
\put(6.000,140.000){\arc{10.000}{1.5708}{4.7124}}
\put(126.000,140.000){\arc{10.000}{4.7124}{7.8540}}
\put(126.000,100.000){\arc{10.000}{4.7124}{7.8540}}
\put(6.000,100.000){\arc{10.000}{1.5708}{4.7124}}
\put(326.000,150.000){\arc{10.000}{4.7124}{7.8540}}
\put(206.000,150.000){\arc{10.000}{1.5708}{4.7124}}
\put(206.000,110.000){\arc{10.000}{1.5708}{4.7124}}
\put(326.000,110.000){\arc{10.000}{4.7124}{7.8540}}
\put(326.000,70.000){\arc{10.000}{4.7124}{7.8540}}
\put(206.000,70.000){\arc{10.000}{1.5708}{4.7124}}
\dashline{4.000}(6,185)(126,185)
\path(6,175)(126,175)
\path(6,135)(126,135)
\dashline{4.000}(6,145)(126,145)
\dashline{4.000}(6,105)(126,105)
\path(6,95)(126,95)
\path(26,175)(26,55)
\path(66,175)(66,55)
\path(106,175)(106,55)
\dashline{4.000}(206,155)(326,155)
\path(206,145)(326,145)
\path(206,105)(326,105)
\dashline{4.000}(206,115)(326,115)
\dashline{4.000}(206,75)(326,75)
\path(206,65)(326,65)
\path(226,185)(226,65)
\path(266,185)(266,65)
\path(306,185)(306,65)
\put(-20,-10){\makebox(0,0)[lb]{\raisebox{0pt}[0pt][0pt]
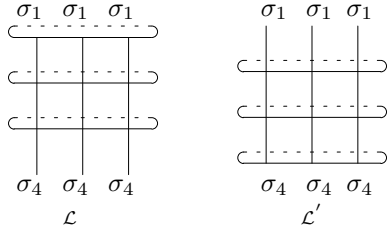
{\shortstack[l]{\footnotesize {\bf Figure(1)} Geometry of the lattice and 
its dual}}}}
\put(256,12){\makebox(0,0)[lb]{\raisebox{0pt}[0pt][0pt]
{\shortstack[l]{${\scriptstyle {\cal L}^{'}}$}}}}
\put(48,12){\makebox(0,0)[lb]{\raisebox{0pt}[0pt][0pt]
{\shortstack[l]{${\scriptstyle {\cal L}}$}}}}
\put(261,41){\makebox(0,0)[lb]{\raisebox{0pt}[0pt][0pt]
{\shortstack[l]{$\sigma_{4}$}}}}
\put(296,41){\makebox(0,0)[lb]{\raisebox{0pt}[0pt][0pt]
{\shortstack[l]{$\sigma_{4}$}}}}
\put(221,41){\makebox(0,0)[lb]{\raisebox{0pt}[0pt][0pt]
{\shortstack[l]{$\sigma_{4}$}}}}
\put(8,41){\makebox(0,0)[lb]{\raisebox{0pt}[0pt][0pt]
{\shortstack[l]{$\sigma_{4}$}}}}
\put(48,41){\makebox(0,0)[lb]{\raisebox{0pt}[0pt][0pt]
{\shortstack[l]{$\sigma_{4}$}}}}
\put(88,41){\makebox(0,0)[lb]{\raisebox{0pt}[0pt][0pt]
{\shortstack[l]{$\sigma_{4}$}}}}
\put(296,194){\makebox(0,0)[lb]{\raisebox{0pt}[0pt][0pt]
{\shortstack[l]{$\sigma_{1}$}}}}
\put(261,194){\makebox(0,0)[lb]{\raisebox{0pt}[0pt][0pt]
{\shortstack[l]{$\sigma_{1}$}}}}
\put(221,194){\makebox(0,0)[lb]{\raisebox{0pt}[0pt][0pt]
{\shortstack[l]{$\sigma_{1}$}}}}
\put(8,194){\makebox(0,0)[lb]{\raisebox{0pt}[0pt][0pt]
{\shortstack[l]{$\sigma_{1}$}}}}
\put(48,194){\makebox(0,0)[lb]{\raisebox{0pt}[0pt][0pt]
{\shortstack[l]{$\sigma_{1}$}}}}
\put(88,194){\makebox(0,0)[lb]{\raisebox{0pt}[0pt][0pt]
{\shortstack[l]{$\sigma_{1}$}}}}
\end{picture}
\end{center}
   
The duality transformation
\begin{equation}
x_{1}  \longleftrightarrow  x_{2}^{-1}    \label{eq:map}
\end{equation}               
relates high temperature to low temperature phase. Ignoring the 
difference in the boundary, the model is self-dual at 
\begin{equation}  
x_{1}x_{2}=1\;,
\end{equation}
and by standard argument\cite{bax} this is a line of phase transition 
(first order fo $Q>4$, and second order for $Q\leq4$).
                                                                    
It is well known that the model can be mapped to the six-vertex
( referred to as  $\Gamma_{1}$ here ) model by 
assigning arrows on the surrounding polygons of the clusters formed 
by Potts variables $\sigma_{i}$ that have the same colors
\cite{bax,lib} in the high temperature expansion. The procedure is 
more transparent from an algebraic point of view. Consider the tensor 
product of $2l$ copies of spin-$\frac{1}{2}$ representations of 
${\rm U}_{q}{\rm sl}(2)$, then $e_{i}$ can be represented in this 
vector space as
\begin{equation}
    e_{i}={\bf 1}\otimes \cdots \otimes \left(\begin{array}{cccc}
 0&0&0&0\\
 0&q^{-1}&-1&0\\  
 0&-1&q&0\\
 0&0&0&0 \end{array}\right) \otimes \cdots \otimes{\bf 1}\;\;\;
 \in {\bf C}^{\otimes 2l}\;,            \label{eq:em}
\end{equation}
where the matrix is proportional to the spin-$0$ projector of the 
$i^{\rm th}$ and $(i+1)^{\rm th}$ spin-$\frac{1}{2}$ representation, 
it is easy to verify that the above indeed satisfies 
$(\!~\ref{eq:TL}\,)$ with
           \[Q^{1/2}=q+q^{-1}\;,\]
and the trace properties $(\!~\ref{eq:tr}\,)$ can be reproduced by 
introducing the Markov trace defined as
\begin{equation} 
\left\{ \begin{array}{rll}  
{\rm tr}\left[q^{2S^{Z}}w(1,e_{1},\ldots,e_{i-1})e_{i}\right]&=&\left
(q+q^{-1}\right)^{-1}{\rm tr}
\left[q^{2S^{Z}}w(1,e_{1},\ldots,e_{i-1})\right]\;,\\
       {\rm tr}\left(q^{2S^{Z}}e_{i}\right)& = &\left(q+
q^{-1}\right)^{-1}{\rm tr}\left(q^{2S^{Z}}{\bf 1}\right)\;,\\            
       {\rm tr}\left(q^{2S^{Z}}{\bf 1}\right)& = &
\left(q+q^{-1}\right)^{2l}\;,
\end{array} \right.  \label{eq:mk}
\end{equation}
where  \[S^{Z}=\sum_{i=1}^{2l}\sigma^{Z}_{i}\]
and 
  \[\sigma_{i}^{Z}={\bf 1}\otimes \cdots \otimes 
\left(\begin{array}{cc}
 \frac{1}{2}&0\\
 0&-\frac{1}{2} \end{array}\right)\otimes \cdots \otimes{\bf 1}\;.\]
The operator $X_{i}$, with $e_{i}$ given as $(\!~\ref{eq:em}\,)$, 
defines the vertex interaction of the staggered six-vertex model, that 
is its matrix elements encode the Boltzmann weights of the vertices 
with various colors for incoming and outgoing lines. The partition 
function is then given by  
\begin{eqnarray}
       Z_{{\rm vertex}}& = &{\rm tr}\left(q^{2S^{Z}}
\tau_{{\rm vertex}}\right)\;,  \label{eq:partition}\\   
             \tau_{{\rm vertex}}& = &\left(q+q^{-1}\right)^{l}
\prod_{i=1}^{l}X_{2i-1}\prod_{i=1}^{l}X_{2i}\;,  \label{eq:6vertex}
\end{eqnarray}
with $e_{i}$'s in the above defined by $(\!~\ref{eq:em}\,)$.
Because of the same trace properties of the two representations, 
$Z_{{\rm vertex}}$ and $Z_{{\rm Potts}}$ are equal for integer $Q$. 
However, the former is defined  as well for Q real (and coincides 
with the geometrical definition based on high temperature expansion).
 The Temperley Lieb generators commute with ${\rm U}_{q}{\rm sl}(2)$,
so the six-vertex model has ${\rm U}_{q}{\rm sl}(2)$ symmetry. The 
mapping between vertex and Potts 
models is not one to one when $q$ is a root of unity due
to  the boundary operator $q^{2S^{Z}}$ in the trace and the 
quantum group symmetry. It is actually one to one 
between the Potts model and only the subset of type II 
representations of ${\rm U}_{q}{\rm sl}(2)$ provided by the 
six-vertex model\cite{pqs,slb}.

\section{Potts model formulation of the $\Gamma_{k}$ vertex model}
 
\subsection{The fusion procedure and mapping to Potts model}     
 
\hspace{5 mm}
    We shall construct a family of Potts models which are related 
to the ${\rm U}_{q}{\rm su}(2)$ invariant 
vertex models based on  spin-$\frac{k}{2}$. We call the latter
\mbox{\boldmath $\Gamma_{k}$} {\bf vertex models}. This name
 is non standard, and we do not know any more standard name to use
instead (in \cite{fran} $\Gamma_{k}$ refers to the number of allowed 
vertices, 6 for $k=1$, 19
for $k=2,\ldots$). For the moment we therefore decide only 
about the  symmetry, not the particular set of interactions. 
The construction uses then ideas of the  
 "fusion procedure" \cite{krs} (properly generalized) 
to  reexpress each 
 ${\rm U}_{q}{\rm su}(2)$ spin-$k/2$ 
in terms of $k$ copies of spin-$1/2$.
A pairs of such $1/2$ spins interact at vertices of   
the six-vertex model ($\Gamma_{1}$ vertex model). 
Each such vertex is then translated into a  Potts model interaction  
using the results of section 2. This provides finally a
Potts model with complicated inhomogeneous interactions 
which  we  call   the 
\mbox{\boldmath $\Gamma_{k}$} {\bf Potts model}.

This construction is best explained by explicit computation. First, 
the Boltzmann weights  of a particular vertex with two incoming and 
two outgoing legs  carrying ${\rm U}_{q}{\rm su}(2)$ 
spin-$k/2$ variables are encoded in a matrix which we call for
simplicity the $\Gamma_{k}$ vertex as well.  It is an operator that 
acts on ${\bf C}^{k+1}\otimes{\bf C}^{k+1}$. The corresponding    
spin-$\frac{k}{2}$ irreducible representations  
can be obtained from the $q$-symmetric tensor product of $k$ copies 
of the spin-$\frac{1}{2}$ one. This is conveniently done using the  
$q$-symmetrizer defined by\cite{jim}
\begin{equation}
       S_{k}=\frac{q^{k(k-1)/2}}{[k]_{q}!}\sum_{\sigma}q^{-|I_{\sigma}|}
\prod_{i\in I_{\sigma}}s(i)\;,
\end{equation}
where
      \[s(i)=q^{-1}{\bf 1}-e_{i}\;.\]
In the above formula, $I_{\sigma}$ denotes the collections of indices in 
the nonreducible decomposition of the symmetric group element 
$\sigma$ in terms of  transposition of neighbors $\tau_{i,i+1}$, 
$|I_{\sigma}|$ indicates the cardinality of the collections and we 
also introduced the $q$-factorial where 
\begin{eqnarray}
 [n]_{q}!&=&(n)_{q}(n-1)_{q}\ldots(1)_{q}\\
{\rm and }\;\;\; (n)_{q}&=&\frac{q^{n}-q^{-n}}{q-q^{-1}}\;.\nonumber
\end{eqnarray}
The symmetrizer can alternatively be written recursively 
as\cite{saa,wen} 
\begin{equation}
 S_{k}=S_{k-1}\left(1-\frac{(k-1)_{q}}{(k)_{q}}e_{k-1}\right)
S_{k-1}\;,
\label{eq:sym}
\end{equation}
which expresses $S_{k}$ solely in terms of products of $\Gamma_{1}$ 
vertices.

The $\Gamma_{k}$ vertex written in terms of  $\Gamma_{1}$
reads then
\begin{equation}
    S_{k}S_{k}[r_{k}(u_{1})r_{k+1}(u_{2})\cdots r_{2k-1}(u_{k})]
  [r_{k-1}(u_{k+1})\cdots r_{2k-2}(u_{2k})]\cdots[r_{1}(u_{k^{2}-k+1})\cdots 
   r_{k}(u_{k^{2}})]S_{k}S_{k}\;,
                   \label{eq:Xu}   
\end{equation}                                          
where the first and last two $S_{k}$'s act respectively on the two 
in- and out- states, and we encoded the  $\Gamma_{1}$  weights in 
the matrix
\begin{equation}
   r_{i}(u_{j})={\bf 1}+\frac{\sin u_{j}}{\sin (\gamma-u_{j})}e_{i} 
\label{eq:r}   
\end{equation}          
with $q=e^{i\gamma}$. Note that $r_{i}(u_{j})$ is identical with 
 the local operator
 $X_{i}$ introduced in the previous section 
. The construction $(\!~\ref{eq:Xu}\,)$ is illustrated
graphically in fig.(2), which shows that the interaction between the 
two spin-$\frac{k}{2}$ states is replaced by the interactions 
$r_{k}(u_{1}),\cdots,r_{k}(u_{k^{2}})$ between $2k$
spin-$\frac{1}{2}$ states.
It is not difficult to see that the operator $(\!~\ref{eq:Xu}\,)$ has 
${\rm U}_{q}{\rm su}(2)$ symmetry since each of the $\Gamma_{1}$ 
vertices regarded as an operator in ${\bf C}^{k+1}\otimes
{\bf C}^{k+1}$ also has ${\rm U}_{q}{\rm su}(2)$ symmetry. This 
together with the fact that $S_{k}$'s project onto  the spin-
$\frac{k}{2}$ irreducible representation, implies that the operator 
$(\!~\ref{eq:Xu}\,)$ is indeed a  $\Gamma_{k}$ vertex . It can 
therefore be written as a linear combination of the projectors 
$P_{j}\;,j=0,1,\cdots,k$. Conversely it can be shown ( see also 
appendix B) that any $\Gamma_{k}$ vertex weight can be written as 
the above using a particular set of  $u_{i}$'s. The number of 
independent parameters for the $\Gamma_{k}$ vertex is equal to $k$ 
(we factored out  an irrelevant overall scale), which is much less 
than the number of   $u_{i}$'s.  But these are the most convenient 
parameters.

\begin{center}  
\setlength{\unitlength}{0.01in}
\begin{picture}(475,344)(0,-25)
\path(163,168)(190,137)
\dashline{4.000}(136,197)(163,168)
\dashline{4.000}(81,137)(108,108)
\path(27,197)(81,137)
\path(108,108)(136,79)
\path(95,93)(122,64)
\path(14,183)(68,123)
\dashline{4.000}(68,123)(95,93)
\path(81,79)(108,50)
\path(0,168)(54,108)
\dashline{4.000}(54,108)(81,79)
\dashline{4.000}(136,108)(163,137)
\path(163,137)(190,168)
\path(81,50)(136,108)
\dashline{4.000}(81,168)(108,197)
\path(108,197)(136,227)
\path(27,108)(81,168)
\path(41,93)(95,152)
\path(122,183)(150,212)
\dashline{4.000}(95,152)(122,183)
\path(14,123)(68,183)
\path(0,137)(54,197)
\path(95,212)(122,241)
\path(81,227)(108,256)
\dashline{4.000}(68,183)(95,212)
\dashline{4.000}(54,197)(81,227)
\path(81,256)(136,197)
\thicklines
\path(153,263)(234,263)
\path(218.000,259.000)(234.000,263.000)(218.000,267.000)
\thinlines
\path(312,92)(271,48)(258,63)
\path(325,77)(298,48)(271,77)
\path(339,63)(325,48)(285,92)
\dashline{4.000}(339,121)(312,92)
\dashline{4.000}(352,107)(325,77)
\dashline{4.000}(366,92)(339,63)
\path(407,136)(366,92)
\path(394,150)(352,107)
\path(380,165)(339,121)
\path(461,77)(434,48)(421,63)
\path(394,92)(339,150)
\path(475,63)(461,48)(434,77)
\path(407,107)(352,165)
\dashline{4.000}(421,63)(394,92)
\dashline{4.000}(434,77)(407,107)
\dashline{4.000}(434,268)(407,239)
\dashline{4.000}(421,283)(394,253)
\path(407,239)(352,180)
\path(475,283)(461,297)(434,268)
\path(394,253)(339,195)
\path(461,268)(434,297)(421,283)
\path(380,180)(339,224)
\path(394,195)(352,239)
\path(407,210)(366,253)
\dashline{4.000}(366,253)(339,283)
\dashline{4.000}(352,239)(325,268)
\dashline{4.000}(339,224)(312,253)
\path(339,283)(325,297)(285,253)
\path(325,268)(298,297)(271,268)
\path(312,253)(271,297)(258,283)
\put(0,0){\makebox(0,0)[lb]{\raisebox{0pt}[0pt][0pt]
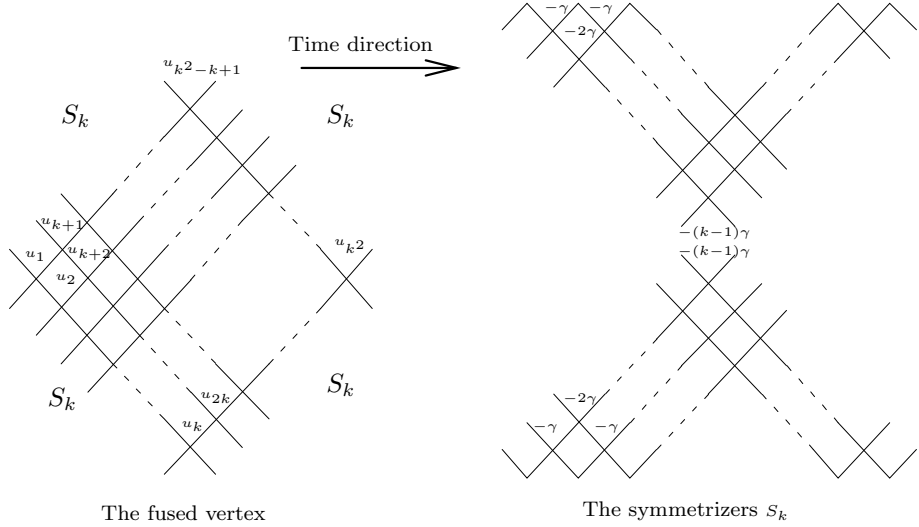
{\shortstack[l]{\footnotesize {\bf Figure(2)} The fused $\Gamma_{k}$ vertex; 
The parameters $u_{1},\cdots,u_{k^{2}}$, etc appearing in the figure}}}}
\put(60,-12){\makebox(0,0)[lb]{\raisebox{0pt}[0pt][0pt]
{\shortstack[l]{\footnotesize are associated with the $\Gamma_{1}$ vertices 
that make up the $\Gamma_{k}$ vertex. The four $S_{k}$'s that}}}} 
\put(60,-24){\makebox(0,0)[lb]{\raisebox{0pt}[0pt][0pt]
{\shortstack[l]{\footnotesize  act on the in- and 
out- states are shown on the rhs.      }}}}
\put(16,181){\makebox(0,0)[lb]{\raisebox{0pt}[0pt][0pt]
{\shortstack[l]{${\scriptscriptstyle u_{k+1}}$}}}}
\put(80,264){\makebox(0,0)[lb]{\raisebox{0pt}[0pt][0pt]
{\shortstack[l]{${\scriptscriptstyle u_{k^{2}-k+1}}$}}}}
\put(24,152){\makebox(0,0)[lb]{\raisebox{0pt}[0pt][0pt]
{\shortstack[l]{${\scriptscriptstyle u_{2}}$}}}}
\put(30,166){\makebox(0,0)[lb]{\raisebox{0pt}[0pt][0pt]
{\shortstack[l]{${\scriptscriptstyle u_{k+2}}$}}}}
\put(90,75){\makebox(0,0)[lb]{\raisebox{0pt}[0pt][0pt]
{\shortstack[l]{${\scriptscriptstyle u_{k}}$}}}}
\put(100,90){\makebox(0,0)[lb]{\raisebox{0pt}[0pt][0pt]
{\shortstack[l]{${\scriptscriptstyle u_{2k}}$}}}}
\put(170,170){\makebox(0,0)[lb]{\raisebox{0pt}[0pt][0pt]
{\shortstack[l]{${\scriptscriptstyle u_{k^{2}}}$}}}}
\put(146,272){\makebox(0,0)[lb]{\raisebox{0pt}[0pt][0pt]
{\shortstack[l]{\footnotesize Time direction}}}}
\put(27,234){\makebox(0,0)[lb]{\raisebox{0pt}[0pt][0pt]
{\shortstack[l]{$S_{k}$}}}}
\put(166,234){\makebox(0,0)[lb]{\raisebox{0pt}[0pt][0pt]
{\shortstack[l]{$S_{k}$}}}}
\put(166,93){\makebox(0,0)[lb]{\raisebox{0pt}[0pt][0pt]
{\shortstack[l]{$S_{k}$}}}}
\put(20,86){\makebox(0,0)[lb]{\raisebox{0pt}[0pt][0pt]
{\shortstack[l]{$S_{k}$}}}}
\put(48,26){\makebox(0,0)[lb]{\raisebox{0pt}[0pt][0pt]
{\shortstack[l]{\footnotesize The fused vertex}}}}
\put(300,28){\makebox(0,0)[lb]{\raisebox{0pt}[0pt][0pt]
{\shortstack[l]{\footnotesize The symmetrizers ${\scriptstyle S_{k}}$}}}}
\put(8,164){\makebox(0,0)[lb]{\raisebox{0pt}[0pt][0pt]
{\shortstack[l]{${\scriptscriptstyle u_{1}}$}}}}
\put(350,165){\makebox(0,0)[lb]{\raisebox{0pt}[0pt][0pt]
{\shortstack[l]{${\scriptscriptstyle -(k-1)\gamma}$}}}}
\put(306,73){\makebox(0,0)[lb]{\raisebox{0pt}[0pt][0pt]
{\shortstack[l]{${\scriptscriptstyle -\gamma}$}}}}
\put(290,88){\makebox(0,0)[lb]{\raisebox{0pt}[0pt][0pt]
{\shortstack[l]{${\scriptscriptstyle -2\gamma}$}}}}
\put(274,73){\makebox(0,0)[lb]{\raisebox{0pt}[0pt][0pt]
{\shortstack[l]{${\scriptscriptstyle -\gamma}$}}}}
\put(350,175){\makebox(0,0)[lb]{\raisebox{0pt}[0pt][0pt]
{\shortstack[l]{${\scriptscriptstyle -(k-1)\gamma}$}}}}
\put(290,280){\makebox(0,0)[lb]{\raisebox{0pt}[0pt][0pt]
{\shortstack[l]{${\scriptscriptstyle -2\gamma}$}}}}
\put(303,293){\makebox(0,0)[lb]{\raisebox{0pt}[0pt][0pt]
{\shortstack[l]{${\scriptscriptstyle -\gamma}$}}}}
\put(280,293){\makebox(0,0)[lb]{\raisebox{0pt}[0pt][0pt]
{\shortstack[l]{${\scriptscriptstyle -\gamma}$}}}}
\end{picture}
\end{center}  

Next, having written the $\Gamma_{k}$ vertex in terms of $\Gamma_{1}$
 vertices and thus the Temperley-Lieb generators, we can use the 
various realizations of  the Temperley-Lieb algebra. In particular, 
we shall consider the Potts model realization, which produces the 
$\Gamma_{k}$ Potts model we set out to construct. The Potts model 
lattice corresponding to the $\Gamma_{k}$ vertex is built up as 
follows: To each $\Gamma_{1}$ vertex $r_{i}$ in $(\!~\ref{eq:Xu}\,)$,
 we substitute the expression $(\!~\ref{eq:e}\,)$ or 
$(\!~\ref{eq:ep}\,)$, and call the operator $(\!~\ref{eq:Xu}\,)$ 
in this representation \mbox{\boldmath $W(u)$}, and the matrix 
elements of $W(u)$ induced by the in- and out- states of the vertex 
will become the Boltzmann weights of the $\Gamma_{k}$ Potts model. 
Graphically we assign either a horizontal or vertical link to each 
spin-$\frac{1}{2}$ vertex $r_{j}$ as shown in fig.(3), $W(u)$ 
therefore corresponds to an operator that acts on $k$ or $k+1$ 
{\bf Potts variables}, and is represented graphically as a connected 
collection of horizontal and vertical links, which shall be referred 
to as the {\bf fundamental block} \mbox{\boldmath ${\cal G}_{k}$} 
hereafter. The lattice ${\cal L}$ is then constructed by replacing 
all the $\Gamma_{k}$ vertices by these fundamental blocks. we shall 
elaborate on this point in next few sections. Note that there are 
exactly two possible mappings to the Potts model links which
originate from the two possible choices of assigning links to the 
generator, namely $(\!~\ref{eq:e}\,)$ and $(\!~\ref{eq:ep}\,)$. 
When $(\!~\ref{eq:e}\,)$ is used, a horizontal ( vertical ) link is 
associated to $r_{j}$ with odd (even) subscript, whereas when 
$(\!~\ref{eq:ep}\,)$ is used, the roles of vertical and horizontal 
links are reversed. We shall denote these two choices of mapping
as {\bf convention A and B} respectively. It will be shown that 
they are related by duality transformation which is inherited from 
that of the $\Gamma_{1}$ Potts model. The relation between the
$\Gamma_{k}$ vertex and Potts models generalizes that of the $k=1$ 
case. In particular, for the vertex model whose lattice has the 
geometry of a cylinder with periodic time boundary condition, the 
partition function defined as in $(\!~\ref{eq:partition}\,)$ is 
equal to that of the corresponding Potts model when $Q$ is an 
integer. The Markov trace in $(\!~\ref{eq:partition}\,)$ again 
restricts the domain of the mapping  to  type II representations   
of ${\rm U}_{q}{\rm su(2)}$ provided by the vertex model.

\begin{center}  
\setlength{\unitlength}{0.008in}
\begin{picture}(585,260)(0,-50)
\put(293,145){\circle*{6}}
\put(333,145){\circle*{6}}
\put(333,105){\circle*{6}}
\put(292,105){\circle*{6}}
\put(493,165){\circle*{6}}
\put(533,165){\circle*{6}}
\put(492,125){\circle*{6}}
\put(532,125){\circle*{6}}
\path(135,115)(225,115)
\path(217.000,113.000)(225.000,115.000)(217.000,117.000)
\path(55,165)(155,65)
\path(95,165)(35,105)
\path(115,145)(35,65)
\dashline{4.000}(35,105)(0,70)
\dashline{4.000}(35,65)(15,45)
\dashline{4.000}(115,185)(95,165)
\dashline{4.000}(135,165)(115,145)
\dashline{4.000}(155,65)(165,55)
\dashline{4.000}(55,165)(35,185)
\path(55,165)(155,65)
\path(95,165)(35,105)
\path(115,145)(35,65)
\dashline{4.000}(35,105)(0,70)
\dashline{4.000}(35,65)(15,45)
\dashline{4.000}(115,185)(95,165)
\dashline{4.000}(135,165)(115,145)
\dashline{4.000}(155,65)(165,55)
\dashline{4.000}(55,165)(35,185)
\dottedline{5}(335,165)(275,105)
\dottedline{5}(355,145)(275,65)
\dottedline{5}(295,165)(395,65)
\dashline{4.000}(295,165)(275,185)
\dashline{4.000}(395,65)(405,55)
\dashline{4.000}(375,165)(355,145)
\dashline{4.000}(355,185)(335,165)
\dashline{4.000}(275,65)(255,45)
\dashline{4.000}(275,105)(240,70)
\dottedline{5}(335,165)(275,105)
\dottedline{5}(355,145)(275,65)
\dottedline{5}(295,165)(395,65)
\dashline{4.000}(295,165)(275,185)
\dashline{4.000}(395,65)(405,55)
\dashline{4.000}(375,165)(355,145)
\dashline{4.000}(355,185)(335,165)
\dashline{4.000}(275,65)(255,45)
\dashline{4.000}(275,105)(240,70)
\thicklines
\path(255,145)(365,145)
\path(255,105)(365,105)
\path(295,185)(295,65)
\path(335,185)(335,65)
\thinlines
\dashline{4.000}(455,105)(420,70)
\dashline{4.000}(455,65)(435,45)
\dashline{4.000}(535,185)(515,165)
\dashline{4.000}(555,165)(535,145)
\dashline{4.000}(575,65)(585,55)
\dashline{4.000}(475,165)(455,185)
\dottedline{5}(475,165)(575,65)
\dottedline{5}(535,145)(455,65)
\dottedline{5}(515,165)(455,105)
\dashline{4.000}(455,105)(420,70)
\dashline{4.000}(455,65)(435,45)
\dashline{4.000}(535,185)(515,165)
\dashline{4.000}(555,165)(535,145)
\dashline{4.000}(575,65)(585,55)
\dashline{4.000}(475,165)(455,185)
\dottedline{5}(475,165)(575,65)
\dottedline{5}(535,145)(455,65)
\dottedline{5}(515,165)(455,105)
\thicklines
\path(445,165)(560,165)
\path(495,185)(495,65)
\path(445,125)(560,125)
\path(535,185)(535,65)
\put(0,0){\makebox(0,0)[lb]{\raisebox{0pt}[0pt][0pt]
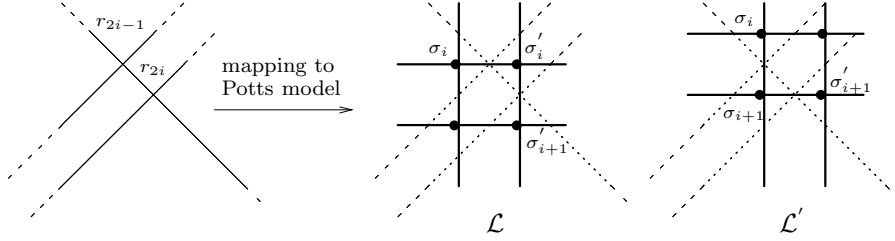
{\shortstack[l]{\footnotesize {\bf Figure(3)} Mapping of the $\Gamma_{1}$ 
vertices to Potts model links. In the first figure on the rhs, }}}}
\put(60,-15){\makebox(0,0)[lb]{\raisebox{0pt}[0pt][0pt]
{\shortstack[l]{\footnotesize $r_{2i-1} ( r_{2i} )$ is replaced by the 
horizontal ( vertical ) Potts model links that
connects  }}}}
\put(60,-30){\makebox(0,0)[lb]{\raisebox{0pt}[0pt][0pt]
{\shortstack[l]{\footnotesize $\sigma_{i}$ and $\sigma_{i}^{'}$ 
( $\sigma_{i}^{'}$ and $\sigma_{i+1}^{'}$ ), the dual Potts model is given 
by the second figure on}}}}
\put(60,-45){\makebox(0,0)[lb]{\raisebox{0pt}[0pt][0pt]
{\shortstack[l]{\footnotesize  the rhs where the above vertex is mapped to 
a vertical ( horizontal ) links }}}}
\put(60,-60){\makebox(0,0)[lb]{\raisebox{0pt}[0pt][0pt]
{\shortstack[l]{\footnotesize connecting $\sigma_{i}$ and $\sigma_{i+1}$ 
( $\sigma_{i+1}$ and $\sigma_{i+1}^{'}$ ) }}}}
\put(95,140){\makebox(0,0)[b]{\raisebox{0pt}[0pt][0pt]
{\shortstack{${\scriptstyle r_{2i}}$}}}}
\put(74,170){\makebox(0,0)[b]{\raisebox{0pt}[0pt][0pt]
{\shortstack{${\scriptstyle r_{2i-1}}$}}}}
\put(140,140){\makebox(0,0)[lb]{\raisebox{0pt}[0pt][0pt]
{\shortstack[l]{\footnotesize mapping to}}}}
\put(140,125){\makebox(0,0)[lb]{\raisebox{0pt}[0pt][0pt]
{\shortstack[l]{\footnotesize Potts model}}}}
\put(282,152){\makebox(0,0)[b]{\raisebox{0pt}[0pt][0pt]
{\shortstack{${\scriptstyle \sigma_{i}}$}}}}
\put(345,152){\makebox(0,0)[b]{\raisebox{0pt}[0pt][0pt]
{\shortstack{${\scriptstyle \sigma_{i}^{'}}$}}}}
\put(339,90){\makebox(0,0)[lb]{\raisebox{0pt}[0pt][0pt]
{\shortstack[l]{${\scriptstyle \sigma_{i+1}^{'}}$}}}}
\put(313,35){\makebox(0,0)[lb]{\raisebox{0pt}[0pt][0pt]
{\shortstack[l]{${\cal L}$}}}}
\put(482,172){\makebox(0,0)[b]{\raisebox{0pt}[0pt][0pt]
{\shortstack{${\scriptstyle \sigma_{i}}$}}}}
\put(482,110){\makebox(0,0)[b]{\raisebox{0pt}[0pt][0pt]
{\shortstack{${\scriptstyle \sigma_{i+1}}$}}}}
\put(552,130){\makebox(0,0)[b]{\raisebox{0pt}[0pt][0pt]
{\shortstack{${\scriptstyle \sigma_{i+1}^{'}}$}}}}
\put(505,35){\makebox(0,0)[lb]{\raisebox{0pt}[0pt][0pt]
{\shortstack[l]{${\cal L}^{'}$}}}}
\end{picture}
\end{center}  

\subsection{The local interaction and its dual}
 
\hspace{5 mm}
 
    We shall first work out explicitly the weight $W(u)$ for $k=2$ 
and $3$, and construct graphically the Potts model representation 
of $W(u)$, ie. the fundamental blocks ${\cal G}_{2}$ and 
${\cal G}_{3}$. The construction of the fundamental block is then 
generalized to arbitrary $k$.
 
For $k=2$, the vertex is 
\begin{equation}
        r_{1}(-\gamma)r_{3}(-\gamma)r_{2}(u_{1})r_{3}(u_{2})
  r_{1}(u_{3})r_{2}(u_{4})r_{1}(-\gamma)r_{3}(-\gamma)\;,            
       \label{eq:Xu2}                                    
\end{equation}          
where $r_{1}(-\gamma)$ is the symmetrizer $S_2$. Substituting 
$(\!~\ref{eq:r}\,)$ into the above, we obtain a sum of products of
Temperley-Lieb generators $e_{i}\;,i=1,2,3$. Mapping to Potts model 
is done by replacing $e_{i}$'s with $(\!~\ref{eq:e}\,)$ or 
$(\!~\ref{eq:ep}\,)$, ie. convention A or B (see fi.(4)).

With convention A, the corresponding operator in the Potts language 
has matrix elements
\begin{eqnarray}
      W_{abcd}(u)& = &\sum_{\alpha,\beta,\gamma,\delta}S_{a\alpha}
S_{b\beta}
\left(1+Q^{1/2}x_{1}\delta_{\alpha\beta}\right)
\left(Q^{-1/2}x_{2}+\delta_{\alpha\gamma}\right)  \label{eq:w2}\\            
                 &   &
\left(Q^{-1/2}x_{3}+\delta_{\beta\delta}\right)       
\left(1+Q^{1/2}x_{4}\delta_{\gamma\delta}\right)
S_{\gamma c}S_{\delta d}\;,   \nonumber
\end{eqnarray}  
where
$a,b,\ldots$ are Potts model variables taking values from 1 to $Q$,  
\[   x_{i}=\frac{\sin u_{i}}{\sin (\gamma-u_{i})}   \]
and
\[S_{\alpha\beta}=-Q^{-1}+\delta_{\alpha \beta}\]
is the symmetrizer $S_{2}$ in the Potts model language. Using the 
fact that
\[\sum_{\beta}S_{\alpha\beta}=0\;,\]
the weight $W$ can be simplified as 
\begin{equation}
W_{abcd}(f_{0},f_{1})=S_{ac}S_{bd}+f_{0}S_{ab}S_{cd}+
\sum_{\alpha}Q^{1/2}f_{1}S_{a\alpha}S_{b\alpha}S_{\alpha c}
S_{\alpha d}   
\end{equation}
with   
\begin{equation}
\begin{array}{lll}
     f_{0}& = &x_{1}x_{2}x_{3}x_{4}\;,\\
     f_{1}& = &x_{1}+x_{4}+x_{1}x_{4}\left(Q^{1/2}+x_{2}+
x_{3}\right)\;.
\end{array}         \label{eq:f}                                      
\end{equation}                               
The three terms in the above expression have origin in the 
geometrical description of the $\Gamma_{2}$ model\cite{sal1}. They 
correspond to the weights of the strand configurations 
shown in fig.(B1)( see appendix B ). More precisely, the algebra 
generated by the corresponding operators is identical to that 
obtained from the strand configurations with algebraic multiplication
 given by appending one picture to another. Further, they are 
related to the projectors $P_{i}$ for $i=0,1,2$ as
\begin{equation}
\begin{array}{rcl}
({\bf 1})_{abcd}&=&S_{ac}S_{bd}\\
(Q-1)(P_{0})_{abcd}&=&S_{ab}S_{cd}\\
{\rm and \;\;} (Q-2)(P_{1})_{abcd}&=&Q\sum_{\alpha}S_{a\alpha}
S_{b\alpha}S_{\alpha c}S_{\alpha d}-S_{ab}S_{cd}\;.    
\end{array}
\end{equation}
Therefore we can write
\begin{equation}
  W_{abcd}(f_{0},f_{1})=({\bf 1})_{abcd}+(f_{0}+Q^{-1/2}f_{1})(Q-1)
(P_{0})_{abcd}+Q^{-1/2}f_{1}(Q-2)(P_{1})_{abcd}     \label{eq:pro}
\end{equation}
 The weight can
also be rewritten in the more physical fashion 
\begin{equation}
 W_{abcd}(f_{0},f_{1})=\exp\left[{\cal K}_{0}(Q-1)P_{0}+
{\cal K}_{1}(Q-2)P_{1}\right]
\end{equation} 
with
\begin{equation}
\begin{array}{rcl}
\exp({\cal K}_{0}(Q-1))-1&=&(f_{0}+Q^{-1/2}f_{1})(Q-1)\;,\\
\exp({\cal K}_{1}(Q-2))-1&=&Q^{-1/2}f_{1}(Q-2)\;.
\end{array}    \label{eq:coupling}
\end{equation}
This  shows that the Potts model we have built is physical for
\[\begin{array}{lrll}
&(f_{0}+Q^{-1/2}f_{1})(Q-1)&>&-1\\
\mbox{ and }\;\;\;&Q^{-1/2}f_{1}(Q-2)&>&-1\;,
\end{array}\]    
so in particular for $f_0,f_1>0$ and $Q>2$. It presents 
a mixture of  ferromagnetic and antiferromagnetic interactions that 
will lead to multicritical behaviors (see section 5).
It is also interesting to express $W$ in terms of Kronecker delta 
only. It reads
\begin{eqnarray}
       W_{abcd}(f_{0},f_{1})& = &\frac{1-3Q^{-1/2}f_{1}+f_{0}}{Q^{2}}
+\frac{Q^{-1/2}f_{1}-f_{0}}{Q}
\left(\delta_{ab}+\delta_{cd}\right)+\frac{Q^{-1/2}f_{1}-1}{Q}
\left(\delta_{ac}+\delta_{bd}\right)  \label{eq:weight2}\\       
                    &   &\mbox{}+\frac{Q^{-1/2}f_{1}}{Q}
\left(\delta_{ad}+\delta_{bc}\right)-Q^{-1/2}f_{1}
\left(\delta_{abc}+\delta_{bcd}+\delta_{abd}+\delta_{acd}\right)
\nonumber\\
        &   &\mbox{}+\delta_{ac}\delta_{bd}+f_{0}\delta_{ab}
     \delta_{cd}+Q^{1/2}f_{1}\delta_{abcd}   \nonumber
\end{eqnarray}  
where
\[\delta_{a_{1}\cdots a_{n}}=\left\{\begin{array}{ll}
    1&\;\;\;\mbox{if}\;\;a_{1}=a_{2}=\cdots=a_{n}\;,\\
    0&\;\;\;\mbox{otherwise}\;. 
        \end{array}\right.\]
The procedure of writing $(\!~\ref{eq:Xu2}\,)$ in terms of Potts 
model variables is depicted in fig.(4a) which shows that the local 
interaction $W_{abcd}(f_{0},f_{1})$ in $(\!~\ref{eq:weight2}\,)$ 
corresponds to the fundamental block ${\cal G}_{2}$
which is a square with $a\:b\:c\:d$ located at the four corners. 
Note that the interaction  between the four sites are rather 
complicated as given in $(\!~\ref{eq:weight2}\,)$.

On the other hand, working with convention B, we have the Potts 
model weight
\begin{eqnarray}
  W_{abcd}^{'}(u)& = &\sum_{\alpha}S^{'}_{ab}S^{'}_{bd}
\left(Q^{-1/2}x_{1}+\delta_{b \alpha}\right)   
\left(1+Q^{1/2}x_{2}\delta_{\alpha d}\right) \label{eq:w2'}   \\
&  & \left(1+Q^{1/2}x_{3}\delta_{\alpha a}\right)   
  \left(Q^{-1/2}x_{4}+\delta_{\alpha c}\right)S^{'}_{ac}S^{'}_{dc} 
\nonumber   
\end{eqnarray}
where
 \[S{'}_{ab}=1-\delta_{ab}\]       
is the $q$-symmetrizer, which  in this representation simply 
constrains  neighboring sites to have different colors, in which 
case, it has value one.    
The weight can likewise be written in terms of projectors as 
$(\!~\ref{eq:pro}\,)$ but with the projectors now given by
\begin{equation}
\begin{array}{rcl}
({\bf 1})_{abcd}&=&\delta_{ad} {\cal S}^{'}\\ 
(Q-1)(P_{0})_{abcd}&=&{\cal S}^{'}\\
(Q-2)(P_{1})_{abcd}&=&(1-\delta_{bc}){\cal S}^{'}
\end{array}
\end{equation}
where
\[{\cal S}^{'}=S{'}_{ab}S{'}_{ac}S{'}_{bd}S{'}_{cd}\]
It is then instructive to factor out ${\cal S}^{'}$ in the weight 
leaving
\begin{equation}
W^{'}_{abcd}(f_{0},f_{1})={\cal S}^{'}(Q^{-1/2}f_{1}+\delta_{ad}
+f_{0}\delta_{bc})\;,
\end{equation}
this is in fact the Potts model considered in \cite{nih}. In this 
form the roles of $f_{1}$ and $f_{0}$ are more transparent; $f_{0}$ 
can be perceived as the parameter which controls the anisotropy 
while $f_{1}$ controls the four sites interaction induced by the 
constraint ${\cal S}^{'}$. This constraint imposed by the 
symmetrizers on the sites is however nontrivial as can be seen in 
the expansion
\begin{eqnarray}
 W_{abcd}^{'}(f_{0},f_{1})& = &Q^{-1/2}f_{1}-Q^{-1/2}f_{1}
\left(\delta_{ab}+\delta_{cd}+\delta_{ac}+\delta_{bd}\right)
+\delta_{ad}+f_{0}\delta_{bc} \label{eq:weight2'}  \\   
     &   &\mbox{}+\left(Q^{-1/2}f_{1}-f_{0}\right)\left(\delta_{abc}
+\delta_{bcd}\right)+\left(Q^{-1/2}f_{1}-1\right)\left(\delta_{abd}
+\delta_{acd}\right) \nonumber\\              
      &   &\mbox +Q^{-1/2}f_{1}\left(\delta_{ac}\delta_{bd}
+\delta_{ab}\delta_{cd}\right)+\left(f_{0}-3Q^{-1/2}f_{1}
+1\right)\delta_{abcd}\;.  \nonumber  
\end{eqnarray}           
The mapping to Potts model using convention B is also shown in 
fig.(4b). The weight $(\!~\ref{eq:w2'}\,)$ corresponds in the 
figure to a collection of vertical and horizontal links. Notice that 
the lattice sites on the top (respectively bottom) rows are 
identified since they have no $\Gamma_{1}$ vertex between them, and
 therefore carry the same color $a$ ( or $d$ ). This gives rise 
to the $45^{\rm o}$ rotated square on the rhs of fig (4b.).
 This rotated square with the
variable $\alpha$ summed over is the fundamental block 
${\cal G}_{2}^{'}$ and corresponds to the weight 
$(\!~\ref{eq:weight2'}\,)$.

\begin{center}  
\setlength{\unitlength}{0.008in}
\begin{picture}(641,390)(0,-100)
\put(359.000,94.750){\arc{14.500}{3.9514}{5.4734}}
\put(400.000,94.750){\arc{14.500}{3.9514}{5.4734}}
\put(359.000,32.667){\arc{11.334}{0.4900}{2.6516}}
\put(400.000,32.667){\arc{11.334}{0.4900}{2.6516}}
\put(359.500,256.000){\arc{12.042}{3.8682}{5.5565}}
\put(399.000,254.750){\arc{14.500}{3.9514}{5.4734}}
\put(359.500,191.833){\arc{9.718}{0.3868}{2.7548}}
\put(399.000,192.667){\arc{11.334}{0.4900}{2.6516}}
\put(210.000,115.250){\arc{14.500}{0.8098}{2.3318}}
\put(169.000,115.250){\arc{14.500}{0.8098}{2.3318}}
\put(210.000,175.750){\arc{14.500}{3.9514}{5.4734}}
\put(169.000,175.750){\arc{14.500}{3.9514}{5.4734}}
\path(548,66)(529,48)
\path(539,76)(517,54)
\path(530,80)(514,64)
\path(544,225)(531,212)
\path(541,234)(521,215)
\path(530,234)(521,225)
\path(439,65)(490,65)
\path(482.000,63.000)(490.000,65.000)(482.000,67.000)
\path(439,224)(490,224)
\path(482.000,222.000)(490.000,224.000)(482.000,226.000)
\path(250,145)(300,145)(300,65)(315,65)
\path(307.000,63.000)(315.000,65.000)(307.000,67.000)
\path(250,145)(300,145)(300,225)(315,225)
\path(307.000,223.000)(315.000,225.000)(307.000,227.000)
\path(55,145)(118,145)
\path(110.000,143.000)(118.000,145.000)(110.000,147.000)
\thicklines
\path(508,243)(549,243)(549,203)
        (508,203)(508,243)
\path(529,94)(499,62)
\path(529,32)(499,62)
\path(529,32)(561,62)
\path(529,94)(561,62)
\thinlines
\dashline{4.000}(379,85)(329,35)
\dashline{4.000}(379,85)(430,35)
\dashline{4.000}(339,85)(329,75)
\dashline{4.000}(329,55)(339,45)
\dashline{4.000}(420,45)(430,55)
\dashline{4.000}(329,95)(379,45)
\dashline{4.000}(420,85)(430,75)
\dashline{4.000}(379,45)(430,95)
\dashline{4.000}(339,85)(354,100)
\dashline{4.000}(364,100)(379,85)
\dashline{4.000}(405,100)(420,85)
\dashline{4.000}(379,85)(395,100)
\dashline{4.000}(364,30)(379,45)
\dashline{4.000}(339,45)(354,30)
\dashline{4.000}(405,30)(420,45)
\dashline{4.000}(379,45)(395,30)
\thicklines
\path(339,105)(339,25)
\path(379,105)(379,25)
\path(420,105)(420,25)
\path(339,65)(420,65)
\thinlines
\dashline{4.000}(379,245)(329,195)
\dashline{4.000}(379,245)(429,195)
\dashline{4.000}(339,245)(329,235)
\dashline{4.000}(329,215)(339,205)
\dashline{4.000}(419,205)(429,215)
\dashline{4.000}(329,255)(379,205)
\dashline{4.000}(419,245)(429,235)
\dashline{4.000}(379,205)(429,255)
\dashline{4.000}(339,245)(355,260)
\dashline{4.000}(364,260)(379,245)
\dashline{4.000}(404,260)(419,245)
\dashline{4.000}(379,245)(394,260)
\dashline{4.000}(364,190)(379,205)
\dashline{4.000}(339,205)(355,190)
\dashline{4.000}(404,190)(419,205)
\dashline{4.000}(379,205)(394,190)
\thicklines
\path(319,245)(439,245)
\path(319,205)(439,205)
\path(399,245)(399,205)
\path(359,245)(359,205)
\thinlines
\path(189,125)(205,110)
\path(215,110)(230,125)
\path(149,125)(164,110)
\path(174,110)(189,125)
\path(189,166)(205,181)
\path(215,181)(230,166)
\path(174,181)(189,166)
\path(149,166)(164,181)
\path(189,125)(240,176)
\path(230,166)(240,156)
\path(139,176)(189,125)
\path(230,125)(240,135)
\path(139,135)(149,125)
\path(149,166)(139,156)
\path(189,166)(240,115)
\path(189,166)(139,115)
\thicklines
\path(0,176)(59,117)
\path(0,118)(59,177)
\put(490,0){\makebox(0,0)[lb]{\raisebox{0pt}[0pt][0pt]
{\shortstack[l]{\scriptsize fundamental blocks}}}}
\put(208,163){\makebox(0,0)[lb]{\raisebox{0pt}[0pt][0pt]
{\shortstack[l]{${\scriptstyle u_{4}}$}}}}
\put(185,179){\makebox(0,0)[lb]{\raisebox{0pt}[0pt][0pt]
{\shortstack[l]{${\scriptstyle u_{3}}$}}}}
\put(185,142){\makebox(0,0)[lb]{\raisebox{0pt}[0pt][0pt]
{\shortstack[l]{${\scriptstyle u_{2}}$}}}}
\put(166,161){\makebox(0,0)[lb]{\raisebox{0pt}[0pt][0pt]
{\shortstack[l]{${\scriptstyle u_{1}}$}}}}
\put(220,140){\makebox(0,0)[lb]{\raisebox{0pt}[0pt][0pt]
{\shortstack[l]{${\scriptstyle -\gamma}$}}}}
\put(140,140){\makebox(0,0)[lb]{\raisebox{0pt}[0pt][0pt]
{\shortstack[l]{${\scriptstyle -\gamma}$}}}}
\put(220,180){\makebox(0,0)[lb]{\raisebox{0pt}[0pt][0pt]
{\shortstack[l]{${\scriptstyle -\gamma}$}}}}
\put(140,178){\makebox(0,0)[lb]{\raisebox{0pt}[0pt][0pt]
{\shortstack[l]{${\scriptstyle -\gamma}$}}}}
\put(519,122){\makebox(0,0)[lb]{\raisebox{0pt}[0pt][0pt]
{\shortstack[l]{${\scriptstyle {\cal G}_{2}^{'}}$}}}}
\put(518,182){\makebox(0,0)[lb]{\raisebox{0pt}[0pt][0pt]
{\shortstack[l]{${\scriptstyle {\cal G}_{2}}$}}}}
\put(440,131){\makebox(0,0)[lb]{\raisebox{0pt}[0pt][0pt]
{\shortstack[l]{\scriptsize variables}}}}
\put(440,145){\makebox(0,0)[lb]{\raisebox{0pt}[0pt][0pt]
{\shortstack[l]{\scriptsize internal}}}}
\put(440,159){\makebox(0,0)[lb]{\raisebox{0pt}[0pt][0pt]
{\shortstack[l]{\scriptsize summing}}}}
\put(200,45){\makebox(0,0)[lb]{\raisebox{0pt}[0pt][0pt]
{\shortstack[l]{\scriptsize convention B}}}}
\put(200,236){\makebox(0,0)[lb]{\raisebox{0pt}[0pt][0pt]
{\shortstack[l]{\scriptsize convention A}}}}
\put(320,130){\makebox(0,0)[lb]{\raisebox{0pt}[0pt][0pt]
{\shortstack[l]{\scriptsize Potts model}}}}
\put(320,146){\makebox(0,0)[lb]{\raisebox{0pt}[0pt][0pt]
{\shortstack[l]{\scriptsize mapping to}}}}
\put(159,93){\makebox(0,0)[lb]{\raisebox{0pt}[0pt][0pt]
{\shortstack[l]{\scriptsize fused vertex}}}}
\put(9,96){\makebox(0,0)[lb]{\raisebox{0pt}[0pt][0pt]
{\shortstack[l]{\scriptsize $\Gamma_{2}$ vertex}}}}
%\put(56,130){\makebox(0,0)[lb]{\raisebox{0pt}[0pt][0pt]
%{\shortstack[l]{\scriptsize procedure}}}}
\put(50,0){\makebox(0,0)[lb]{\raisebox{0pt}[0pt][0pt]
{\shortstack[l]{\scriptsize decomposition}}}}
\put(130,126){\makebox(0,0)[lb]{\raisebox{0pt}[0pt][0pt]
{\shortstack[l]{\scriptsize 3}}}}
\put(130,145){\makebox(0,0)[lb]{\raisebox{0pt}[0pt][0pt]
{\shortstack[l]{\scriptsize 2}}}}
\put(105,165){\makebox(0,0)[lb]{\raisebox{0pt}[0pt][0pt]
{\shortstack[l]{\scriptsize $i=$ 1}}}}
\put(398,176){\makebox(0,0)[lb]{\raisebox{0pt}[0pt][0pt]
{\shortstack[l]{${\scriptstyle \delta}$}}}}
\put(358,175){\makebox(0,0)[lb]{\raisebox{0pt}[0pt][0pt]
{\shortstack[l]{${\scriptstyle \beta}$}}}}
\put(399,265){\makebox(0,0)[lb]{\raisebox{0pt}[0pt][0pt]
{\shortstack[l]{${\scriptstyle \gamma}$}}}}
\put(359,265){\makebox(0,0)[lb]{\raisebox{0pt}[0pt][0pt]
{\shortstack[l]{${\scriptstyle \alpha}$}}}}
\put(440,185){\makebox(0,0)[lb]{\raisebox{0pt}[0pt][0pt]
{\shortstack[l]{\footnotesize d}}}}
\put(320,184){\makebox(0,0)[lb]{\raisebox{0pt}[0pt][0pt]
{\shortstack[l]{\footnotesize b}}}}
\put(438,266){\makebox(0,0)[lb]{\raisebox{0pt}[0pt][0pt]
{\shortstack[l]{\footnotesize c}}}}
\put(319,265){\makebox(0,0)[lb]{\raisebox{0pt}[0pt][0pt]
{\shortstack[l]{\footnotesize a}}}}
\put(418,5){\makebox(0,0)[lb]{\raisebox{0pt}[0pt][0pt]
{\shortstack[l]{\footnotesize d}}}}
\put(380,5){\makebox(0,0)[lb]{\raisebox{0pt}[0pt][0pt]
{\shortstack[l]{\footnotesize d}}}}
\put(339,6){\makebox(0,0)[lb]{\raisebox{0pt}[0pt][0pt]
{\shortstack[l]{\footnotesize d}}}}
\put(420,115){\makebox(0,0)[lb]{\raisebox{0pt}[0pt][0pt]
{\shortstack[l]{\footnotesize a}}}}
\put(379,116){\makebox(0,0)[lb]{\raisebox{0pt}[0pt][0pt]
{\shortstack[l]{\footnotesize a}}}}
\put(338,116){\makebox(0,0)[lb]{\raisebox{0pt}[0pt][0pt]
{\shortstack[l]{\footnotesize a}}}}
\put(529,15){\makebox(0,0)[lb]{\raisebox{0pt}[0pt][0pt]
{\shortstack[l]{\footnotesize d}}}}
\put(558,76){\makebox(0,0)[lb]{\raisebox{0pt}[0pt][0pt]
{\shortstack[l]{\footnotesize c}}}}
\put(498,76){\makebox(0,0)[lb]{\raisebox{0pt}[0pt][0pt]
{\shortstack[l]{\footnotesize b}}}}
\put(529,105){\makebox(0,0)[lb]{\raisebox{0pt}[0pt][0pt]
{\shortstack[l]{\footnotesize a}}}}
\put(558,194){\makebox(0,0)[lb]{\raisebox{0pt}[0pt][0pt]
{\shortstack[l]{\footnotesize d}}}}
\put(497,195){\makebox(0,0)[lb]{\raisebox{0pt}[0pt][0pt]
{\shortstack[l]{\footnotesize b}}}}
\put(559,256){\makebox(0,0)[lb]{\raisebox{0pt}[0pt][0pt]
{\shortstack[l]{\footnotesize c}}}}
\put(498,254){\makebox(0,0)[lb]{\raisebox{0pt}[0pt][0pt]
{\shortstack[l]{\footnotesize a}}}}
\put(608,66){\makebox(0,0)[lb]{\raisebox{0pt}[0pt][0pt]
{\shortstack[l]{\footnotesize {\bf 4b}}}}}
\put(608,225){\makebox(0,0)[lb]{\raisebox{0pt}[0pt][0pt]
{\shortstack[l]{\footnotesize {\bf 4a}}}}}
\put(0,-30){\makebox(0,0)[lb]{\raisebox{0pt}[0pt][0pt]
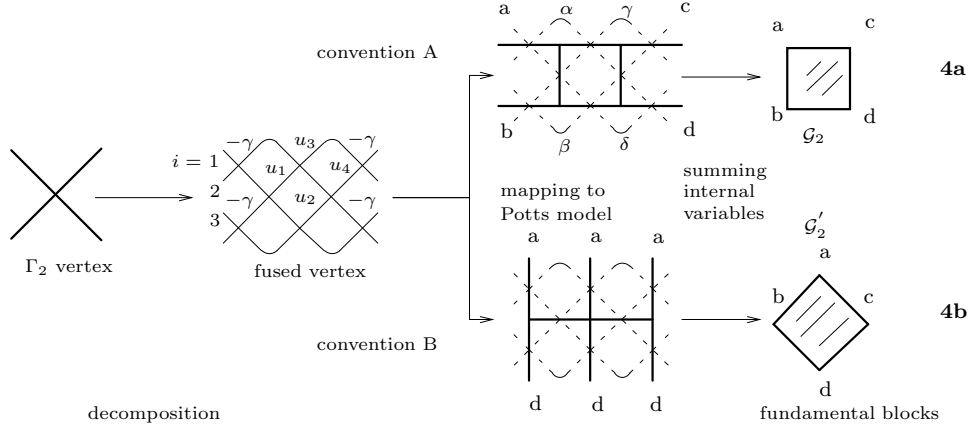
{\shortstack[l]{\footnotesize {\bf Figure(4)} The construction of the
fundamental blocks ${\cal G}_{2}^{(')}$; $\Gamma_{2}$ vertex is first
decomposed into $\Gamma_{1}$ vertices}}}}
\put(70,-45){\makebox(0,0)[lb]{\raisebox{0pt}[0pt][0pt]
{\shortstack[l]{\footnotesize   which occupy row $i=1,2$ and 3, the
variables $-\gamma,u_{1},\cdots$ shown next to the vertices are parameters }}}}
\put(70,-60){\makebox(0,0)[lb]{\raisebox{0pt}[0pt][0pt]
{\shortstack[l]{\footnotesize  that occur in $(\!~\ref{eq:Xu2}\,)$. When 
convention A ( B ) is used, vertices on row $i=1,3$ are mapped to}}}}
\put(70,-75){\makebox(0,0)[lb]{\raisebox{0pt}[0pt][0pt]
{\shortstack[l]{\footnotesize  horizontal ( vertical ) links, 
and that on row $i=2$ are mapped to vertical ( horizontal ) links. These }}}}
\put(70,-90){\makebox(0,0)[lb]{\raisebox{0pt}[0pt][0pt]
{\shortstack[l]{\footnotesize  two sets of links give rise to the fundamental 
blocks upon summing the internal sites variables }}}}
\put(70,-105){\makebox(0,0)[lb]{\raisebox{0pt}[0pt][0pt]
{\shortstack[l]{\footnotesize 
labelled by $\alpha,\beta,\cdots$. }}}}
\end{picture}
\end{center}  

The above discussion shows that the Potts model representation for 
the $\Gamma_{2}$ vertex model is achieved by replacing the spin-1 
vertex either by ${\cal G}_{2}$ that corresponds to 
$(\!~\ref{eq:weight2}\,)$ or ${\cal G}_{2}^{'}$ that corresponds 
to $(\!~\ref{eq:weight2'}\,)$. This is an extension of the 
spin-$1/2$ case where the $\Gamma_1$ vertex is replaced by a 
horizontal ${\cal G}_{1}$ or vertical ${\cal G}_{1}^{'}$ link.            
      
For $k=3$ the vertex is
\begin{equation}
     S_{3}S_{3}r_{3}(u_{1})r_{4}(u_{2})r_{5}(u_{3})r_{2}(u_{4})
r_{3}(u_{5})r_{4}(u_{6})r_{1}(u_{7})r_{2}(u_{8})r_{3}(u_{9})S_{3}
S_{3}
\end{equation}
with the symmetrizer $S_{3}$  given by
\begin{equation}
\begin{array}{rcl}   
        S_{3}& = &r_{1}(-\gamma)r_{2}(-2\gamma)r_{1}(-\gamma)\\ 
    \tilde{S}_{3} & = &r_{5}(-\gamma)r_{4}(-2\gamma)r_{5}(-\gamma)\;.
\end{array}
\end{equation}
The corresponding Potts model Boltzmann weight obtained by convention A has
expression
\begin{equation}
     W(u)_{aebcfd}=\sum_{\alpha,\beta,\gamma,\delta}S_{ae\alpha}
S_{be\beta}\left(\delta_{\alpha\gamma}\delta_{ef}\delta_{\beta\delta}
+Q^{-1/2}f_{0}\delta_{\alpha\beta}\delta_{\gamma\delta}
+f_{1}\delta_{\alpha\beta\gamma\delta}+Q^{-1/2}f_{2}
\delta_{\alpha\gamma}\delta_{\beta\delta}\right)S_{\gamma fc}
S_{\delta fd}   \label{eq:w}
\end{equation}                                                                 
where 
      \[S_{abc}=-\frac{1}{Q-1}\left(1-\delta_{ab}-\delta_{bc}-(Q-1)
\delta_{ac}+Q\delta_{abc}\right)\]           \nonumber
is the contribution from the symmetrizer and
\begin{equation}
\begin{array}{lll}       
    f_{0}& = &\prod_{i=1}^{9}x_{i}\\
    f_{1}& = &x_{1}x_{2}x_{4}[x_{5}+x_{9}+x_{5}x_{9}(Q^{1/2}+x_{6}
+x_{8})+x_{8}x_{9}(Q^{1/2}+x_{7})+x_{6}x_{9}(Q^{1/2}+x_{3})]\\
     &   &\mbox{}+[x_{1}+x_{5}+x_{1}x_{5}(Q^{1/2}+x_{2}+x_{4})
+x_{1}x_{4}(Q^{1/2}+x_{7})+x_{1}x_{2}(Q^{1/2}+x_{3})]x_{6}x_{8}x_{9}\\ 
    &   &\mbox{}+x_{1}x_{2}x_{4}[(Q^{1/2}+x_{5})(Q^{1/2}+x_{3}+x_{7})
+x_{3}x_{7}]x_{6}x_{8}x_{9}+x_{1}x_{9}(x_{2}x_{8}+x_{4}x_{6})\\            
  f_{2}& = &x_{1}+x_{9}+x_{1}x_{9}[Q^{1/2}+x_{4}+x_{8}+x_{4}x_{8}
(Q^{1/2}+x_{7})+x_{2}+x_{6}+x_{2}x_{6}(Q^{1/2}+x_{3})]\\
     &   &\mbox{}+x_{5}[1+x_{1}(Q^{1/2}+x_{2}+x_{4})]
[1+x_{9}(Q^{1/2}+x_{6}+x_{8})]\\    
\end{array}                          \label{eq:fg}   
\end{equation}  
with $x_{i}$'s defined similarly as those in the $k=2$ case. 
The four terms in $(\!~\ref{eq:w}\,)$ are again associated to the 
strand configurations in the geometrical interpretation of the 
$\Gamma_{3}$ vertex model as shown in fig.(B2). The weight has 
complicated dependence on $a,\:b,\;\cdots,\:f$, it involves all 
possible interactions among the six sites. The explicit expression 
is given in appendix A. The construction is illustrated in fig.(5a). 
The top two figures on the rhs show that when the variables labelled 
by Greek letters are summed over, the resulting figure is 
${\cal G}_{3}$, a hexagonal plaquette with $a\cdots d$ occupying the 
six corners. It corresponds to the weight $W_{aebcfd}$ which depends 
only on the variables $a\cdots d$. 

Alternatively, one could map the $\Gamma_{3}$ vertex to the Potts 
model using convention B. This gives the Boltzmann weight
\begin{equation}
     W^{'}(\tilde{u})_{aebcfd}=\sum_{\alpha,\beta,\gamma,\delta}
S_{be\alpha}S_{df\beta}(\delta_{\alpha\gamma}\delta_{\beta\delta}
+\tilde{f}_{0}Q^{1/2}\delta_{\alpha\beta}\delta_{\gamma\delta}
\delta_{ef}+\tilde{f}_{1}\delta_{\alpha\beta}\delta_{\gamma\delta}
+\tilde{f}_{2}Q^{1/2}\delta_{\alpha\gamma\beta\delta})
S_{\gamma ea}S_{\delta fc}   \label{eq:w'}
\end{equation}                                                                 
where $\tilde{f}_{0},\tilde{f}_{1}$ and $\tilde{f}_{2}$ are similarly 
defined in terms of \(\tilde{u}_{i}\) as in $(\!~\ref{eq:fg}\,)$, and they 
are decorated with tilde for reason that will be clear in the next 
few subsections. Graphically, the Potts model weight
 $(\!~\ref{eq:w'}\,)$ corresponds to a hexagon ${\cal G}_{3}^{'}$ 
which differ from ${\cal G}_{3}$ by $90^{\rm o}$ rotation as shown 
in the rhs of fig.(5b). Note that in this case, we have also 
considered the five sites $e$ ( $f$ ) on the top ( bottom ) row as 
a single site for the same reason as in the ${\cal G}_{2}^{'}$ case. 

\begin{center}  
\input{f5}
\end{center}
  
Hitherto, we have demonstrated the construction of the Potts model 
for $k=1$ ( section 2 ), 2 and 3. Unlike the $k=1$ case where the 
$\Gamma_{1}$ vertex is mapped to a Potts model link (horizontal or 
vertical), the $\Gamma_{2}$ and $\Gamma_{3}$ vertices are mapped to 
a set of horizontal and vertical links. These Potts models still 
reside on square lattices and have inhomogeneous nearest neighbor 
interaction such as that given in $(\!~\ref{eq:w2}\,)$. For the 
Boltzmann weight, however, it is more natural to sum over 
the variables associated with the internal sites, such as that 
labelled by Greek letters in $(\!~\ref{eq:w2}\,)$ or fig.(5a) and 
work with less number of variables which are associated with the 
sites on the boundary of the fundamental block. In this case, the 
interactions are no longer restricted to nearest neighbors as can 
be seen, for example, in $(\!~\ref{eq:weight2}\,)$. The corresponding 
fundamental blocks which depend on less Potts variables are more 
appropriately regarded as the plaquettes shown in the rhs of 
figs.(4a),(4b),(5a) and (5b).

These three members of the family of Potts models appear to be quite
different from one another, nevertheless we will show they have 
common symmetry properties . For arbitrary $k$ the Boltzmann weight 
can in principle be written down following the procedure outlined 
for $k=2$ and 3. It is expected to be quite complicated. It depends 
on $k$ parameters and can be regarded as an operator that acts on 
$k$ ( convention A ) or $k+1$ ( convention B ) Potts variables. In 
what follows, we shall mainly discuss the geometrical shape of the 
fundamental block ${\cal G}_{k}$ ( ${\cal G}_{k}^{'}$ ) for 
arbitrary $k$.

For $k$ odd, the $\Gamma_{k}$ vertex maps either to a hexagonal 
plaquette ${\cal G}_{k}$ with four slanted edges and two horizontal 
edges by convention A or to a hexagonal plaquette ${\cal G}_{k}^{'}$ 
that  has four slanted edges and two vertical edges by convention B. 
The number of lattice sites on each slanted edge is equal to 
$(k+1)/2$, while there are two sites on the horizontal and vertical 
edges (see fig.(6)). The case of $k=1$ is a degenerate situation 
where the slanted edge shrinks to a lattice site and the hexagon is 
flattened to become a vertical or horizontal link. 

\bigskip
{\footnotesize 
It is not difficult to arrive at the shape of the fundamental block. 
In the Potts model language, the symmetrizer $S_{k}$, which 
consists of the vertices $r_{i},\cdots,r_{i+k-1}$ involves $(k+1)/2$ 
sites since $k$ is odd. These Potts sites are those residing on the 
slanted edge. The top and bottom rows of $\Gamma_{1}$ vertices have 
odd subscripts. When convention A is used, they are mapped to 
horizontal links which becomes the two horizontal edges of the
hexagon. There are $2k$ Potts sites on each of these horizontal edge, 
but $2k-2$ of them are internal. As an example for $k=3$ ( see 
fig.(5a) ), the four Greek letters label the internal sites. When 
the variables associated with these internal sites are summed over 
in the Boltzmann weight, the horizontal edge effectively carries two 
sites. In the $k=3$ case, they are labelled by $a\:c$ and $b\:d$. 
When convention B is used, the top and bottom rows of $\Gamma_{1}$ 
vertices are mapped to vertical links. The lattice sites on the top 
and bottom rows are respectively identified due to the fact that 
there are no $\Gamma_{1}$ vertices between them ( see for example 
fig.(5b) ). The $\Gamma_{1}$ vertices which are midway from the top 
and bottom rows are mapped to vertical edges, which eventually form 
the vertical edges of the hexagon ${\cal G}_{k}^{'}$.}

\begin{center}  
\setlength{\unitlength}{0.007in}
\begin{picture}(314,174)(0,-10)
\put(30,109){\circle*{6}}
\put(114,109){\circle*{6}}
\put(47,127){\circle*{6}}
\put(98,127){\circle*{6}}
\put(98,76){\circle*{6}}
\put(47,76){\circle*{6}}
\put(114,92){\circle*{6}}
\put(30,92){\circle*{6}}
\put(89,135){\circle*{6}}
\put(55,135){\circle*{6}}
\put(122,100){\circle*{6}}
\put(21,100){\circle*{6}}
\put(55,68){\circle*{6}}
\put(89,68){\circle*{6}}
\put(249,150){\circle*{6}}
\put(249,65){\circle*{6}}
\put(266,133){\circle*{6}}
\put(266,82){\circle*{6}}
\put(214,82){\circle*{6}}
\put(214,133){\circle*{6}}
\put(232,65){\circle*{6}}
\put(232,150){\circle*{6}}
\put(274,90){\circle*{6}}
\put(274,125){\circle*{6}}
\put(241,56){\circle*{6}}
\put(241,158){\circle*{6}}
\put(206,125){\circle*{6}}
\put(206,91){\circle*{6}}
\put(243,157){\circle{2}}
\path(260,101)(238,79)
\path(259,123)(225,90)
\path(244,135)(217,109)
\path(102,101)(80,80)
\path(90,116)(59,85)
\path(76,123)(51,98)
\path(152.000,105.000)(144.000,103.000)(152.000,101.000)
\path(144,103)(179,103)
\path(171.000,101.000)(179.000,103.000)(171.000,105.000)
\thicklines
\path(111,113)(124,100)
\path(22,100)(35,113)
%\blacken
\path(44,122)(57,135)(89,135)(103,122)
\dashline{4.000}(103,122)(111,113)
\dashline{4.000}(44,122)(35,113)
\dashline{4.000}(44,80)(35,88)
\dashline{4.000}(103,80)(111,88)
%\blacken
\path(44,80)(57,68)(89,68)(103,80)
\path(22,100)(35,88)
\path(111,88)(124,100)
\path(253,68)(241,55)
\path(241,158)(253,145)
%\blacken
\path(262,137)(274,123)(274,90)(262,77)
\dashline{4.000}(262,77)(253,68)
\dashline{4.000}(262,137)(253,145)
\dashline{4.000}(219,137)(228,145)
\dashline{4.000}(219,77)(228,68)
%\blacken
\path(219,137)(206,123)(206,90)(219,77)
\path(241,158)(228,145)
\path(228,68)(241,55)
\put(-120,0){\makebox(0,0)[lb]{\raisebox{0pt}[0pt][0pt]
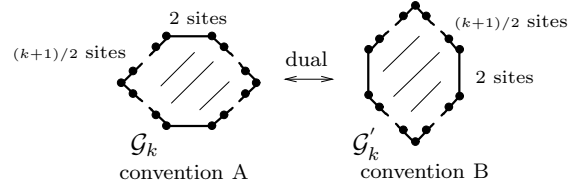
{\shortstack[l]{\footnotesize {\bf Figure(6)} The fundamental 
blocks obtained by convention A and B for odd ${\scriptstyle k}$ model}}}}
\put(200,29){\makebox(0,0)[lb]{\raisebox{0pt}[0pt][0pt]
{\shortstack[l]{\footnotesize convention B}}}}
\put(20,28){\makebox(0,0)[lb]{\raisebox{0pt}[0pt][0pt]
{\shortstack[l]{\footnotesize convention A}}}}
\put(286,100){\makebox(0,0)[lb]{\raisebox{0pt}[0pt][0pt]
{\shortstack[l]{\scriptsize 2 sites}}}}
\put(270,142){\makebox(0,0)[lb]{\raisebox{0pt}[0pt][0pt]
{\shortstack[l]{\scriptsize ${\scriptstyle (k+1)/2}$ sites}}}}
\put(-60,119){\makebox(0,0)[lb]{\raisebox{0pt}[0pt][0pt]
{\shortstack[l]{\scriptsize ${\scriptstyle (k+1)/2}$ sites}}}}
\put(57,144){\makebox(0,0)[lb]{\raisebox{0pt}[0pt][0pt]
{\shortstack[l]{\scriptsize 2 sites}}}}
\put(194,48){\makebox(0,0)[lb]{\raisebox{0pt}[0pt][0pt]
{\shortstack[l]{${\cal G}_{k}^{'}$}}}}
\put(28,48){\makebox(0,0)[lb]{\raisebox{0pt}[0pt][0pt]
{\shortstack[l]{${\cal G}_{k}$}}}}
\put(144,113){\makebox(0,0)[lb]{\raisebox{0pt}[0pt][0pt]
{\shortstack[l]{\footnotesize dual}}}}
\end{picture}
\end{center}  

For $k$ even, the fundamental block ${\cal G}_{k}$ obtained using 
convention A is an octagon with two vertical and horizontal edges 
each carrying two Potts sites, and four slanted edges with $k/2$ 
Potts sites on each of them (see fig.(7)). For $k=2$, the slanted 
edge reduces to a single Potts site and the octagon becomes a square. 
When convention B is used, the fundamental block ${\cal G}_{k}^{'}$ 
is a $45^{\rm o}$ rotated square with $k/2$ sites on every edge. 

\bigskip

{\footnotesize
In both conventions, the slanted edges are originated from the
symmetrizers $S_{k}$, which acts on $(k+2)/2$ or $k/2$ lattice sites 
since $k$ is even. For convention A, vertices at the top and bottom 
rows are associated with vertical edges, while vertices midway from 
the top and bottom rows are associated with horizontal edges, these 
edges are the two vertical and horizontal ones of the octagon. For 
convention B, all Potts sites on the top and bottom rows are 
respectively identified for the same reason given in the odd $k$ 
case, they therefore become the top and bottom corners of the 
$45^{\rm o}$ rotated square. The other two corners of the rotated 
square come from vertices midway from the top and bottom rows, 
which are associated with horizontal edges. Finally in arriving at 
the shape of the fundamental blocks, we have assumed that internal 
sites variables are summed over.}    

\begin{center}  
\setlength{\unitlength}{0.007in}
\begin{picture}(305,208)(0,-10)
\put(104,107){\circle*{6}}
\put(118,119){\circle*{6}}
\put(111,113){\circle*{6}}
\put(71,73){\circle*{6}}
\put(65,66){\circle*{6}}
\put(77,79){\circle*{6}}
\put(52,80){\circle*{6}}
\put(65,66){\circle*{6}}
\put(58,73){\circle*{6}}
\put(19,113){\circle*{6}}
\put(13,120){\circle*{6}}
\put(25,107){\circle*{6}}
\put(104,132){\circle*{6}}
\put(117,119){\circle*{6}}
\put(111,126){\circle*{6}}
\put(71,165){\circle*{6}}
\put(64,172){\circle*{6}}
\put(77,159){\circle*{6}}
\put(25,132){\circle*{6}}
\put(13,119){\circle*{6}}
\put(19,126){\circle*{6}}
\put(58,165){\circle*{6}}
\put(65,172){\circle*{6}}
\put(52,159){\circle*{6}}
\put(213,98){\circle*{6}}
\put(208,103){\circle*{6}}
\put(213,149){\circle*{6}}
\put(208,144){\circle*{6}}
\put(280,165){\circle*{6}}
\put(275,170){\circle*{6}}
\put(230,165){\circle*{6}}
\put(235,170){\circle*{6}}
\put(301,143){\circle*{6}}
\put(297,149){\circle*{6}}
\put(301,103){\circle*{6}}
\put(297,98){\circle*{6}}
\put(235,76){\circle*{6}}
\put(230,82){\circle*{6}}
\put(275,76){\circle*{6}}
\put(280,82){\circle*{6}}
\path(91,119)(64,92)
\path(87,140)(46,99)
\path(65,141)(41,117)
\dashline{4.000}(83,84)(100,101)
\thicklines
\path(118,119)(100,101)
\path(65,66)(83,84)
\path(65,66)(48,84)
\path(13,120)(30,102)
\thinlines
\dashline{4.000}(48,85)(30,102)
\thicklines
\path(117,119)(99,137)
\path(64,172)(82,155)
\thinlines
\dashline{4.000}(99,137)(82,155)
\dashline{4.000}(47,155)(29,137)
\thicklines
\path(13,119)(30,137)
\path(65,172)(48,155)
\dashline{4.000}(228,83)(214,97)
\dashline{4.000}(295,150)(282,164)
\dashline{4.000}(295,97)(282,83)
\dashline{4.000}(228,164)(214,150)
\path(208,103)(214,97)
\path(208,144)(214,150)
\path(208,143)(208,103)
\path(275,170)(282,164)
\path(235,170)(228,164)
\path(235,170)(275,170)
\path(301,143)(301,103)
\path(301,143)(295,150)
\path(301,103)(295,97)
\path(235,76)(275,76)
\path(235,76)(228,83)
\path(275,76)(282,83)
\thinlines
\path(249,154)(219,125)
\path(273,147)(231,105)
\path(289,126)(254,91)
\put(-100,0){\makebox(0,0)[lb]{\raisebox{0pt}[0pt][0pt]
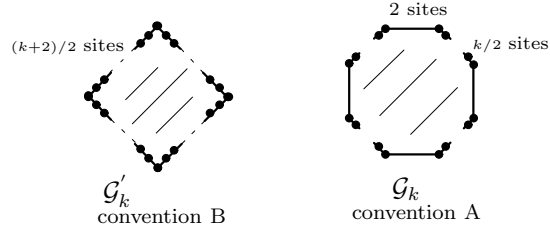
{\shortstack[l]{\footnotesize {\bf Figure(7)} The fundamental blocks obtained
by convention A and B for even $k$ model}}}}
\put(20,25){\makebox(0,0)[lb]{\raisebox{0pt}[0pt][0pt]
{\shortstack[l]{\footnotesize convention B}}}}
\put(24,42){\makebox(0,0)[lb]{\raisebox{0pt}[0pt][0pt]
{\shortstack[l]{${\cal G}_{k}^{'}$}}}}
\put(-45,153){\makebox(0,0)[lb]{\raisebox{0pt}[0pt][0pt]
{\shortstack[l]{\scriptsize ${\scriptstyle (k+2)/2}$ sites}}}}
\put(300,155){\makebox(0,0)[lb]{\raisebox{0pt}[0pt][0pt]
{\shortstack[l]{\scriptsize ${\scriptstyle k/2}$ sites}}}}
\put(238,180){\makebox(0,0)[lb]{\raisebox{0pt}[0pt][0pt]
{\shortstack[l]{\scriptsize 2 sites}}}}
\put(240,45){\makebox(0,0)[lb]{\raisebox{0pt}[0pt][0pt]
{\shortstack[l]{${\cal G}_{k}$}}}}
\put(210,28){\makebox(0,0)[lb]{\raisebox{0pt}[0pt][0pt]
{\shortstack[l]{\footnotesize convention A}}}}
\end{picture}
\end{center}  

\subsection{The patching }

\hspace{5mm}

The previous subsection deals with the construction of the local 
weight $W^{(')}$ of the Potts model and also the fundamental block 
${\cal G}_{k}^{(')}$. These fundamental blocks are the building 
block of the lattice just as in the $\Gamma_{1}$ Potts model the 
lattice is constructed from vertical and horizontal links, which, 
in our notation, are ${\cal G}_{1}$ and ${\cal G}_{1}^{'}$. Since 
there are two conventions ( A or B ) to be used in getting the Potts 
model, two lattices can be constructed. They are denoted as 
\mbox{\boldmath ${\cal L}$ and ${\cal L}^{'}$}. Detailed analysis 
shows a splitting between $k$ even and $k$ odd cass.
 
For $k$ even, we use the $\Gamma_{2}$ vertex model as an example. 
Beginning with the vertex model lattice, we replace each of the 
$\Gamma_{2}$ vertices by the fundamental block. For convention A, 
this gives rise to a Potts model lattice ${\cal L}$ which resembles 
a square check board \cite{syo}. First,  each $\Gamma_{2}$ vertex 
is replaced by the fused vertex $(\!~\ref{eq:Xu2}\,)$ shown in 
fig.(2), and subsequently this fused vertex is mapped using 
convention A to the square ${\cal G}_{2}$ which has only lattice 
sites on the four corners. Neighboring ${\cal G}_{2}$'s are connected 
along the NE-SW or NW-SE direction by sharing a lattice site on 
their common corner. For convention B, the lattice ${\cal L}^{'}$ is 
a $45^{\rm o}$ rotated square lattice. This is obtained by replacing 
each fused vertex with the $45^{\rm o}$ rotated square plaquette 
${\cal G}_{2}^{'}$ and glueing neighboring  ${\cal G}_{2}^{'}$ along 
the slanted edge. In this case, ${\cal L}$ and ${\cal L}^{'}$ are 
both square lattices, however, the former is check board like and 
only alternate squares are given Boltzmann weight $W$, while the 
later has all squares associated with the weight $W^{'}$. See 
fig.(8) for the construction of the two lattices. 

\begin{center}  
\input{f8}
\end{center}  

It is straight forward to extend the above to arbitrary even $k$. For 
convention A, we replace each of the $\Gamma_{k}$ vertices by the 
octagon ${\cal G}_{k}$, and neighboring octagons are connected along 
the NE-SW or NW-SE direction by glueing together  pairs of slanted 
edges. The resulting lattice ${\cal L}$ is shown in fig.(9a) 
( for $k=4$ ) where the filled circles denote the lattice 
sites. For convention B, the $\Gamma_{k}$ vertices are mapped to 
$45^{\rm o}$ rotated square plaquettes ${\cal G}_{k}^{'}$, the 
resulting lattice ${\cal L}^{'}$ is a $45^{\rm o}$ rotated square 
lattice just like the $k=2$ lattice ${\cal L}^{'}$, however, there 
are $(k+2)/2$ lattice sites on every edge (see fig.(9b) for $k=4$).

\begin{center}  
\setlength{\unitlength}{0.006in}
\begin{picture}(524,346)(0,-10)
\put(39,132){\circle*{6}}
\put(13,158){\circle*{6}}
\put(13,185){\circle*{6}}
\put(39,212){\circle*{6}}
\put(93,185){\circle*{6}}
\put(93,158){\circle*{6}}
\put(66,132){\circle*{6}}
\put(66,212){\circle*{6}}
\put(173,212){\circle*{6}}
\put(173,132){\circle*{6}}
\put(199,158){\circle*{6}}
\put(199,185){\circle*{6}}
\put(146,212){\circle*{6}}
\put(119,185){\circle*{6}}
\put(119,158){\circle*{6}}
\put(146,132){\circle*{6}}
\put(146,238){\circle*{6}}
\put(119,265){\circle*{6}}
\put(119,292){\circle*{6}}
\put(146,318){\circle*{6}}
\put(199,292){\circle*{6}}
\put(199,265){\circle*{6}}
\put(173,238){\circle*{6}}
\put(173,318){\circle*{6}}
\put(119,265){\circle*{6}}
\put(119,185){\circle*{6}}
\put(146,212){\circle*{6}}
\put(146,238){\circle*{6}}
\put(93,265){\circle*{6}}
\put(66,238){\circle*{6}}
\put(66,212){\circle*{6}}
\put(93,185){\circle*{6}}
\put(39,238){\circle*{6}}
\put(13,265){\circle*{6}}
\put(13,292){\circle*{6}}
\put(39,318){\circle*{6}}
\put(93,292){\circle*{6}}
\put(93,265){\circle*{6}}
\put(66,238){\circle*{6}}
\put(66,318){\circle*{6}}
\put(386,185){\circle*{6}}
\put(406,205){\circle*{6}}
\put(346,225){\circle*{6}}
\put(366,245){\circle*{6}}
\put(346,145){\circle*{6}}
\put(366,165){\circle*{6}}
\put(326,205){\circle*{6}}
\put(306,185){\circle*{6}}
\put(406,165){\circle*{6}}
\put(326,245){\circle*{6}}
\put(366,205){\circle*{6}}
\put(366,285){\circle*{6}}
\put(346,305){\circle*{6}}
\put(306,265){\circle*{6}}
\put(326,285){\circle*{6}}
\put(466,185){\circle*{6}}
\put(446,205){\circle*{6}}
\put(386,265){\circle*{6}}
\put(406,245){\circle*{6}}
\put(426,225){\circle*{6}}
\put(446,165){\circle*{6}}
\put(426,145){\circle*{6}}
\put(446,245){\circle*{6}}
\put(406,285){\circle*{6}}
\put(326,165){\circle*{6}}
\put(426,305){\circle*{6}}
\put(446,285){\circle*{6}}
\put(466,265){\circle*{6}}
\path(79,172)(53,145)
\path(66,172)(39,145)
\path(66,185)(39,158)
\path(53,185)(26,158)
\path(53,198)(26,172)
\thicklines
\path(39,212)(66,212)(93,185)
        (93,158)(66,132)(39,132)
        (13,158)(13,185)(39,212)
\path(146,212)(173,212)(199,185)
        (199,158)(173,132)(146,132)
        (119,158)(119,185)(146,212)
\thinlines
\path(159,198)(133,172)
\path(159,185)(133,158)
\path(173,185)(146,158)
\path(173,172)(146,145)
\path(186,172)(159,145)
\path(186,278)(159,252)
\path(173,278)(146,252)
\path(173,292)(146,265)
\path(159,292)(133,265)
\path(159,305)(133,278)
\thicklines
\path(146,318)(173,318)(199,292)
        (199,265)(173,238)(146,238)
        (119,265)(119,292)(146,318)
\path(93,265)(119,265)(146,238)
        (146,212)(119,185)(93,185)
        (66,212)(66,238)(93,265)
\thinlines
\path(106,252)(79,225)
\path(106,238)(79,212)
\path(119,238)(93,212)
\path(119,225)(93,198)
\path(133,225)(106,198)
\path(79,278)(53,252)
\path(66,278)(39,252)
\path(66,292)(39,265)
\path(53,292)(26,265)
\path(53,305)(26,278)
\thicklines
\path(39,318)(66,318)(93,292)
        (93,265)(66,238)(39,238)
        (13,265)(13,292)(39,318)
\thinlines
\path(371,185)(346,160)
\path(366,195)(341,170)
\path(356,200)(331,175)
\path(351,210)(326,185)
\path(431,210)(406,185)
\path(436,200)(411,175)
\path(446,195)(421,170)
\path(451,185)(426,160)
\path(411,225)(386,200)
\path(406,235)(381,210)
\path(396,240)(371,215)
\path(391,250)(366,225)
\path(431,290)(406,265)
\path(436,280)(411,255)
\path(446,275)(421,250)
\path(451,265)(426,240)
\path(351,290)(326,265)
\path(356,280)(331,255)
\path(366,275)(341,250)
\path(371,265)(346,240)
\thicklines
\path(426,305)(466,265)(346,145)
        (306,185)(386,265)(466,185)
        (426,145)(306,265)(346,305)
        (386,265)(426,305)
\put(360,0){\makebox(0,0)[lb]{\raisebox{0pt}[0pt][0pt]
{\shortstack[l]{\footnotesize lattice from convention B}}}}
\put(290,20){\makebox(0,0)[lb]{\raisebox{0pt}[0pt][0pt]
{\shortstack[l]{\footnotesize {\bf Figure(9b)} The ${\scriptstyle \Gamma_{4}}$ 
Potts model}}}}
\put(50,0){\makebox(0,0)[lb]{\raisebox{0pt}[0pt][0pt]
{\shortstack[l]{\footnotesize lattice from convention A}}}}
\put(-20,20){\makebox(0,0)[lb]{\raisebox{0pt}[0pt][0pt]
{\shortstack[l]{\footnotesize {\bf Figure(9a)} The ${\scriptstyle \Gamma_{4}}$ 
Potts model}}}}
\put(227,85){\makebox(0,0)[lb]{\raisebox{0pt}[0pt][0pt]
{\shortstack[l]{${\scriptstyle k=4}$}}}}
\put(340,50){\makebox(0,0)[lb]{\raisebox{0pt}[0pt][0pt]
{\shortstack[l]{\scriptsize convention B}}}}
\put(366,85){\makebox(0,0)[lb]{\raisebox{0pt}[0pt][0pt]
{\shortstack[l]{${\scriptstyle {\cal L}^{'}}$}}}}
\put(58,50){\makebox(0,0)[lb]{\raisebox{0pt}[0pt][0pt]
{\shortstack[l]{\scriptsize convention A}}}}
\put(66,85){\makebox(0,0)[lb]{\raisebox{0pt}[0pt][0pt]
{\shortstack[l]{${\scriptstyle {\cal L}}$}}}}
\end{picture}
\end{center}  

For $k$ odd, we use  the $\Gamma_{3}$ model as an example. Recall 
that ${\cal G}_{3}$ and ${\cal G}_{3}^{'}$ differ by $90^{\rm o}$ 
rotation, we expect therefore the lattices ${\cal L}$ and 
${\cal L}^{'}$ to have similar symmetry properties. 
\begin{center} 
\setlength{\unitlength}{0.006in}
\begin{picture}(695,440)(0,-125)
\put(57.000,155.000){\arc{10.000}{0.6435}{2.4981}}
\put(313.000,141.000){\arc{10.000}{3.7851}{5.6397}}
\put(233.000,189.000){\arc{10.000}{3.7851}{5.6397}}
\put(201.000,189.000){\arc{10.000}{3.7851}{5.6397}}
\put(297.000,189.000){\arc{10.000}{3.7851}{5.6397}}
\put(265.000,189.000){\arc{10.000}{3.7851}{5.6397}}
\put(201.000,107.000){\arc{10.000}{0.6435}{2.4981}}
\put(233.000,107.000){\arc{10.000}{0.6435}{2.4981}}
\put(297.000,107.000){\arc{10.000}{0.6435}{2.4981}}
\put(265.000,107.000){\arc{10.000}{0.6435}{2.4981}}
\put(121.000,155.000){\arc{10.000}{0.6435}{2.4981}}
\put(153.000,155.000){\arc{10.000}{0.6435}{2.4981}}
\put(89.000,155.000){\arc{10.000}{0.6435}{2.4981}}
\put(121.000,237.000){\arc{10.000}{3.7851}{5.6397}}
\put(153.000,237.000){\arc{10.000}{3.7851}{5.6397}}
\put(57.000,237.000){\arc{10.000}{3.7851}{5.6397}}
\put(89.000,237.000){\arc{10.000}{3.7851}{5.6397}}
\put(169.000,189.000){\arc{10.000}{3.7851}{5.6397}}
\put(417.000,226.000){\arc{10.000}{0.6435}{2.4981}}
\put(671.000,212.000){\arc{10.000}{3.7851}{5.6397}}
\put(592.000,255.500){\arc{17.000}{4.2224}{5.2023}}
\put(560.000,255.500){\arc{17.000}{4.2224}{5.2023}}
\put(655.000,255.500){\arc{17.000}{4.2224}{5.2023}}
\put(623.500,257.500){\arc{13.038}{4.1457}{5.2791}}
\put(560.000,178.000){\arc{10.000}{0.6435}{2.4981}}
\put(592.000,178.000){\arc{10.000}{0.6435}{2.4981}}
\put(655.000,178.000){\arc{10.000}{0.6435}{2.4981}}
\put(623.500,177.000){\arc{8.062}{0.5191}{2.6224}}
\put(480.000,226.000){\arc{10.000}{0.6435}{2.4981}}
\put(512.000,226.000){\arc{10.000}{0.6435}{2.4981}}
\put(449.000,226.000){\arc{10.000}{0.6435}{2.4981}}
\put(480.000,307.000){\arc{10.000}{3.7851}{5.6397}}
\put(512.000,307.000){\arc{10.000}{3.7851}{5.6397}}
\put(417.000,307.000){\arc{10.000}{3.7851}{5.6397}}
\put(449.000,307.000){\arc{10.000}{3.7851}{5.6397}}
\put(528.000,255.500){\arc{17.000}{4.2224}{5.2023}}
\put(544.000,95.500){\arc{17.000}{4.2224}{5.2023}}
\put(464.500,148.000){\arc{8.062}{3.6607}{5.7640}}
\put(433.000,147.000){\arc{10.000}{3.7851}{5.6397}}
\put(528.000,147.000){\arc{10.000}{3.7851}{5.6397}}
\put(496.000,147.000){\arc{10.000}{3.7851}{5.6397}}
\put(464.500,65.000){\arc{8.062}{0.5191}{2.6224}}
\put(528.000,66.000){\arc{10.000}{0.6435}{2.4981}}
\put(496.000,66.000){\arc{10.000}{0.6435}{2.4981}}
\put(639.000,18.000){\arc{10.000}{0.6435}{2.4981}}
\put(671.000,18.000){\arc{10.000}{0.6435}{2.4981}}
\put(608.000,18.000){\arc{10.000}{0.6435}{2.4981}}
\put(576.000,18.000){\arc{10.000}{0.6435}{2.4981}}
\put(639.000,95.500){\arc{17.000}{4.2224}{5.2023}}
\put(671.000,95.500){\arc{17.000}{4.2224}{5.2023}}
\put(576.000,95.500){\arc{17.000}{4.2224}{5.2023}}
\put(608.000,95.500){\arc{17.000}{4.2224}{5.2023}}
\put(433.000,66.000){\arc{10.000}{0.6435}{2.4981}}
\path(325,160)(360,160)(360,60)(395,60)
\path(387.000,58.000)(395.000,60.000)(387.000,62.000)
\path(325,160)(360,160)(360,260)(395,260)
\path(387.000,258.000)(395.000,260.000)(387.000,262.000)
\path(305,156)(269,192)
\path(321,188)(237,104)
\path(301,192)(321,172)
\path(293,192)(205,104)
\path(205,192)(293,104)
\path(309,144)(269,104)
\path(165,192)(125,152)
\path(61,240)(149,152)
\path(149,240)(61,152)
\path(85,240)(49,204)
\path(85,152)(49,188)
\path(33,236)(117,152)
\path(117,240)(33,156)
\path(33,172)(53,152)
\path(53,240)(33,220)
\path(157,240)(177,220)
\path(177,236)(93,152)
\path(161,204)(125,240)
\path(177,124)(197,104)
\path(177,188)(261,104)
\path(197,192)(157,152)
\path(261,192)(177,108)
\path(93,240)(229,104)
\path(237,192)(321,108)
\path(321,124)(301,104)
\path(229,192)(177,140)
\dashline{4.000}(663,227)(627,263)
\dashline{4.000}(679,259)(596,175)
\dashline{4.000}(659,263)(679,243)
\dashline{4.000}(651,263)(564,175)
\dashline{4.000}(564,263)(651,175)
\dashline{4.000}(667,215)(627,175)
\dashline{4.000}(524,263)(484,223)
\dashline{4.000}(421,310)(508,223)
\dashline{4.000}(508,310)(421,223)
\dashline{4.000}(445,310)(409,275)
\dashline{4.000}(445,223)(409,259)
\dashline{4.000}(393,306)(476,223)
\dashline{4.000}(476,310)(393,227)
\dashline{4.000}(393,243)(413,223)
\dashline{4.000}(413,310)(393,291)
\dashline{4.000}(516,310)(536,291)
\dashline{4.000}(536,306)(453,223)
\dashline{4.000}(520,275)(484,310)
\dashline{4.000}(536,195)(556,175)
\dashline{4.000}(536,259)(620,175)
\dashline{4.000}(556,263)(516,223)
\dashline{4.000}(620,263)(536,179)
\dashline{4.000}(453,310)(588,175)
\dashline{4.000}(596,263)(679,179)
\dashline{4.000}(679,195)(659,175)
\dashline{4.000}(588,263)(536,211)
\thicklines
\path(385,298)(544,298)
\path(417,298)(417,235)
\path(449,298)(449,235)
\path(480,298)(480,235)
\path(512,298)(512,235)
\path(417,267)(512,267)
\path(385,235)(544,235)
\path(528,251)(687,251)
\path(560,251)(560,187)
\path(592,251)(592,187)
\path(624,251)(624,187)
\path(655,251)(655,187)
\path(528,187)(687,187)
\path(560,219)(655,219)
\path(560,11)(560,43)(687,43)(687,11)
\path(560,107)(560,75)
\path(655,107)(655,11)
\path(624,107)(624,11)
\path(592,107)(592,11)
\path(401,75)(687,75)(687,107)
\path(433,107)(528,107)
\path(528,138)(528,75)
\path(496,138)(496,75)
\path(465,138)(465,75)
\path(433,138)(433,75)
\path(401,138)(560,138)
\thinlines
\dashline{4.000}(604,103)(552,51)
\dashline{4.000}(695,35)(675,15)
\dashline{4.000}(612,103)(695,19)
\dashline{4.000}(468,150)(604,15)
\dashline{4.000}(635,103)(552,19)
\dashline{4.000}(572,103)(532,63)
\dashline{4.000}(552,99)(635,15)
\dashline{4.000}(552,35)(572,15)
\dashline{4.000}(536,115)(500,150)
\dashline{4.000}(552,146)(468,63)
\dashline{4.000}(532,150)(552,131)
\dashline{4.000}(429,150)(409,131)
\dashline{4.000}(409,83)(429,63)
\dashline{4.000}(492,150)(409,67)
\dashline{4.000}(409,146)(492,63)
\dashline{4.000}(461,63)(425,99)
\dashline{4.000}(461,150)(425,115)
\dashline{4.000}(524,150)(437,63)
\dashline{4.000}(437,150)(524,63)
\dashline{4.000}(540,103)(500,63)
\dashline{4.000}(683,55)(643,15)
\dashline{4.000}(580,103)(667,15)
\dashline{4.000}(667,103)(580,15)
\dashline{4.000}(675,103)(695,83)
\dashline{4.000}(695,99)(612,15)
\dashline{4.000}(679,67)(643,103)
\put(17,110){\makebox(0,0)[lb]{\raisebox{0pt}[0pt][0pt]
{\scriptsize \shortstack[l]{8}}}}
\put(17,130){\makebox(0,0)[lb]{\raisebox{0pt}[0pt][0pt]
{\scriptsize \shortstack[l]{7}}}}
\put(17,145){\makebox(0,0)[lb]{\raisebox{0pt}[0pt][0pt]
{\scriptsize \shortstack[l]{6}}}}
\put(17,160){\makebox(0,0)[lb]{\raisebox{0pt}[0pt][0pt]
{\scriptsize \shortstack[l]{5}}}}
\put(17,175){\makebox(0,0)[lb]{\raisebox{0pt}[0pt][0pt]
{\scriptsize \shortstack[l]{4}}}}
\put(17,190){\makebox(0,0)[lb]{\raisebox{0pt}[0pt][0pt]
{\scriptsize \shortstack[l]{3}}}}
\put(17,205){\makebox(0,0)[lb]{\raisebox{0pt}[0pt][0pt]
{\scriptsize \shortstack[l]{2}}}}
\put(-5,220){\makebox(0,0)[lb]{\raisebox{0pt}[0pt][0pt]
{\scriptsize \shortstack[l]{${\scriptstyle i=}$1}}}}

\put(500,-20){\makebox(0,0)[lb]{\raisebox{0pt}[0pt][0pt]
{\shortstack[l]{\footnotesize {\bf 10b}}}}}
\put(75,-20){\makebox(0,0)[lb]{\raisebox{0pt}[0pt][0pt]
{\shortstack[l]{\footnotesize {\bf 10a}}}}}
\put(-20,-60){\makebox(0,0)[lb]{\raisebox{0pt}[0pt][0pt]
{\shortstack[l]{\footnotesize {\bf Figure(10)} The two possible patching of 
the fundamental blocks from the neigboring }}}}
\put(80,-80){\makebox(0,0)[lb]{\raisebox{0pt}[0pt][0pt]
{\shortstack[l]{\footnotesize  vertices ${\rm V}_{1}$ and 
${\rm V}_{2}$. The top figure on the rhs uses convention A for the }}}}
\put(80,-100){\makebox(0,0)[lb]{\raisebox{0pt}[0pt][0pt]
{\shortstack[l]{\footnotesize  neighboring $\Gamma_{3}$ vertices, the 
resulting Potts lattices are not compatible. }}}}
\put(80,-120){\makebox(0,0)[lb]{\raisebox{0pt}[0pt][0pt]
{\shortstack[l]{\footnotesize The bottom figure shows the correct mapping 
where different conventions}}}}
\put(80,-140){\makebox(0,0)[lb]{\raisebox{0pt}[0pt][0pt]
{\shortstack[l]{\footnotesize  are used on neighboring 
$\Gamma_{3}$ vertices. }}}}
%\put(105,-40){\makebox(0,0)[lb]{\raisebox{0pt}[0pt][0pt]
%{\footnotesize \shortstack[l]{vertices ${\rm V}_{1}$ and ${\rm V}_{2}$}}}}
%\put(69,-20){\makebox(0,0)[lb]{\raisebox{0pt}[0pt][0pt]
%{\shortstack[l]{\footnotesize {\bf Figure(10a)} Neighboring 
%$\Gamma_{3}$}}}}
\put(205,75){\makebox(0,0)[lb]{\raisebox{0pt}[0pt][0pt]
{\footnotesize \shortstack[l]{${\rm V}_{2}$}}}}
\put(80,115){\makebox(0,0)[lb]{\raisebox{0pt}[0pt][0pt]
{\footnotesize \shortstack[l]{${\rm V}_{1}$}}}}
\put(245,45){\makebox(0,0)[lb]{\raisebox{0pt}[0pt][0pt]
{\footnotesize \shortstack[l]{convention B}}}}
\put(235,60){\makebox(0,0)[lb]{\raisebox{0pt}[0pt][0pt]
{\footnotesize \shortstack[l]{convention A}}}}
\put(245,260){\makebox(0,0)[lb]{\raisebox{0pt}[0pt][0pt]
{\footnotesize \shortstack[l]{convention A}}}}
\put(235,275){\makebox(0,0)[lb]{\raisebox{0pt}[0pt][0pt]
{\footnotesize \shortstack[l]{convention A}}}}
\end{picture}
\end{center} 

\bigskip

{\footnotesize
As before we first replace each $\Gamma_{3}$ vertex by the fused 
vertex $(\!~\ref{eq:Xu2}\,)$ shown in fig.(2) and consider two 
neighboring fused vertices connected along the NW-SE direction. Let 
us denote them as ${\rm V}_{1}$ and ${\rm V}_{2}$ respectively 
(see fig.(10a)) where ${\rm V}_{1}$ is on the upper left corner of 
${\rm V}_{2}$. The $\Gamma_{1}$ vertices that belong to 
${\rm V}_{1}$ are $r_{1},\cdots,r_{5}$ and that belong to 
${\rm V}_{2}$ are $r_{4},\cdots,r_{8}$. The symmetrizers $S_{3}$'s 
that connect ${\rm V}_{1}$ and ${\rm V}_{2}$ occupy rows $i=4$ and 5, 
ie. they are made out of $r_{4}$ and $r_{5}$. Suppose ${\rm V}_{1}$ 
is mapped to the Potts model fundamental block ${\cal G}_{3}$ using 
convention A, the $\Gamma_{1}$ vertices $r_{4}$ and $r_{5}$ will be 
mapped respectively to a vertical and a horizontal link. However, 
this implies that the $\Gamma_{1}$ vertices on the top row of 
${\rm V}_{2}$, which are $r_{4}$, have to be mapped accordingly
to vertical links so that the two set of Potts model links obtained 
from ${\rm V}_{1}$ and ${\rm V}_{2}$ are compatible. In other words, 
convention B has to be used on ${\rm V}_{2}$ which replaces 
${\rm V}_{2}$ by ${\cal G}_{3}^{'}$. This  point is illustrated 
in fig.(10b).}

\bigskip

 Hence, neighboring $\Gamma_{3}$ vertices are to be mapped to
Potts model links using different conventions. The resulting lattice 
${\cal L}$  has the geometry shown in the lhs of fig.(11). If the 
mapping of ${\rm V}_{1}$ is done with convention B, the lattice 
${\cal L}^{'}$ obtained (see rhs of fig.(11)) is related to 
${\cal L}$ by replacing all ${\cal G}_{3}$ by ${\cal G}_{3}^{'}$ and
vice versa. This  feature is present in all odd $k$ models. For the 
$\Gamma_{1}$ Potts model, vertical ${\cal G}_{1}^{'}$ and horizontal 
${\cal G}_{1}$ links are associated respectively to neighboring 
vertices connected along the NE-SW or NW-SE direction. For higher 
odd $k$, the Potts model lattice ${\cal L}$ or ${\cal L}^{'}$ have 
to be constructed with both ${\cal G}_{k}$ and ${\cal G}_{k}^{'}$,  
for the same reason as in the   $k=3$ case, ie. the symmetrizer 
$S_{k}$ contains even number of rows of $\Gamma_{1}$ vertices.
Since ${\cal G}_{k}^{(')}$ has the same hexagonal shape for $k\geq3$, 
the lattice ${\cal L}$ and ${\cal L}^{'}$ are identical to that of 
the $k=3$ model except that there are $(k+1)/2$ lattices on the 
slanted edge.

\begin{center} 
\setlength{\unitlength}{0.006in}
\begin{picture}(432,273)(0,-10)
\put(36,78){\circle*{6}}
\put(12,102){\circle*{6}}
\put(84,78){\circle*{6}}
\put(108,102){\circle*{6}}
\put(84,126){\circle*{6}}
\put(36,126){\circle*{6}}
\put(132,78){\circle*{6}}
\put(108,102){\circle*{6}}
\put(180,78){\circle*{6}}
\put(204,102){\circle*{6}}
\put(180,126){\circle*{6}}
\put(132,126){\circle*{6}}
\put(84,222){\circle*{6}}
\put(108,198){\circle*{6}}
\put(36,222){\circle*{6}}
\put(12,198){\circle*{6}}
\put(36,174){\circle*{6}}
\put(84,174){\circle*{6}}
\put(84,126){\circle*{6}}
\put(84,174){\circle*{6}}
\put(108,198){\circle*{6}}
\put(132,174){\circle*{6}}
\put(108,102){\circle*{6}}
\put(132,126){\circle*{6}}
\put(132,174){\circle*{6}}
\put(108,198){\circle*{6}}
\put(180,174){\circle*{6}}
\put(204,198){\circle*{6}}
\put(180,222){\circle*{6}}
\put(132,222){\circle*{6}}
\put(420,126){\circle*{6}}
\put(420,78){\circle*{6}}
\put(396,54){\circle*{6}}
\put(372,78){\circle*{6}}
\put(396,150){\circle*{6}}
\put(372,126){\circle*{6}}
\put(324,126){\circle*{6}}
\put(300,150){\circle*{6}}
\put(372,126){\circle*{6}}
\put(396,150){\circle*{6}}
\put(372,174){\circle*{6}}
\put(324,174){\circle*{6}}
\put(372,174){\circle*{6}}
\put(372,222){\circle*{6}}
\put(396,246){\circle*{6}}
\put(420,222){\circle*{6}}
\put(396,150){\circle*{6}}
\put(420,174){\circle*{6}}
\put(324,126){\circle*{6}}
\put(324,78){\circle*{6}}
\put(300,54){\circle*{6}}
\put(276,78){\circle*{6}}
\put(300,150){\circle*{6}}
\put(276,126){\circle*{6}}
\put(324,222){\circle*{6}}
\put(324,174){\circle*{6}}
\put(300,150){\circle*{6}}
\put(276,174){\circle*{6}}
\put(300,246){\circle*{6}}
\put(276,222){\circle*{6}}
\thicklines
\path(84,78)(36,78)(12,102)
        (36,126)(84,126)(108,102)(84,78)
\thinlines
\path(72,90)(90,108)
\path(54,87)(78,111)
\path(42,90)(66,114)
\path(30,96)(48,114)
\thicklines
\path(180,78)(132,78)(108,102)
        (132,126)(180,126)(204,102)(180,78)
\thinlines
\path(168,90)(186,108)
\path(150,87)(174,111)
\path(138,90)(162,114)
\path(126,96)(144,114)
\thicklines
\path(36,222)(84,222)(108,198)
        (84,174)(36,174)(12,198)(36,222)
\thinlines
\path(48,210)(30,192)
\path(66,213)(42,189)
\path(78,210)(54,186)
\path(90,204)(72,186)
\path(114,120)(96,138)
\path(120,132)(96,156)
\path(123,144)(99,168)
\path(120,162)(102,180)
\thicklines
\path(132,174)(132,126)(108,102)
        (84,126)(84,174)(108,198)(132,174)
\path(180,174)(132,174)(108,198)
        (132,222)(180,222)(204,198)(180,174)
\thinlines
\path(168,186)(186,204)
\path(150,183)(174,207)
\path(138,186)(162,210)
\path(126,192)(144,210)
\path(390,132)(408,114)
\path(384,120)(408,96)
\path(381,108)(405,84)
\path(384,90)(402,72)
\thicklines
\path(372,78)(372,126)(396,150)
        (420,126)(420,78)(396,54)(372,78)
\path(372,126)(324,126)(300,150)
        (324,174)(372,174)(396,150)(372,126)
\thinlines
\path(360,138)(378,156)
\path(342,135)(366,159)
\path(330,138)(354,162)
\path(318,144)(336,162)
\path(402,168)(384,186)
\path(408,180)(384,204)
\path(411,192)(387,216)
\path(408,210)(390,228)
\thicklines
\path(420,222)(420,174)(396,150)
        (372,174)(372,222)(396,246)(420,222)
\thinlines
\path(294,132)(312,114)
\path(288,120)(312,96)
\path(285,108)(309,84)
\path(288,90)(306,72)
\thicklines
\path(276,78)(276,126)(300,150)
        (324,126)(324,78)(300,54)(276,78)
\thinlines
\path(294,228)(312,210)
\path(288,216)(312,192)
\path(285,204)(309,180)
\path(288,186)(306,168)
\thicklines
\path(276,174)(276,222)(300,246)
        (324,222)(324,174)(300,150)(276,174)
\put(80,0){\makebox(0,0)[lb]{\raisebox{0pt}[0pt][0pt]
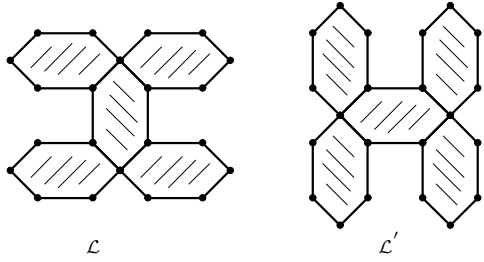
{\shortstack[l]{\footnotesize {\bf Figure(11)} The $\Gamma_{3}$ 
Potts model lattices}}}}
\put(78,30){\makebox(0,0)[lb]{\raisebox{0pt}[0pt][0pt]
{\shortstack[l]{${\scriptstyle {\cal L}}$}}}}
\put(333,30){\makebox(0,0)[lb]{\raisebox{0pt}[0pt][0pt]
{\shortstack[l]{${\scriptstyle {\cal L}^{'}}$}}}}
\end{picture}
\end{center} 

Graphically, it is easy to see that four neighboring blocks 
${\cal G}_{k}^{(')}$ of the $\Gamma_{k}$ Potts model can be combined 
in a natural way to form the fundamental block ${\cal G}_{2k}$ or 
${\cal G}_{2k}^{'}$ of the $\Gamma_{2k}$ Potts model (see fig.(12)).
Nonetheless, this  does not imply that a  $\Gamma_{k}$ Potts model 
can also be considered,after summation over the appropriate 
variables, as  $\Gamma_{2k}$ Potts model. Although the geometry is 
the same, the full definition of the $\Gamma_k$ Potts models implies 
that they are build from a vertex model with spin $k/2$ 
representation of ${\rm U}_{q}{\rm su(2)}$, and this constrains the 
form of the Boltzmann weights. Figure (12) looks like one step in 
real space renormalization group. The fact that starting from a 
$\Gamma_k$ Potts model one does not get a $\Gamma_{2k}$  suggests 
that  the $\Gamma_k$ Potts models for different values
of $k$ belong a priori to different universality classes.

\begin{center} 
\setlength{\unitlength}{0.007in}
\begin{picture}(533,280)(0,-35)
\put(112,208){\circle*{6}}
\put(111,183){\circle*{6}}
\put(61,233){\circle*{6}}
\put(86,233){\circle*{6}}
\put(86,208){\circle*{6}}
\put(62,208){\circle*{6}}
\put(86,183){\circle*{6}}
\put(86,158){\circle*{6}}
\put(64,158){\circle*{6}}
\put(61,183){\circle*{6}}
\put(38,183){\circle*{6}}
\put(37,208){\circle*{6}}
\put(257,182){\circle*{6}}
\put(257,207){\circle*{6}}
\put(233,234){\circle*{6}}
\put(205,234){\circle*{6}}
\put(178,207){\circle*{6}}
\put(180,182){\circle*{6}}
\put(206,157){\circle*{6}}
\put(233,155){\circle*{6}}
\put(108,72){\circle*{6}}
\put(47,47){\circle*{6}}
\put(69,26){\circle*{6}}
\put(26,69){\circle*{6}}
\put(47,90){\circle*{6}}
\put(69,111){\circle*{6}}
\put(90,90){\circle*{6}}
\put(70,70){\circle*{6}}
\put(91,49){\circle*{6}}
\put(244,49){\circle*{6}}
\put(243,90){\circle*{6}}
\put(221,111){\circle*{6}}
\put(200,90){\circle*{6}}
\put(179,69){\circle*{6}}
\put(221,26){\circle*{6}}
\put(200,47){\circle*{6}}
\put(261,72){\circle*{6}}
\put(464,229){\circle*{6}}
\put(487,229){\circle*{6}}
\put(448,214){\circle*{6}}
\put(500,214){\circle*{6}}
\put(513,202){\circle*{6}}
\put(512,176){\circle*{6}}
\put(500,163){\circle*{6}}
\put(438,202){\circle*{6}}
\put(436,177){\circle*{6}}
\put(451,163){\circle*{6}}
\put(462,152){\circle*{6}}
\put(486,150){\circle*{6}}
\put(482,112){\circle*{6}}
\put(516,79){\circle*{6}}
\put(525,68){\circle*{6}}
\put(493,101){\circle*{6}}
\put(439,67){\circle*{6}}
\put(450,57){\circle*{6}}
\put(452,79){\circle*{6}}
\put(472,101){\circle*{6}}
\put(517,57){\circle*{6}}
\put(493,36){\circle*{6}}
\put(473,36){\circle*{6}}
\put(482,24){\circle*{6}}
\put(353,36){\circle*{6}}
\put(332,36){\circle*{6}}
\put(342,24){\circle*{6}}
\put(342,112){\circle*{6}}
\put(377,79){\circle*{6}}
\put(385,68){\circle*{6}}
\put(353,101){\circle*{6}}
\put(354,57){\circle*{6}}
\put(342,67){\circle*{6}}
\put(332,79){\circle*{6}}
\put(310,57){\circle*{6}}
\put(332,57){\circle*{6}}
\put(313,79){\circle*{6}}
\put(332,101){\circle*{6}}
\put(355,79){\circle*{6}}
\put(378,57){\circle*{6}}
\put(299,67){\circle*{6}}
\put(326,229){\circle*{6}}
\put(350,202){\circle*{6}}
\put(323,202){\circle*{6}}
\put(349,229){\circle*{6}}
\put(311,214){\circle*{6}}
\put(362,214){\circle*{6}}
\put(375,202){\circle*{6}}
\put(374,176){\circle*{6}}
\put(350,177){\circle*{6}}
\put(362,163){\circle*{6}}
\put(300,202){\circle*{6}}
\put(298,177){\circle*{6}}
\put(323,176){\circle*{6}}
\put(313,163){\circle*{6}}
\put(322,152){\circle*{6}}
\put(348,150){\circle*{6}}
\thicklines
\path(62,233)(62,208)(87,208)
	(87,233)(62,233)
\path(37,208)(37,183)(62,183)
	(62,208)(37,208)
\path(62,183)(62,158)(87,158)
	(87,183)(62,183)
\path(87,208)(87,183)(112,183)
	(112,208)(87,208)
\path(207,234)(234,234)(259,207)
	(259,182)(234,155)(207,155)
	(182,182)(182,207)(207,234)
\thinlines
\dashline{4.000}(182,207)(259,207)
\dashline{4.000}(234,234)(234,155)
\dashline{4.000}(207,234)(207,155)
\dashline{4.000}(182,182)(259,182)
\thicklines
\path(90,91)(47,49)
\path(69,113)(111,70)
\path(90,49)(111,70)
\path(69,27)(47,49)
\path(69,27)(90,49)
\path(69,70)(90,49)
\thinlines
\path(47,91)(69,70)
\thicklines
\path(47,49)(69,70)
\path(47,91)(26,70)
\path(47,49)(26,70)
\path(69,70)(47,91)
\path(69,113)(47,91)
\path(221,113)(200,91)
\path(200,49)(179,70)
\path(200,91)(179,70)
\path(221,27)(243,49)
\path(221,27)(200,49)
\path(243,49)(264,70)
\path(221,113)(264,70)
\thinlines
\dashline{4.000}(200,91)(243,49)
\dashline{4.000}(243,91)(200,49)
\dashline{4.000}(463,176)(489,176)
\dashline{4.000}(489,201)(489,176)(501,163)
\dashline{4.000}(463,201)(489,201)(501,214)
\dashline{4.000}(450,163)(463,176)(463,201)(450,214)
\thicklines
\path(463,227)(489,227)(515,201)
	(515,176)(489,150)(463,150)
	(438,176)(438,201)(463,227)
\thinlines
\dashline{4.000}(451,78)(473,78)
\dashline{4.000}(451,56)(473,56)
\dashline{4.000}(495,56)(483,67)(473,56)(473,35)
\dashline{4.000}(516,56)(495,56)(495,35)
\dashline{4.000}(516,78)(495,78)
\dashline{4.000}(473,100)(473,78)(483,67)
	(495,78)(495,100)
\thicklines
\path(483,111)(439,67)(483,23)
	(527,67)(483,111)
\path(343,67)(354,56)(354,35)
	(343,23)(332,35)(332,56)(343,67)
\path(343,111)(354,100)(354,78)
	(343,67)(332,78)(332,100)(343,111)
\path(343,67)(354,78)(377,78)
	(387,67)(377,56)(354,56)(343,67)
\path(300,67)(311,78)(332,78)
	(343,67)(332,56)(311,56)(300,67)
\path(312,214)(325,227)(351,227)
	(363,214)(351,201)(325,201)(312,214)
\path(312,214)(325,201)(325,176)
	(312,163)(299,176)(299,201)(312,214)
\path(363,214)(377,201)(377,176)
	(363,163)(351,176)(351,201)(363,214)
\path(312,163)(325,176)(351,176)
	(363,163)(351,150)(325,150)(312,163)
\thinlines
\path(370,125)(450,125)
\path(442.000,123.000)(450.000,125.000)(442.000,127.000)
\path(95,125)(180,125)
\path(172.000,123.000)(180.000,125.000)(172.000,127.000)
\put(0,-10){\makebox(0,0)[lb]{\raisebox{0pt}[0pt][0pt]
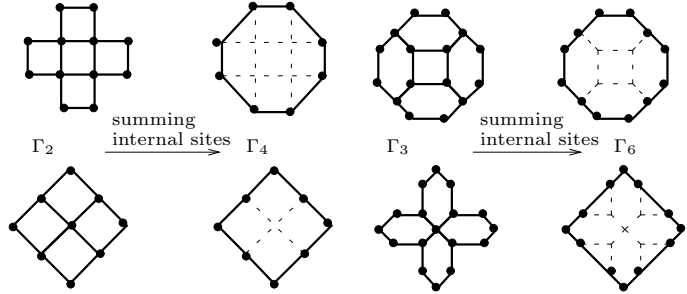
{\shortstack[l]{\footnotesize {\bf Figure(12)} Relation between 
${\cal G}_{k}$ and ${\cal G}_{2k}$. The cases of $k=2,3$ are illustrated}}}}
\put(80,-30){\makebox(0,0)[lb]{\raisebox{0pt}[0pt][0pt]
{\shortstack[l]{\footnotesize  which shows that four 
neighboring ${\cal G}_{k}$ have the geometry of a ${\cal G}_{2k}$.}}}}
\put(375,145){\makebox(0,0)[lb]{\raisebox{0pt}[0pt][0pt]
{\shortstack[l]{\scriptsize summing }}}}
\put(375,130){\makebox(0,0)[lb]{\raisebox{0pt}[0pt][0pt]
{\shortstack[l]{\scriptsize internal sites}}}}
\put(100,130){\makebox(0,0)[lb]{\raisebox{0pt}[0pt][0pt]
{\shortstack[l]{\scriptsize internal sites}}}}
\put(100,145){\makebox(0,0)[lb]{\raisebox{0pt}[0pt][0pt]
{\shortstack[l]{\scriptsize summing }}}}
%\put(5,90){\makebox(0,0)[lb]{\raisebox{0pt}[0pt][0pt]
%{\shortstack[l]{\scriptsize Dual}}}}
\put(40,125){\makebox(0,0)[lb]{\raisebox{0pt}[0pt][0pt]
{\shortstack[l]{${\scriptstyle \Gamma_{2}}$}}}}
\put(200,125){\makebox(0,0)[lb]{\raisebox{0pt}[0pt][0pt]
{\shortstack[l]{${\scriptstyle \Gamma_{4}}$}}}}
\put(305,125){\makebox(0,0)[lb]{\raisebox{0pt}[0pt][0pt]
{\shortstack[l]{${\scriptstyle \Gamma_{3}}$}}}}
\put(475,125){\makebox(0,0)[lb]{\raisebox{0pt}[0pt][0pt]
{\shortstack[l]{${\scriptstyle \Gamma_{6}}$}}}}
\end{picture}
\end{center} 

\subsection{Isotropy of the Potts model}

\hspace{5mm}

 We consider now  some symmetries of these Potts  models.  First, it 
 is clear that for arbitrary $k$ the weight $W$ is invariant under 
reflection of the the fundamental block about a horizontal or 
vertical line drawn across its center. As an example, the weight 
$W$ of the $k=2$ model is invariant under the following reflections;
\begin{equation}
\begin{array}{ll}
\mbox{reflection about vertical line}&\left\{\begin{array}{llll}
(a\:b\:c\:d)&\longrightarrow&(c\:d\:a\:b)&\;\;\;{\cal G}_{2}\;,\\
(a\:b\:c\:d)&\longrightarrow&(a\:c\:b\:d)&\;\;\;{\cal G}_{2}^{'}\;,
\end{array}\right.\\
\mbox{reflection about horizontal line}&\left\{\begin{array}{llll}
(a\:b\:c\:d)&\longrightarrow&(b\:a\:d\:c)&\;\;\;{\cal G}_{2}\;,\\
(a\:b\:c\:d)&\longrightarrow&(d\:b\:c\:a)&\;\;\;{\cal G}_{2}^{'}\;,
\end{array}\right.
\end{array}
\end{equation}
where $a,\cdots,d$ are Potts model variables associated to the 
lattice sites on the boundary of ${\cal G}_{2}$ and 
${\cal G}_{2}^{'}$ as given in figs.(4a) and (4b).

In addition, for $k$ even, the fundamental block has $90^{\rm o}$ 
rotational symmetry about its center. The weight associated to the 
fundamental block however does not remain unchanged under such a 
rotation. We first consider the $k=2$ model. The $90^{\rm o}$ 
rotation is given, using the notation of figs.(4a) and (4b), by
\begin{equation}
(a\:b\:c\:d)\longrightarrow(c\:a\:d\:b)
\end{equation}
for both ${\cal G}_{2}$ and ${\cal G}_{2}^{'}$. Under rotation, 
the corresponding weights undergo the transformation;
\begin{equation}
\begin{array}{lll}
W(f_{0},f_{1})&\longrightarrow&f_{0}W_{90^{\rm o}}(f_{0}^{-1},
f_{0}^{-1}f_{1})\;,\\ 
W^{'}(f_{0},f_{1})&\longrightarrow&f_{0}W^{'}_{90^{\rm o}}
(f_{0}^{-1},f_{0}^{-1}f_{1})\;, 
\end{array}
\end{equation}
and for $f_0=1$ they are invariant. We shall refer to this symmetry
as {\bf face isotropy}. Face isotropy is also present in higher 
even $k$ models.  To examine the behavior of the weight under this 
rotation, we recall that the underlying $\Gamma_{k}$ vertex  has 
$k$ independent parameters which can be taken as the coefficients of 
the projectors $P_{j}\;;j=1,\cdots,k$. However, a more convenient 
choice for our purpose is provided by the weights of the $k+1$ 
strand configurations from the loop model realization of the 
$\Gamma_{k}$ vertex model as explained in appendix B. This set of 
configurations for $k$ even contains a special element which is 
invariant under $90^{\rm o}$ rotation (for the case of $k=2$, it is 
given by the last picture in fig.(B1)). The others group naturally 
into $k/2$ pairs such that configurations that belong to the same 
pair differ only by $90^{\rm o}$ rotation (again, if we refer to 
fig.(B1), the first and second pictures belong to the same pair). 
The parameters of the Boltzmann weight of the $\Gamma_{k}$ Potts 
model can now be taken as the weights of these strand configurations. 
They are denoted as $(f_{0},1),(f_{1},f_{k-1}),\ldots,(f_{(k-2)/2},
f_{(k+2)/2})$ and $f_{k/2}$ where the brackets enclose the weights 
of the paired up configurations. Rotation of ${\cal G}_{k}$ or 
${\cal G}_{k}^{'}$ about its center corresponds to rotation of the 
strands configuration and is thus given by the following 
transformation of the parameters;
\begin{equation}
\begin{array}{lll}
f_{0}&\longrightarrow&f_{0}^{-1}\;,\\ 
f_{i}f_{0}^{-1/2}&\longrightarrow&f_{k-i}f_{0}^{-1/2}\;\;\;;
i=1,\ldots,k-1\;,
\end{array}
\end{equation}                                                                
which is obtained by rescaling the weight $W^{(')}$ by $f_{0}$ and 
using the fact that configurations that belong to the same pair are 
interchanged by the $90^{\rm o}$ rotation. From the above it is clear 
that face isotropy is achieved when
\begin{equation}
\begin{array}{lll}
f_{0}&=&1\;,\\ 
f_{i}&=&f_{k-i}\;\;\;;i=1,\ldots,(k-2)/2\;,
\end{array}
\end{equation}                                                                
which reduces the number of parameters to \mbox{\boldmath $k/2$}.

For $k$ odd, there is no face isotropy since the hexagonal 
fundamental blocks are not invariant under $90^{\rm o}$ rotation.

Next, we examine the behavior of the Potts model under $90^{\rm o}$ 
rotation of the lattice and shall refer to such a symmetry as 
{\bf lattice isotropy}. We shall study the {\bf staggered} case 
where weights of ${\cal G}_{k}$ that are neighboring to each other 
have independent sets of parameters, so the model depends on 
\mbox{\boldmath $2k$} parameters (in the following, staggered and 
homogeneous refer implicitely to the underlying vertex model). 
Lattice isotropy can then exist in both even and odd $k$ models. In 
fact, eventhough the lattice structures are very different between 
the odd and even $k$ models as pointed out earlier, their parameters 
basically behave the same under rotation of lattice. As an example, 
we first consider the staggered $\Gamma_{3}$ Potts model. Recall that 
the lattice is constructed out of ${\cal G}_{3}$ and 
${\cal G}_{3}^{'}$, and since the model is staggered, the two sets 
of parameters $f_{0},f_{1},f_{2}$ and $\tilde{f}_{0},\tilde{f}_{1},
\tilde{f}_{2}$ that belong respectively to the weights 
$(\!~\ref{eq:w}\,)$, $(\!~\ref{eq:w'}\,)$ of ${\cal G}_{3}$'s and 
${\cal G}_{3}^{'}$'s are considered to be independent. To examine 
the transformation of the weights, we first rewrite them as
\begin{eqnarray}
     W^{'}(\tilde{f}_{0},\tilde{f}_{1},\tilde{f}_{2})_{aebcfd}&=&
Q^{1/2}\tilde{f}_{0}\sum_{\alpha,\beta,\gamma,\delta}S_{be\alpha}
S_{df\beta}(\delta_{\alpha\beta}\delta_{\gamma\delta}\delta_{ef}
+Q^{-1/2}\tilde{f}_{0}^{-1}\delta_{\alpha\gamma}\delta_{\beta\delta}
+\tilde{f}_{2}\tilde{f}_{0}^{-1}\delta_{\alpha\gamma\beta\delta}
 \label{eq:w3}\\ 
  & &\mbox{}+Q^{-1/2}\tilde{f}_{1}\tilde{f}_{0}^{-1}
\delta_{\alpha\beta}\delta_{\gamma\delta} )S_{\gamma ea}
S_{\delta fc}\;, \nonumber  \\
     W(f_{0},f_{1},f_{2})_{aebcfd}&=&Q^{-1/2}f_{0}
\sum_{\alpha,\beta,\gamma,\delta}S_{ae\alpha}S_{be\beta}
(\delta_{\alpha\beta}\delta_{\gamma\delta}+
f_{0}^{-1}Q^{1/2}\delta_{\alpha\gamma}\delta_{ef}\delta_{\beta\delta}
+f_{2}f_{0}^{-1}\delta_{\alpha\gamma}\delta_{\beta\delta} 
\label{eq:w3'} \\ 
& &\mbox{}+f_{1}f_{0}^{-1}Q^{1/2}\delta_{\alpha\beta\gamma\delta})
S_{\gamma fc}S_{\delta fd}\;.  \nonumber 
\end{eqnarray}                                    
After the rotation of the lattice, $(\!~\ref{eq:w3}\,)$ and 
$(\!~\ref{eq:w3'}\,)$ correspond to weights of ${\cal G}_{3}$ and 
${\cal G}_{3}^{'}$ respectively, we therefore compare 
$(\!~\ref{eq:w3}\,)$ with $(\!~\ref{eq:w}\,)$ and 
$(\!~\ref{eq:w3'}\,)$ with $(\!~\ref{eq:w'}\,)$. This implies the 
following relation between the partition functions before and 
after the rotation 
\begin{equation}
Z(f_{0},f_{1},f_{2};\tilde{f}_{0},\tilde{f}_{1},\tilde{f}_{2})=
(f_{0}\tilde{f}_{0})^{N}Z_{90^{\rm o}}
(\tilde{f}_{0}^{-1},\tilde{f}_{2}\tilde{f}_{0}^{-1},\tilde{f}_{1}
\tilde{f}_{0}^{-1};f_{0}^{-1},f_{2}f_{0}^{-1},f_{1}f_{0}^{-1})\;,
\end{equation}
where $N$ denotes the total number of underlying $\Gamma_{3}$ 
vertices of the model. Clearly, lattice isotropy is given by
\begin{equation}
\begin{array}{rll}
f_{0}&=&\tilde{f}_{0}^{-1}\;,\\
f_{1}f_{0}^{-1/2}&=&\tilde{f}_{2}\tilde{f}_{0}^{-1/2}\;,\\
f_{2}f_{0}^{-1/2}&=&\tilde{f}_{1}\tilde{f}_{0}^{-1/2}\;.
\end{array}
\end{equation}
It is straight forward to extend  the study of the lattice symmetry 
to higher $k$ models. For $k$ odd, we again adopt the parametrization 
offered by the weights of the strand configurations. However, there 
are now two sets of identical strands configurations from neighboring 
${\cal G}_{k}$ and ${\cal G}_{k}^{'}$, the parameters are
$(f_{0},1),(f_{1},f_{k-1}),\ldots,(f_{(k-1)/2},f_{(k+1)/2})$ and
$(\tilde{f}_{0},1),(\tilde{f}_{1},\tilde{f}_{k-1}),\ldots,
(\tilde{f}_{(k-1)/2},\tilde{f}_{(k+1)/2})$ where the first 
and second sets belong respectively to the weights of ${\cal G}_{k}$ 
and ${\cal G}_{k}^{'}$. Parameters are paired up according to the 
same criterion as before. Notice that in this case there is no strand 
configuration which is invariant under $90^{\rm o}$ rotation. For the 
$k=3$, these are precisely the parameters that appear in 
$(\!~\ref{eq:w}\,)$ and$(\!~\ref{eq:w'}\,)$. Since rotation of the 
lattice by $90^{\rm o}$ turns ${\cal G}_{k}$ into ${\cal G}_{k}^{'}$ 
and vice versa, and also interchanges configurations that belong to 
the same pair, we have the following mapping of the parameters
\begin{equation}
\begin{array}{rll}
f_{0}&\longleftrightarrow&\tilde{f}_{0}^{-1}\;,\\ 
f_{i}f_{0}^{-1/2}&\longleftrightarrow&\tilde{f}_{k-i}
\tilde{f}_{0}^{-1/2}\;\;\;;i=1,\ldots,k-1\;.
\end{array}
\end{equation}                                                                
Thus, the model with lattice isotropy has \mbox{\boldmath $k$} 
parameters where
\begin{equation}
\begin{array}{lll}
f_{0}&=&\tilde{f}_{0}^{-1}\;,\\ 
f_{i}f_{0}^{-1/2}&=&\tilde{f}_{k-i}\tilde{f}_{0}^{-1/2}\;\;\;;
i=1,\ldots,k-1\;.
\end{array}    \label{eq:oddlisotropy}
\end{equation} 
                                                               
For even $k$, similar situation occurs. We again use the weights of 
the strand configurations as parameters, which read 
$(f_{0},1),(f_{1},f_{k-1}),\ldots,(f_{(k-2)/2},f_{(k+2)/2}),
f_{k/2}$ and $(\tilde{f}_{0},1),(\tilde{f}_{1},\tilde{f}_{k-1}),
\ldots,(\tilde{f}_{(k-2)/2},\tilde{f}_{(k+2)/2})$, $\tilde{f}_{k/2}$.
These two sets of parameters again correspond respectively to the 
weights of neighboring ${\cal G}_{k}$'s. The only difference between 
this model and the previous one is that the entire lattice is 
constructed out of ${\cal G}_{k}$ or  ${\cal G}^{'}_{k}$ exclusively. 
Under $90^{\rm o}$ rotation of the lattice, the parameters transform 
as
\begin{equation}
\begin{array}{rll}
f_{0}&\longleftrightarrow&\tilde{f}_{0}^{-1}\;,\\ 
f_{i}f_{0}^{-1/2}&\longleftrightarrow&\tilde{f}_{k-i}
\tilde{f}_{0}^{-1/2}\;\;\;;i=1,\ldots,k-1\;,
\end{array}
\end{equation}                                                                
for similar reasons as in the odd $k$ case. The Potts model with 
lattice isotropy is given by 
\begin{equation}
\begin{array}{lll}
f_{0}&=&\tilde{f}_{0}^{-1}\;,\\ 
f_{i}f_{0}^{-1/2}&=&\tilde{f}_{k-i}\tilde{f}_{0}^{-1/2}\;\;\;;
i=1,\ldots,k-1\;,
\end{array}  \label{eq:evenlisotropy}
\end{equation}                                                                
and has \mbox{\boldmath $k$} independent parameters. 

Conditions $(\!~\ref{eq:oddlisotropy}\,)$ and 
$(\!~\ref{eq:evenlisotropy}\,)$ both define staggered models in 
general.

The above transformations of the parameters  show conversely that the 
{\bf homogeneous} model where
\begin{equation}
\begin{array}{lll}
f_{0}&=&\tilde{f}_{0}\;,\\ 
f_{i}&=&\tilde{f}_{i}\;\;\;;i=1,\ldots,k-1\;,
\end{array}
\end{equation}
is in general not invariant under the rotation of the lattice except 
when face isotropy  is present in every ${\cal G}_{k}$'s weight and 
this applies only to even $k$ models.

\subsection{Self duality}

\hspace{5mm}

In the construction of the $\Gamma_{k}$ Potts model, there exist two 
possible conventions (A or B) that can be used. These two choices 
give rise to two lattices. For $k$ odd, the lattices ${\cal L}$ and 
${\cal L}^{'}$ have identical structure due to the fact that both 
${\cal G}_{k}$ and ${\cal G}_{k}^{'}$ have to be used, while for 
even $k$, ${\cal L}$ and ${\cal L}^{'}$ are  different. Let us 
discuss the relation between the models obtained by conventions A 
and B in more details. We consider a general staggered Potts model.
We first look at the $\Gamma_{1}$ Potts model to examine the origin 
of this two choices. We have, in the notation of this section, 
neighboring $\Gamma_{1}$ vertices $X_{2i-1}$ and $X_{2i}$ given by
\begin{equation}
\begin{array}{lll} 
      X_{2i-1}&=&{\bf 1}+f_{0}e_{2i-1}\;,\\                          
      X_{2i}&=&{\bf 1}+\tilde{f}_{0}e_{2i}         
\end{array}               \label{eq:X2}        
\end{equation}                                
where $f_{0}$ and $\tilde{f}_{0}$ are independent (they should be 
identified with $x_{1}^{-1},x_{2}$ defined in section 2).
Mapping to Potts model using convention A, we replace $X_{2i-1}$ and 
$X_{2i}$ respectively by horizontal and vertical links. The vertices 
in terms of Potts model variables read
\begin{equation}
\begin{array}{lllll} 
      (X_{2i-1})_{ab}&=&\delta_{ab}+f_{0}Q^{-1/2}\;,\\                          
      (X_{2i})_{bc}&=&1+\tilde{f}_{0}Q^{1/2}\delta_{bc}\;,         
\end{array}                      \label{eq:r2}
\end{equation}                                                            
where $a\:b$ and $b\:c$ are the sites on the horizontal and vertical 
links. After duality transformation \cite{bax} $X_{2i-1}$ and 
$X_{2i}$ are instead associated to a vertical and horizontal link, 
which corresponds to convention B 
\begin{equation}
\begin{array}{lllll} 
      (X_{2i-1})_{bc}&=&1+f_{0}Q^{1/2}\delta_{bc}\;,\\                          
      (X_{2i})_{ab}&=&\tilde{f}_{0}Q^{-1/2}+\delta_{ab}\;,         
\end{array}                      \label{eq:r2d}
\end{equation}               
where we stick to the convention that $a\:b$ ($b\:c$) are sites on 
the horizontal (vertical) link. This implies, following fusion, that 
for arbitrary $k$, the Potts models obtained by conventions A and B 
are related by duality transformation.

Combining the above argument with result of subsection 3.3 on the 
structure of the lattices ${\cal L}$ and ${\cal L}^{'}$, we see that 
the question of self duality does not arise  for the family of even 
$k$ Potts models since the lattice ${\cal L}$ and its dual 
${\cal L}^{'}$ are not the same. But for $k$ odd, self duality can 
occur since both ${\cal L}$ and ${\cal L}^{'}$ have the same 
structure. The self duality condition can be easily identified for 
the $\Gamma_{1}$ Potts model by comparing $(\!~\ref{eq:r2}\,)$ with 
$(\!~\ref{eq:r2d}\,)$. We see that duality transformation amounts to 
\begin{equation}
f_{0} \longleftrightarrow \tilde{f}_{0}\;,
\end{equation}
which is equivalent to  $(\!~\ref{eq:map}\,)$. For higher odd $k$ 
models, similar transformation of the parameters can be deduced by 
using the fact that duality map is equivalent to interchanging 
convention A with convention B. Thus using the same set of 
parameters as in the previous subsection, the duality transformation 
 is given by
\begin{equation}
\begin{array}{lll}
f_{0}&\longleftrightarrow&\tilde{f}_{0}\;,\\               
f_{i}&\longleftrightarrow&\tilde{f}_{i}\;\;\;;i=1,\ldots,k-1\;.
\end{array}
\end{equation}                                     
And self duality is given by the condition
\begin{equation}
\begin{array}{lll}
f_{0}&=&\tilde{f}_{0}\;,\\               
f_{i}&=&\tilde{f}_{i}\;\;\;;i=1,\ldots,k-1\;.
\end{array}
\end{equation}                                     
which means that the model becomes {\bf homogeneous} with 
\mbox{\boldmath $k$} parameters. If the $\Gamma_{k}$ Potts model has 
lattice isotropy to begin with, then self duality map is given by 
\begin{equation}
\begin{array}{lll}
f_{0}&\longleftrightarrow&f^{-1}_{0}\;,\\               
f_{i}f_{0}^{-1/2}&\longleftrightarrow&f_{k-i}f_{0}^{-1/2}\;\;\;
i=1,\ldots,k-1\;.
\end{array}    \label{eq:l&d}
\end{equation}                                   
where the relation $(\!~\ref{eq:oddlisotropy}\,)$ has been used. 
Thus imposing the conditions of self duality and lattice isotropy, 
the number of parameters reduces to \mbox{\boldmath $(k-1)/2$} since 
we now have
\begin{equation}
\begin{array}{lllllll}
f_{0}&=&\tilde{f}_{0}&=&1& &\;,\\               
f_{i}&=&f_{k-i}&=&\tilde{f}_{i}&=&\tilde{f}_{k-i}\;\;\;;i=1,
\ldots,(k-1)/2\;.
\end{array}    \label{eq:gk}
\end{equation}  
Recall that in the case $k=1$ these conditions determined completely 
the Potts model interaction.

\subsection{Integrability}

\hspace{5mm}

We  conclude this section by discussing some special cases of the 
Potts models which are known to be integrable. For simplicity, we 
restrict ourselves to the homogeneous models. In this restricted 
class, the $\Gamma_{1}$ distinguishes itself from the rest in that 
it is integrable for any $f_{0}$. The corresponding $R$ matrix can 
be written as
\begin{equation}
\check{R}(u)=1+\frac{\sin u}{\sin(\gamma-u)}e\;,
\end{equation}
where
\[f_{0}=\frac{\sin u}{\sin(\gamma-u)}\;.\]
For higher $k$, only subsets of the full parameter space are 
integrable. A standard such case is obtained from the special choice 
in $(\!~\ref{eq:Xu}\,)$
\begin{equation}
\left\{ \begin{array}{rcll}
    u_{j}&=&u_{j+1}-\gamma&;\;\; (i-1)k <j\leq ik<k^{2}\;,\\   
    u_{ik}&=&u+(i-1)\gamma&\;\;\;;i=1,\ldots,k\;,      
\end{array} \right.      \label{eq:u}
\end{equation}
where $u$ is the spectral parameter of the $R$ matrix . We shall 
refer to this as the {\bf JB} integrable line as this is constructed 
by Jimbo in \cite{krs}. The $R$ matrix for this $\Gamma_{k}$ model 
can also be written as a linear combination of projectors;
\begin{equation}                                       
    \check{R}(u)=P_{j=k}+\frac{y^{2}-q^{2k}}{1-y^{2}q^{2k}}P_{j=k-1}
+ \ldots +\frac{y^{2}-q^{2k}}{1-y^{2}q^{2k}}\frac{y^{2}-q^{2(k-1)}}
{1-y^{2}q^{2(k-1)}}\ldots \frac{y^{2}-q^{2}}{1-y^{2}q^{2}}P_{j=0}\;,
\end{equation}
where $y=\exp(-iu)$ is the multiplicative spectral parameter.
The corresponding Potts model is also integrable since the fact that 
Yang Baxter equation is satisfied is expressed algebraically,
without reference to a particular representation. In terms of the 
parameters $f_{i}$'s of the $\Gamma_{2}$ and $\Gamma_{3}$ Potts 
models the JB integrable line is given by the following formula
\begin{equation}
\begin{array}{ll}
\Gamma_{2}&\left\{\begin{array}{lll}
          f_{0}& = &\frac{\textstyle \sin (u)\,\sin (\gamma+u)}
{\textstyle \sin (2\gamma-u)\,
                    \sin (\gamma-u)}\;,\\
          f_{1}& = &\frac{\textstyle Q^{1/2}\,\sin u}
{\textstyle \sin (2\gamma-u)}\;,
\end{array}\right.      \\
\Gamma_{3}&\left\{\begin{array}{lll}
f_{0}&=&\frac{\textstyle \sin u\,\sin (u+\gamma)\,\sin (u+2\gamma)}
{\textstyle \sin (\gamma-u)\,\sin (2\gamma-u)\,\sin (3\gamma-u)}\;,\\
f_{1}&=&\frac{\textstyle (Q-1)\,\sin u\,\sin (u+\gamma)}
{\textstyle \sin (3\gamma-u)\,\sin (2\gamma-u)}\;,\\ 
f_{2}&=&\frac{\textstyle (Q-1)\,\sin u}{\textstyle \sin (3\gamma-u)}\;.
\end{array}\right.      
\end{array}         \label{eq:FZline}
\end{equation}                                              
In terms of the spectral parameter $u$, rotation of the lattice 
by $90^{\rm o}$ is given by
\[u\longrightarrow \gamma-u\;.\]
This transformation of $u$ actually applies to all the $\Gamma_{k}$ 
models on the JB integrable line. The condition for lattice isotropy 
$f_0=1$ is therefore equivalent to  $u=\gamma/2$, and the parameters 
$f_{i}$'s read
\begin{equation}
\begin{array}{ll}
\Gamma_{2}&\left\{\begin{array}{lll}
          f_{0}& = &1\;,\\ 
          f_{1}& = &\frac{Q^{1/2}}{Q^{1/2}+1}\;,
\end{array}\right.\\
\Gamma_{3}&\left\{\begin{array}{lllll}
          f_{0}&=&1& &\;,\\
          f_{1}&=&f_{2}&=&\frac{(Q-1)}{Q+Q^{1/2}-1}.          
\end{array}\right. \label{eq:in}  
\end{array} 
\end{equation}                                              

Besides the JB integrable line there is another integrable line in 
all $\Gamma_{k}$ models, which is related to the Temperley-Lieb 
algebra. For arbitrary $k$, since the projector $(k+1)_{q}P_{0}$
satisfies the Temperley-Lieb algebra\cite{saa} with
\begin{equation}
\begin{array}{rll}
e&=&(k+1)_{q}P_{0}\\
\mbox{ and }\;\;\;e^{2}&=&(k+1)_{q}e\;,
\end{array}
\end{equation}
the $R$ matrix
\begin{equation}
\check{R}(u)={\bf 1}+\frac{\sin u}{\sin(\gamma-u)}(k+1)_{q}P_{0} 
\label{eq:TLintegrable}
\end{equation}
satisfies the Yang-Baxter equation with $u$ being the spectral 
parameter. We shall refer to this line as the {\bf TL} integrable 
line. Obviously, for $k=1$, the TL and JB lines coincide. The Potts 
model which corresponds to the above $R$ matrix 
$(\!~\ref{eq:TLintegrable}\,)$ has parameters given by
\begin{equation}
\begin{array}{lll}
f_{0}&=&\frac{\textstyle \sin u}{\textstyle \sin(\gamma-u)}\;,\\
f_{i}&=&0\;\;\;;i=1,\cdots,k-1\;.
\end{array} \label{eq:TLline}
\end{equation}

For $k\geq2$, these two lines JB and TL do not exhaust all the 
integrable cases. As an example, for the $\Gamma_{2}$ model, there 
is an additional integrable line given by
\begin{equation}
\begin{array}{lll}
     f_{0}& = &-\frac{\textstyle \sin (u)\,\cos (\gamma-u)}
{\textstyle \sin (2\gamma-u)\,\cos (3\gamma-u)}\;,\\
     f_{1}& = &\frac{\textstyle Q^{1/2}\,\sin u}
{\textstyle \sin (2\gamma-u)}\;, 
     \end{array}                     \label{eq:IKline}
\end{equation}                   
which is related to the Izergin-Korepin  model\cite{ik}. In this case
the model with lattice isotropy is given by
\begin{equation}
u=3\gamma/2+\pi/4
\end{equation}
or
\begin{equation}
\begin{array}{lll}
f_{0}&=&1\;,\\
f_{1}&=&-(q+q^{-1})(1+i(q-q^{-1}))\;.
\end{array}
\end{equation}

\section {The Spontaneous Magnetization}
 
\subsection{ Definitions }
 
\hspace{5mm}
 
The spontaneous magnetization of the usual ($\Gamma_1$) Potts model 
on the first order transition line has been computed exactly by 
Baxter\cite{bax1} using the method of corner transfer matrix (CTM). 
We will show in this section that using the fusion procedure, 
spontaneous magnetizations of the homogeneous $\Gamma_{k}$ Potts 
model can similarly be computed. As for $k=1$ the model is expected 
 to undergo a first order phase transition along the JB integrable 
line for $Q>4$ \cite{bax},\cite{djkm}. In what follows, we shall 
restrict $Q$ to this range and replace $\gamma$ by $-i\lambda$  
with $\lambda$ being real, thus
\begin{equation}
\sqrt{Q}=2\cosh\lambda
\end{equation}
and work exclusively on the JB integrable line.
 
To define the spontaneous magnetization for the fused Potts model, 
we generalize the work of \cite{bax1} and consider some fundamental 
block ${\cal G}_{k}$( or ${\cal G}_{k}^{'}$ ) sufficiently 
remote from the boundary and fix a site $\sigma_{\rm o}$ 
which we shall refer to as central site. The spontaneous 
magnetization is then defined as
\begin{equation}
M=\frac{Q\langle \delta_{\sigma_{\rm o},1}\rangle-1}{Q-1}
\end{equation}
where   
\[ \langle \delta_{\sigma_{\rm o},1}\rangle=Z_{\rm Potts}^{-1}
\sum_{\{\sigma\}}\delta_{\sigma_{\rm o},1}\prod_{{\cal G}_{k}^{(')}}
W({\cal G}_{k}^{(')})\;.   \]

\begin{center} 
\setlength{\unitlength}{0.008in}
\begin{picture}(318,340)(0,-40)
\put(165,151){\circle*{6}}
\path(191,280)(139,280)
\path(147.000,282.000)(139.000,280.000)(147.000,278.000)
\path(139,280)(139,229)
\path(137.000,237.000)(139.000,229.000)(141.000,237.000)
\path(139,229)(88,229)
\path(96.000,231.000)(88.000,229.000)(96.000,227.000)
\path(88,229)(88,176)
\path(86.000,184.000)(88.000,176.000)(90.000,184.000)
\path(88,176)(36,176)
\path(44.000,178.000)(36.000,176.000)(44.000,174.000)
\path(36,176)(36,125)
\path(34.000,133.000)(36.000,125.000)(38.000,133.000)
\path(36,125)(88,125)
\path(80.000,123.000)(88.000,125.000)(80.000,127.000)
\path(88,125)(88,73)
\path(86.000,81.000)(88.000,73.000)(90.000,81.000)
\path(88,73)(139,73)
\path(131.000,71.000)(139.000,73.000)(131.000,75.000)
\path(139,73)(139,22)
\path(137.000,30.000)(139.000,22.000)(141.000,30.000)
\path(139,22)(191,22)
\path(183.000,20.000)(191.000,22.000)(183.000,24.000)
\path(191,22)(191,73)
\path(193.000,65.000)(191.000,73.000)(189.000,65.000)
\path(191,73)(242,73)
\path(234.000,71.000)(242.000,73.000)(234.000,75.000)
\path(242,73)(242,125)
\path(244.000,117.000)(242.000,125.000)(240.000,117.000)
\path(242,125)(295,125)
\path(287.000,123.000)(295.000,125.000)(287.000,127.000)
\path(295,125)(295,176)
\path(297.000,168.000)(295.000,176.000)(293.000,168.000)
\path(295,176)(242,176)
\path(250.000,178.000)(242.000,176.000)(250.000,174.000)
\path(242,176)(242,229)
\path(244.000,221.000)(242.000,229.000)(240.000,221.000)
\path(242,229)(191,229)
\path(199.000,231.000)(191.000,229.000)(199.000,227.000)
\path(191,229)(191,280)
\path(193.000,272.000)(191.000,280.000)(189.000,272.000)
\path(139,229)(191,229)(191,73)
	(139,73)(139,229)
\path(88,176)(242,176)(242,125)
	(88,125)(88,176)
\dashline{4.000}(139,254)(114,229)(114,203)
	(139,203)(165,229)(165,254)(139,254)
\dashline{4.000}(139,151)(114,176)(114,203)
	(139,203)(165,176)(165,151)(139,151)
\dashline{4.000}(88,203)(62,176)(62,151)
	(88,151)(114,176)(114,203)(88,203)
\dashline{4.000}(88,99)(62,125)(62,151)
	(88,151)(114,125)(114,99)(88,99)
\dashline{4.000}(139,151)(114,125)(114,99)
	(139,99)(165,125)(165,151)(139,151)
\dashline{4.000}(191,151)(217,125)(217,99)
	(191,99)(165,125)(165,151)(191,151)
\dashline{4.000}(191,151)(217,176)(217,203)
	(191,203)(165,176)(165,151)(191,151)
\dashline{4.000}(191,254)(217,229)(217,203)
	(191,203)(165,229)(165,254)(191,254)
\dashline{4.000}(242,151)(217,176)(217,203)
	(242,203)(268,176)(268,151)(242,151)
\dashline{4.000}(242,151)(217,125)(217,99)
	(242,99)(268,125)(268,151)(242,151)
\dashline{4.000}(139,48)(114,73)(114,99)
	(139,99)(165,73)(165,48)(139,48)
\dashline{4.000}(191,48)(217,73)(217,99)
	(191,99)(165,73)(165,48)(191,48)
\put(230,37){\makebox(0,0)[lb]{\raisebox{0pt}[0pt][0pt]
{\shortstack[l]{\footnotesize ${\scriptstyle N=3\;\in\;}$odd}}}}
\put(170,159){\makebox(0,0)[lb]{\raisebox{0pt}[0pt][0pt]
{\shortstack[l]{${\scriptstyle \sigma_{\rm o}}$}}}}
\put(144,134){\makebox(0,0)[lb]{\raisebox{0pt}[0pt][0pt]
{\shortstack[l]{${\scriptstyle \alpha_{1}}$}}}}
\put(148,31){\makebox(0,0)[lb]{\raisebox{0pt}[0pt][0pt]
{\shortstack[l]{${\scriptstyle \alpha_{N}}$}}}}
\put(62,206){\makebox(0,0)[lb]{\raisebox{0pt}[0pt][0pt]
{\shortstack[l]{\footnotesize c}}}}
\put(6,155){\makebox(0,0)[lb]{\raisebox{0pt}[0pt][0pt]
{\shortstack[l]{\footnotesize c}}}}
\put(100,206){\makebox(0,0)[lb]{\raisebox{0pt}[0pt][0pt]
{\shortstack[l]{\footnotesize b}}}}
\put(50,155){\makebox(0,0)[lb]{\raisebox{0pt}[0pt][0pt]
{\shortstack[l]{\footnotesize b}}}}
\put(148,164){\makebox(0,0)[lb]{\raisebox{0pt}[0pt][0pt]
{\shortstack[l]{\footnotesize a}}}}
\put(-60,0){\makebox(0,0)[lb]{\raisebox{0pt}[0pt][0pt]
{\shortstack[l]{\footnotesize {\bf Figure(13)} The geometry of the
$\Gamma_{k}$ Potts model lattice; $a,b,c$ are the face variables,}}}}
\put(20,-20){\makebox(0,0)[lb]{\raisebox{0pt}[0pt][0pt]
{\shortstack[l]{\footnotesize  the central site $\sigma_{\rm o}$ is taken in 
the figure to be the filled circle at the center, }}}}
\put(20,-40){\makebox(0,0)[lb]{\raisebox{0pt}[0pt][0pt]
{\shortstack[l]{\footnotesize and the spin variables 
are denoted by the $\alpha_{i}$ where $i=1$ and $N$ are shown.  }}}}
\end{picture}
\end{center} 

For the sake of computation, we consider the underlying $\Gamma_{k}$ 
vertex model lattice to be an $l\times l$ square as in fig.(13)
 where we show the $\Gamma_{k}$( $k$ odd in this case ) vertices and 
the respective fundamental blocks ${\cal G}_{k}$, the $\alpha_{i}$'s 
denote the spin-$\frac{k}{2}$ arrows which can have $k+1$ states 
$\frac{k}{2},\frac{k}{2}-1,\ldots,-\frac{k}{2}$, and $i$
takes values from 1 to $N$ with $N$ taken to be odd always. As for 
$k=1$ the boundary conditions  are conveniently defined  in terms of 
the vertex degrees of freedom: we require that  the spin arrows 
along the perimeter all have the same state.  This provides  $k+1$ 
boundary conditions of which we assume (more later) that they 
select different phases of the $\Gamma_{k}$ Potts model. 

Recall that each spin-$\frac{k}{2}$ arrow can be regarded from the 
fusion point of view as being made up of $k$ spin-$\frac{1}{2}$
arrows. These spin-$\frac{1}{2}$ arrows form surrounding polygons 
for the Potts model links, in particular, due to the boundary 
condition, the perimeter can be viewed as $k$  polygons enclosing 
the Potts model.  We then define the central site $\sigma_{\rm o}$ 
to be connected to the boundary if there is no spin-$\frac{1}{2}$ 
surrounding polygon enclosing it other than those from the perimeter. 
With this definition, we can now relate the spontaneous 
magnetization to the percolation probability $P$ defined as the
probability of $\sigma_{\rm o}$ being connected to the boundary. It 
is easy to see that 
\begin{equation}
\langle \delta_{\sigma_{\rm o},1}\rangle=Q^{-1}(1-P)+P
\end{equation}
and 
\begin{equation}
M=P\;.
\end{equation}
        
As in the case of $\Gamma_{1}$ Potts model, the percolation 
probability can be expressed in terms of variables of the vertex
model. Referring to fig.(13), we define the following quantity
\cite{kel}
\begin{equation}
S(\alpha)={\rm e}^{-(i\pi+2\lambda)(\alpha_{1}+\ldots+\alpha_{N-1})}            
\end{equation}
and its expectation value
\begin{equation}
\langle S(\alpha) \rangle=Z^{-1}_{\rm vertex}\sum_{\{\alpha\}}
S(\alpha)\prod({\rm vertex}\;\;{\rm weight})\;.
\end{equation}
Recall that closed loop formed by the spin-$\frac{1}{2}$ surrounding 
polygon has orientation given by that of the spin arrow, and it 
acquires a weight (${\rm e}^{\lambda}$) ${\rm e}^{-\lambda}$ when 
the direction is (anti-)clockwise. Writing the $\alpha_{i}$'s in 
$S(\alpha)$ as sum of spin-$\frac{1}{2}$ states, it is clear that 
when $\sigma_{\rm o}$ is connected to the boundary, there are as 
many left pointing spin-$\frac{1}{2}$ arrows among $\alpha_{1},
\ldots,\alpha_{N-1}$ as there as right pointing ones, giving 
\[ S(\alpha)=1\;.        \]
On the other hand, if $\sigma_{\rm o}$ is not connected to the 
boundary, there must be some  spin-$\frac{1}{2}$ surrounding 
polygons enclosing the central site in addition to those from the 
boundary. Each of these polygons includes odd number of 
spin-$\frac{1}{2}$ states of $\alpha_{1},\ldots,\alpha_{N-1}$. The
total contribution of each polygon, taking into account the weight 
from its orientation, is
\[ {\rm e}^{\lambda}{\rm e}^{-(i\pi+2\lambda)/2}+
{\rm e}^{-\lambda}{\rm e}^{(i\pi+2\lambda)/2}=0\;.      \]
Thus $S(\alpha)$ counts the number of polygon configurations for 
which $\sigma_{\rm o}$ is connected to the boundary. We therefore 
have
\begin{equation}
\langle S(\alpha) \rangle= P= M\;.
\end{equation}   
 
Having established the above equality we could investigate directly
 the spectrum of the relevant corner transfer matrix. There is 
however a faster way that uses already known results for solid on 
solid models.

\subsection{ The local height probability and spontaneous 
magnetization }
 
\hspace{5mm}
  
The mapping between $\Gamma_{k}$ vertex model and sos\cite{bei} 
is standard. Height variables $l_{i}$'s are assigned to faces 
separated by the arrow spins (see fig.(14)). The heights are given 
integer values consistent with 
\begin{equation}
l_{i}-l_{i-1}=2\beta_{i}\;\;\;\in\;\;\{k,k-2,\ldots,-k\}  \label{eq:c}
\end{equation}
where $l_{i}$ and $l_{i-1}$ are heights of faces separated the spin 
arrow $\beta_{i}$. Each vertex is replaced by a square face with 
heights attached to the four corners and contributes to the 
partition function a weight $W(l_{i-1},l_{i},l_{i+1},l_{i}^{'})$ 
which is set equal to the weight of the underlying vertex.

\begin{center} 
\setlength{\unitlength}{0.008in}
\begin{picture}(260,152)(0,-10)
\put(75,118){\circle*{6}}
\put(41,80){\circle*{6}}
\put(75,43){\circle*{6}}
\put(110,78){\circle*{6}}
\path(39,115)(110,44)
\path(110,115)(39,44)
\dashline{4.000}(75,115)(39,80)(75,44)
	(110,80)(75,115)
\path(136,97)(216,97)
\path(208.000,95.000)(216.000,97.000)(208.000,99.000)
\put(57,128){\makebox(0,0)[lb]{\raisebox{0pt}[0pt][0pt]
{\shortstack[l]{${\scriptstyle l_{i-1}}$}}}}
\put(57,22){\makebox(0,0)[lb]{\raisebox{0pt}[0pt][0pt]
{\shortstack[l]{${\scriptstyle l_{i+1}}$}}}}
\put(123,75){\makebox(0,0)[lb]{\raisebox{0pt}[0pt][0pt]
{\shortstack[l]{${\scriptstyle l_{i}^{'}}$}}}}
\put(0,75){\makebox(0,0)[lb]{\raisebox{0pt}[0pt][0pt]
{\shortstack[l]{${\scriptstyle l_{i}}$}}}}
\put(20,124){\makebox(0,0)[lb]{\raisebox{0pt}[0pt][0pt]
{\shortstack[l]{${\scriptstyle \beta_{i}}$}}}}
\put(114,124){\makebox(0,0)[lb]{\raisebox{0pt}[0pt][0pt]
{\shortstack[l]{${\scriptstyle \beta_{i}^{'}}$}}}}
\put(114,36){\makebox(0,0)[lb]{\raisebox{0pt}[0pt][0pt]
{\shortstack[l]{${\scriptstyle \beta_{i+1}^{'}}$}}}}
\put(10,36){\makebox(0,0)[lb]{\raisebox{0pt}[0pt][0pt]
{\shortstack[l]{${\scriptstyle \beta_{i+1}}$}}}}
\put(141,102){\makebox(0,0)[lb]{\raisebox{0pt}[0pt][0pt]
{\shortstack[l]{\footnotesize Time direction}}}}
\put(-60,0){\makebox(0,0)[lb]{\raisebox{0pt}[0pt][0pt]
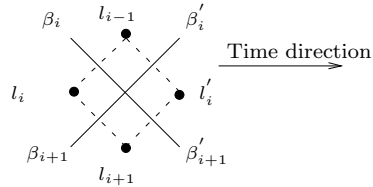
{\shortstack[l]{\footnotesize {\bf Figure(14)} Vertex and sos correspondance}}}}
\end{picture}
\end{center} 
 
This sos model is a special case of the fused eight-vertex restricted 
solid on solid model(rsos)\cite{abf,djkm}.  The rsos model is 
integrable and is equivalent to the fusion of the eight-vertex model 
where heights can assume values from 1 to $L-1$ besides the above 
constraint $(\!~\ref{eq:c}\,)$. In the limit\cite{frt}  
\begin{equation}
\begin{array}{rcl}
L& \longrightarrow &\infty\;\;\;{\rm and}\\
l_{i}& \longrightarrow& \infty\;\;\; {\rm for\;\; all }\;\; i\;,
\end{array}         \label{eq:l}
\end{equation}
such that relative heights of neighbouring faces remains unchanged, 
the sos model can be recovered from the rsos model. More precisely, 
the regime III of the rsos model corresponds to the $Q>4$ range of 
the sos model in the above limit\cite{sal2}. The Local height 
probability defined as\cite{abf}
\begin{equation}
P(a/b,c)=\frac{Z(a/b,c)}{\sum_{a=1}^{L-1}Z(a/b,c)}   \label{eq:p}
\end{equation}
has been computed exactly for the rsos model in the thermodynamics 
limit using the method of corner transfer matrix\cite{djkm}. In this 
formula, $Z(a/b,c)$ denotes the rsos partition function with central 
height given by $a$, and $b\:c$ are heights of the faces at the 
boundary of the lattice that determine the state of the arrow spin 
on the perimeter (see fig.(13)). The lattice we considered has the 
$b$ face separated from the $a$ face (the central face) by even 
number of steps (see fig.(13)). In regime III, the rsos model has 
ground state configurations such that all faces separated by even 
steps assume the same value while all other faces assume another 
fixed value. The sos model has also the same ground state 
configurations since relative heights are not affected by the limit. 
A ground state is selceted by fixing the  heights $b\:c$ at the 
boundary.
 
   Taking the above limit $(\!~\ref{eq:l}\,)$, the local height 
probability in the the thermodynamics limit has the expression 
\begin{equation}
P(a/b,c)=\frac{x^{((b+c)/2-a)^{2}/2k}C^{l}_{m}}{\sum_{a=0}^{\infty}
x^{((b+c)/2-a)^{2}/2k}C^{l}_{m}}
\end{equation}
where
\[\begin{array}{lll}
l&=&(c-b+k)/2\;,\\         
m&=&(b+c)/2-a+k/2 \bmod 2k\;,\\      
x&=&{\rm e}^{-2 \lambda}\;,    
\end{array}\]     
and $C^{l}_{m}(x)$ is the SU(2) level-$k$ string
 function\cite{kac}, which depends 
on $\lambda$ and has the following properties
\begin{equation}
\left\{ \begin{array}{rcccl}
    C^{l}_{m} & = & 0 &  & {\rm if }\;\; l \neq m \bmod 2\;,\\ \\
    C^{l}_{m} & = & C^{l}_{m+2k} &=& C^{l}_{-m}\;, \\  \\
    C^{l}_{m} & = & C^{k-l}_{k-m} &=& C^{k-l}_{k+m}\;. 
\end{array}  \right.   \label{eq:prop}
\end{equation} 
Notice that there is no dependence on the spectral parameter $u$, 
and $C^{l}_{m}$ in the above formula is nonvanishing because $a$ 
and $b$ are  on the same sublattice.
 
The spontaneous magnetization can now be computed by noting that the 
exponent in $S(\alpha)$ is 
\[ 2(\alpha_{1}+\ldots+\alpha_{N-1})=b-a \]
and making the substitution
\[ S(\alpha)={\rm e}^{-(\lambda + i\pi/2)(b-a)} \]
which depends solely on the central height for given boundary 
condition. The expectation value therefore becomes
\begin{equation}    
 \langle S(\alpha) \rangle =\sum_{a=0}^{\infty}
{\rm e}^{-(\lambda+i\pi/2)(b-a)}P(a/b,c)\;. 
\end{equation}    
Replacing in the above formula $m$ by $m+2nk$ for 
 \[m=-k+1,-k+2,\ldots,k   \]
and \[ b-a=n\in {\bf Z}\;, \]
the spontaneous magnetization becomes
\begin{equation}
M= \langle S(\alpha) \rangle
=\frac{\sum_{m}\sum_{n}x^{2kn^{2}+2mn+m^{2}/2k+k/8-l/2}
(-1)^{(m-l)/2+nk} C^{l}_{m}}
{\sum_{m}\sum_{n}x^{2kn^{2}+(2m-k)n+(m-k/2)^{2}/2k}C^{l}_{m}}\;, 
\end{equation}
where 
\[l=k,k-1,\ldots,0\]     
specifies the various boundary conditions. The formula can be expressed 
as finite sums of products of elliptic theta functions and string 
functions. Putting $k=1$ into the above formula, we recover 
Baxter's\cite{bax1} results for the spontaneous magnetization of the 
standard Potts model. For given $k$, we find   $k+1$ spontaneous 
magnetizations associated to the various boundary conditions of the 
vertex model. However, it is not difficult to show  using
$(\!~\ref{eq:prop}\,)$ that whenever $l$ is odd, $M=0$. The latter 
condition is equivalent to
\begin{equation}
c-b=\left\{  \begin{array}{ll} 
     k,k-4,\ldots,-k+2\;\;\;&{\rm for\;\;odd\;\; }k\;,\\ 
     k-2,k-6,\ldots,-k+2\;\;\;&{\rm for\;\;even\;\; }k\;,  
\end{array}    \right.             \nonumber
\end{equation}                                                             
 Depending on the parity of $k$, 
there are thus $(k+1)/2$ or $k/2$ vertex boundary conditions that  
actually give rise to vanishing magnetization. $M\neq 0$ for even  
$l$, ie.
\begin{equation}
l=  \left\{   \begin{array}{ll}
    k-1,k-3,\ldots,0&\;\;\;{\rm for\;\; odd\;\;}k\\
    k,k-2,\ldots,0&\;\;\;{\rm for\;\; even\;\;}k
\end{array}   \right.       \nonumber
\end{equation}
One then has  
\begin{equation}                   
M=\frac{\sum_{m=-k+1}^{k}\theta_{\nu}(2mi\lambda,x^{2k})
x^{m^{2}/2k+k/8-l/2}(-1)^{(m-l)/2}C^{l}_{m}}
{\sum_{m=-k+1}^{k}\theta_{3}((2m-k)i\lambda,x^{2k})x^{(m-k/2)^{2}/2k}
C^{l}_{m}}
\label{eq:magnet}
\end{equation}  
with 
\[\nu =  \left\{ \begin{array}{ll}  
          4&\;\;\;{\rm for\;\; odd\;\;}k   \\
          3&\;\;\;{\rm for\;\; even\;\;}k\;,    
\end{array}       \right.      \]
where the elliptic theta functions $\theta_{\nu}$ are defined as
\begin{eqnarray}
\theta_{3}(u,q)&=&\sum_{n \in Z}q^{n^{2}}{\rm e}^{2niu}  \nonumber  \\ 
\theta_{4}(u,q)&=&\sum_{n \in Z}(-1)^{n}q^{n^{2}}{\rm e}^{2niu}\;. 
\nonumber
\end{eqnarray} 
The above expressions are distinct for the various values of $l$. 
 For example when $k=2$, the results for 
$l=0, 2$ have different behaviors in the large $Q$ limit as shown below. 
With the help of the following approximation for the string 
functions\cite{jmo}
\begin{equation}
C^{l}_{l+m}\stackrel{x \rightarrow 0}{\approx}\left\{ 
\begin{array}{lc}
x^{-m^{2}/2k-ml/k}C^{l}_{l}(1+{\rm O}(x^{2}))&m \leq 0\\  \\
x^{-m^{2}/2k-m(k-l)/k}C^{l}_{l}(1+{\rm O}(x^{2}))&m>0
\end{array}  \right.
\end{equation}
valid for $0\leq l \leq k$, one finds that in the limit where $Q$  
approaches infinity, the spontaneous magnetizations for the various 
boundary conditions given by even $l$  approach unity as follows
\begin{equation}
 M  \stackrel{\lambda \rightarrow \infty}{\approx} \left\{ 
\begin{array}{ll}   
1-2x-x^{2}+(x-x^{2}+\delta_{k,1}x^{2})\delta_{l,0}\;\;\;&
\mbox{for odd }k\;,\\  
 1-2x-x^{2}+(x-x^{2})\delta_{l,0}+x\delta_{l,k}\;\;\;&
\mbox{for even }k\;.
\end{array}  \right.   \label{eq:a1}
\end{equation}           
Another limit to consider is $Q=4^{+}$ or $\lambda=0^{+}$ which 
divides the regions of first and second order transitions along the 
integrable line\cite{cns}. The expansion around $Q-4$ can be obtained 
by employing the modular transformation formula of the string functions,
\begin{equation}
C^{l}_{m}(\tau)=\sqrt{\frac{\tau}{ik(k+2)}}\sum_{l^{'}=0}^{k}
\sum_{m^{'}=-k+1}^{k}{\rm e}^{i\pi mm^{'}/k}\sin\left[
\frac{\pi(l+1)(l^{'}+1)}{k+2}\right]C^{l^{'}}_{m^{'}}(-1/\tau)
\end{equation}
and that for the elliptic theta functions, which is standard. 
The parameter of the modular
group in this case is $\tau=2i\lambda/\pi$. Making the transformation 
and taking the limit leads to
\begin{equation}
 M  \stackrel{\lambda \rightarrow 0}{\approx} \left\{ 
\begin{array}{ll}       
\frac{\textstyle {\rm e}^{-(2k+1)\pi^{2}(Q-4)^{-1/2}/8(k+2)}}
{\textstyle 2\sin[\pi(l+1)/2(k+2)]}
   &\;\;\;{\rm for\;\; odd }\;\;k    \\              
\frac{\textstyle {\rm e}^{-k\pi^{2}(Q-4)^{-1/2}/4(k+2)}}
{\textstyle \sin[\pi(l+1)/(k+2)]}
   &\;\;\;{\rm for\;\; even }\;\;k\;.       
\end{array}  \right.       \label{eq:a2}
\end{equation}             
Hence  the magnetization vanishes with an essential singularity 
as $Q \rightarrow 4^{+}$. A numerical calculation of the spontaneous 
magnetization given in $(\!~\ref{eq:magnet}\,)$ shows that for given 
$k$ and $Q$ in the domain $[4,\infty]$, the spontaneous magnetizations 
are bounded from below and above by 0 and 1 as physically expected. 
Moreover they are ordered as follows
\begin{equation}
\begin{array}{llllllllll}
\mbox{$k$ odd}\;\;\;&M_{0}&>&M_{k-1}&>&M_{2}&>&M_{k-3}&>\cdots>&
M_{(k-1)/2}\;,\\  
\mbox{$k$ even}\;\;\;&M_{k}&>&M_{0}&>&M_{k-2}&>&M_{2}&>\cdots>&
M_{k/2}\;.
\end{array}\label{naturalorder}
\end{equation}  
As in the construction of the models, we again observe a natural 
splitting between  $k$ odd and $k$ even.   

\subsection{Conjectured phase diagram}

\hspace{5mm}
 
We have computed spontaneous magnetizations in a rather formal fashion, 
and their meaning  for $k>1$ is not completely clear. The simplest 
possibility is to assume that, as in the $k=1$ case, different vertex 
boundary conditions correspond indeed to different phases of the 
$\Gamma_k$ Potts model. We shall then refer to the cases  where $M=0$  
as disordered phases, although such phases may well have for instance 
antiferromagnetic order. Similarly ordered phase refers to $M\neq 0$. 
The order (\ref{naturalorder}) suggests that such phases can be
characterized by their degree of spin alignment. 

We can then make some conjectures about 
the structure of the phase diagram of the family of $\Gamma_{k}$ Potts 
model in the neighborhood of the JB integrable line 
$(\!~\ref{eq:u}\,)$ for $Q>4$. We shall consider only the staggered 
$\Gamma_{k}$ Potts model with lattice isotropy. This Potts model has 
$k$ parameters $f_{i}$; $i=0,\cdots,k-1$ which are introduced in the 
previous section. Recall that the JB integrable model is homogeneous 
and satisfies the condition $(\!~\ref{eq:u}\,)$, therefore the 
requirment of lattice isotropy fixes the spectral parameter $u$ to 
be $\gamma/2$ and the JB integrable model corresponds to a point 
(denoted as ${\rm P}_{\rm JB}$ in the sequel) in the $k$ dimensional
parameter space. We now wish to build, for fixed $Q>4$, the phase 
diagram in the neighborhood of this integrable point, where the above 
calculation of spontaneous magnetization is performed. We expect that 
there  are $k+1$ distinct phases, while   $k$ parameters are at our 
disposal. As for $k=1$ we expect that these $k+1$ phases coexist only 
at ${\rm P}_{\rm JB}$.  This implies that the phase diagram around 
${\rm P}_{\rm JB}$ has the topology of the dual of a $k$-simplex where 
each phase has a common boundary with any other phases. The boundaries 
that separate the phases can be deduced from the symmetry properties of 
the Potts model. The discussion again split into two cases; $k$ odd and 
$k$ even. For simplicity we just discuss two examples.
 
We start by the $\Gamma_{3}$ Potts model, which has two ordered and two 
disordered phases. The model has a  duality transformation given by
\begin{eqnarray}
f_{0}&\longleftrightarrow&f_{0}^{-1}\;,\\
\mbox{ and }\;\;\;f_{0}^{-1/2}f_{1}&\longleftrightarrow&
f_{0}^{-1/2}f_{2}\;.
\end{eqnarray}
We expect that duality still interchanges respectively ordered and 
disordered phases, and therefore that the boundaries of these four 
phases are invariant surfaces of the duality map. These surfaces have 
to be given  by
\begin{equation}
\begin{array}{rrrll}
{\cal B}_{0}&:&f_{0}-1&=&0\;,\\
{\cal B}_{1}&:&f_{1}-f_{2}&=&0\\
\mbox{ and }\;\;\;\tilde{\cal B}_{12}&:&
F(f_{0}^{-1/2}f_{1},f_{0}^{-1/2}f_{2})&=&0
\end{array}
\end{equation}
where the unknown function $F$ depends only on $f_{0}^{-1/2}f_{1}$ 
and $f_{0}^{-1/2}f_{2}$, and satisfies the following conditions
\[\begin{array}{rll}
F(f_{0}^{-1/2}f_{1},f_{0}^{-1/2}f_{2})&=&
F(f_{0}^{-1/2}f_{2},f_{0}^{-1/2}f_{1})\;,\\
\mbox{ and }\;\;\;F(f_{0}^{-1/2}f_{1},f_{0}^{-1/2}f_{2})&=&0\;\;\;
\mbox{ at }{\rm P}_{\rm JB}\;.
\end{array}\]

\begin{center} 
\setlength{\unitlength}{0.008in}
\begin{picture}(358,380)(0,-50)
\thicklines
\put(148,129){\circle*{6}}

\thinlines
\path(160,200)(260,200)
\path(166,202)(160,200)(166,198)
\path(206.243,95.071)(202.000,88.000)(209.071,92.243)
\path(202,88)(222,108)(267,108)
\path(126.243,230.071)(122.000,223.000)(129.071,227.243)
\path(122,223)(162,263)(222,263)

%\thicklines
%\path(102,268)(102,218)
%\path(82,168)(52,153)
%\dashline{4.000}(142,198)(82,168)

\thinlines
\dottedline{5}(102,218)(102,123)
\dashline{4.000}(149,76)(104,121)

\thicklines
\path(147,223)(102,268)

\thinlines
\path(102,218)(102,308)
\path(104.000,300.000)(102.000,308.000)(100.000,300.000)

\thicklines
\path(147,223)(147,78)
\path(85,190)(85,64)
\path(85,118)(155,48)
\path(185,160)(255,90)
\path(155,48)(255,90)

\thinlines
\dashline{4.000}(185,160)(185,106)
\thicklines
\path(185,233)(185,160)

%\thinlines
%\dashline{4.000}(143,202)(185,160)

\dottedline{5}(100,145)(200,145)
\path(200,145)(277,145)
\path(269.000,143.000)(277.000,145.000)(269.000,147.000)

\dottedline{5}(100,145)(86,134)
%\path(77,125)(12,60)
\path(86,134)(12,60)
\path(16.189,67.103)(12.000,60.000)(19.038,64.296)

\thicklines
\dashline{4.000}(147,129)(85,190)
\path(147,129)(250,26)
\path(89,187)(55,221)
%\path(50,152)(85,118)
\path(85,190)(85,214)
\path(185,256)(185,233)
\path(85,118)   (89.686,117.766)
        (93.721,117.600)
        (97.172,117.499)
        (100.106,117.466)
        (104.687,117.600)
        (108.000,118.000)
 
\path(108,118)  (110.366,118.095)
        (113.218,118.320)
        (116.387,118.650)
        (119.701,119.060)
        (122.989,119.525)
        (126.083,120.020)
        (131.000,121.000)
 
\path(131,121)  (134.674,122.726)
        (139.087,124.989)
        (143.457,127.258)
        (147.000,129.000)
 
\path(147,129)  (151.101,131.470)
        (153.693,132.969)
        (156.443,134.541)
        (159.200,136.107)
        (161.813,137.591)
        (166.000,140.000)
 
\path(166,140)  (168.597,142.813)
        (171.719,146.234)
        (174.732,149.538)
        (177.000,152.000)
 
\path(177,152)  (179.995,154.880)
        (182.142,157.055)
        (185.000,160.000)
 
\path(85,214)   (89.686,213.817)
        (93.721,213.686)
        (97.172,213.608)
        (100.106,213.582)
        (104.687,213.686)
        (108.000,214.000)
 
\path(108,214)  (110.366,214.148)
        (113.218,214.422)
        (116.387,214.811)
        (119.701,215.301)
        (122.989,215.878)
        (126.083,216.529)
        (131.000,218.000)
 
\path(131,218)  (134.674,219.250)
        (139.087,221.354)
        (143.457,223.782)
        (147.000,226.000)
 
\path(147,226)  (151.101,227.994)
        (153.693,229.374)
        (156.443,230.906)
        (159.200,232.512)
        (161.813,234.115)
        (166.000,237.000)
 
\path(166,237)  (168.597,239.321)
        (171.719,242.474)
        (174.732,245.640)
        (177.000,248.000)
 
\path(177,248)  (179.995,250.995)
        (182.142,253.142)
        (185.000,256.000)
 
\thinlines
\path(185,106)  (180.825,101.604)
        (177.747,98.452)
        (174.000,95.000)
 
\path(174,95)   (171.235,92.964)
        (167.827,90.769)
        (164.505,88.690)
        (162.000,87.000)
 
\path(162,87)   (158.355,82.924)
        (155.000,79.000)
 
\path(155,79)   (152.094,78.063)
        (148.654,77.298)
        (145.386,76.634)
        (143.000,76.000)
 
\path(143,76)   (139.843,72.980)
        (135.000,68.000)
 
\thicklines
\path(135,68)   (130.788,66.156)
        (127.697,64.945)
        (124.000,64.000)
 
\path(124,64)   (119.757,63.718)
        (116.000,64.000)
 
\path(116,64)   (112.514,63.870)
        (108.115,63.827)
        (103.659,63.870)
        (100.000,64.000)
 
\path(100,64)   (95.058,63.827)
        (90.830,63.870)
        (88.136,63.924)
        (85.000,64.000)
 
\put(92,148){\makebox(0,0)[lb]{\raisebox{0pt}[0pt][0pt]
{\shortstack[l]{\footnotesize 0}}}}
\put(148,150){\makebox(0,0)[lb]{\raisebox{0pt}[0pt][0pt]
{\shortstack[l]{${\scriptstyle {\rm P}_{JB}}$}}}}
\put(-10,45){\makebox(0,0)[lb]{\raisebox{0pt}[0pt][0pt]
{\shortstack[l]{${\scriptstyle f_{0}^{-1/2}f_{1}}$}}}}
\put(293,145){\makebox(0,0)[lb]{\raisebox{0pt}[0pt][0pt]
{\shortstack[l]{${\scriptstyle f_{0}^{-1/2}f_{2}}$}}}}
\put(283,100){\makebox(0,0)[lb]{\raisebox{0pt}[0pt][0pt]
{\shortstack[l]{${\scriptstyle {\cal B}_{0}}$}}}}
\put(268,195){\makebox(0,0)[lb]{\raisebox{0pt}[0pt][0pt]
{\shortstack[l]{${\scriptstyle \tilde{{\cal B}}_{12}}$}}}}
\put(225,265){\makebox(0,0)[lb]{\raisebox{0pt}[0pt][0pt]
{\shortstack[l]{${\scriptstyle {\cal B}_{1}}$}}}}
\put(107,296){\makebox(0,0)[lb]{\raisebox{0pt}[0pt][0pt]
{\shortstack[l]{\footnotesize $f_{0}$}}}}
\put(108,190){\makebox(0,0)[lb]{\raisebox{0pt}[0pt][0pt]
{\shortstack[l]{\footnotesize 1}}}}
\put(147,238){\makebox(0,0)[lb]{\raisebox{0pt}[0pt][0pt]
{\shortstack[l]{\footnotesize d}}}}
\put(72,182){\makebox(0,0)[lb]{\raisebox{0pt}[0pt][0pt]
{\shortstack[l]{\footnotesize o}}}}
\put(208,171){\makebox(0,0)[lb]{\raisebox{0pt}[0pt][0pt]
{\shortstack[l]{\footnotesize o'}}}}
\put(155,33){\makebox(0,0)[lb]{\raisebox{0pt}[0pt][0pt]
{\shortstack[l]{\footnotesize d'}}}}
\put(-60,0){\makebox(0,0)[lb]{\raisebox{0pt}[0pt][0pt]
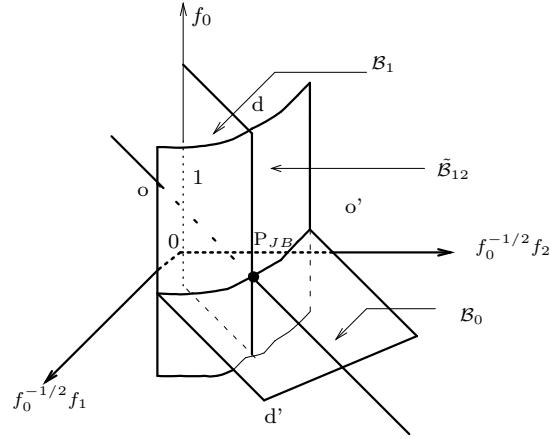
{\shortstack[l]{\footnotesize {\bf Figure(15)} Phase diagram of the 
$\Gamma_{3}$ Potts model in the neighborhood  of the}}}}
\put(20,-16){\makebox(0,0)[lb]{\raisebox{0pt}[0pt][0pt]
{\shortstack[l]{\footnotesize  JB integrable line point. The three 
surfaces $f_{0}=1,f_{1}=f_{2}$ and $F=0$ }}}}
\put(20,-32){\makebox(0,0)[lb]{\raisebox{0pt}[0pt][0pt]
{\shortstack[l]{\footnotesize divide the four phases $o,d,o',d'$. The phase 
diagram has the topology}}}}
\put(20,-48){\makebox(0,0)[lb]{\raisebox{0pt}[0pt][0pt]
{\shortstack[l]{\footnotesize that the four phases meet only at the 
integrable point $P_{\rm JB}$.}}}}
\end{picture}
\end{center} 

Incorporating the fact that the phase diagram has the topology of the 
dual of a (degenerate) 3-simplex, we arrive at the phase diagram shown 
in fig.(15). The phases are grouped into two pairs (o,d) and (o',d') 
where o(') and d(') denote respectively the ordered and disordered 
phases that are exchanged under duality. Since spaces below and above 
${\cal B}_{0}$, and spaces on the left and right of ${\cal B}_{1}$ are 
interchanged by duality transformation, ${\cal B}_{0}$ and ${\cal B}_{1}$ 
are therefore the boundaries that separate  o from d and o' from d' 
respectively. The boundary that divides these two pairs of phases is 
provided by the surface $\tilde{\cal B}_{12}$, which is symmetric in 
$f_{0}^{-1/2}f_{1}$ and $f_{0}^{-1/2}f_{2}$, and contains the integrable 
point ${\rm P}_{\rm JB}$. 

Consider now the example of the $\Gamma_{2}$ Potts model.  The parameters 
are $f_{0}$ and $f_{1}$. Notice first that the model with face isotropy 
has only one parameter, $f_1$, and thus cannot be expected to exhibit 
three phases in the neighborhood of the JB point. This means that face 
isotropy has to be spontaneously broken. There is no self duality, but 
simultaneous rotation of each of the fundamental blocks by $90^0$ about 
its center interchanges ordered and disordered phases. Rotation of the 
fundamental blocks amounts to
\[f_{0}\longleftrightarrow f_{0}^{-1}\;.\]
There are two lines which are invariant under this transformation;
\begin{equation}
\begin{array}{rrrll}
{\cal B}_{0}&:&f_{0}-1&=&0\;,\\
\tilde{\cal B}_{1}&:&G(f_{0},f_{1})&=&0
\end{array}
\end{equation}
where the function $G$ satisfies
\begin{eqnarray*}
G(f_{0},f_{1})&=&G(f_{0}^{-1},f_{1})\;,\\
G(f_{0},f_{1})&=&0\;\;\;\mbox{ at P}_{\rm JB}\;,\\
\mbox{ and }\;\;\;\left.\frac{\partial G}{\partial f_{0}}
\right|_{{\rm P}_{\rm JB}}
&=&0\;.           
\end{eqnarray*}

\begin{center} 
\setlength{\unitlength}{0.006in}
\begin{picture}(359,440)(0,-40)
\path(120,160)(120,150)
\path(184.243,182.071)(180.000,175.000)(187.071,179.243)
\path(180,175)(200,195)(235,195)
\path(124.243,302.071)(120.000,295.000)(127.071,299.243)
\path(120,295)(140,315)(200,315)
\thicklines
\dashline{4.000}(60,55)(60,30)
\dashline{4.000}(270,70)(290,50)
\dashline{4.000}(120,320)(120,360)
\path(120,195)(120,320)
\thinlines
\path(0,155)(320,155)
\path(312.000,153.000)(320.000,155.000)(312.000,157.000)
\path(42.000,347.000)(40.000,355.000)(38.000,347.000)
\path(40,355)(40,35)
\thicklines
\path(60,55)    (60.405,59.008)
        (60.799,62.874)
        (61.183,66.602)
        (61.558,70.195)
        (61.923,73.657)
        (62.280,76.990)
        (62.628,80.199)
        (62.969,83.287)
        (63.302,86.258)
        (63.629,89.114)
        (63.949,91.859)
        (64.263,94.498)
        (64.875,99.465)
        (65.469,104.045)
        (66.048,108.263)
        (66.616,112.149)
        (67.177,115.728)
        (67.734,119.030)
        (68.292,122.080)
        (68.853,124.907)
        (69.421,127.538)
        (70.000,130.000)
 
\path(70,130)   (70.625,132.655)
        (71.298,135.660)
        (72.027,138.963)
        (72.818,142.511)
        (73.679,146.252)
        (74.619,150.131)
        (75.645,154.098)
        (76.764,158.099)
        (77.984,162.080)
        (79.313,165.991)
        (80.759,169.777)
        (82.328,173.386)
        (84.030,176.766)
        (85.870,179.863)
        (87.858,182.626)
        (90.000,185.000)
 
\path(90,185)   (92.807,187.203)
        (96.289,189.135)
        (100.243,190.797)
        (104.467,192.186)
        (108.758,193.302)
        (112.912,194.144)
        (116.727,194.710)
        (120.000,195.000)
 
\path(120,195)  (124.795,194.952)
        (127.525,194.740)
        (130.426,194.416)
        (133.462,193.987)
        (136.594,193.463)
        (139.785,192.853)
        (142.999,192.167)
        (146.196,191.413)
        (149.341,190.600)
        (152.396,189.738)
        (155.322,188.835)
        (158.083,187.901)
        (160.641,186.944)
        (165.000,185.000)
 
\path(165,185)  (167.870,183.372)
        (170.941,181.361)
        (174.178,179.014)
        (177.545,176.382)
        (181.005,173.510)
        (184.524,170.449)
        (188.065,167.245)
        (191.592,163.949)
        (195.069,160.607)
        (198.460,157.267)
        (201.730,153.980)
        (204.842,150.791)
        (207.762,147.751)
        (210.451,144.907)
        (215.000,140.000)
 
\path(215,140)  (217.303,137.367)
        (219.820,134.293)
        (222.518,130.840)
        (225.367,127.071)
        (228.336,123.049)
        (231.393,118.836)
        (234.508,114.494)
        (237.650,110.088)
        (240.786,105.678)
        (243.887,101.329)
        (246.921,97.102)
        (249.857,93.060)
        (252.664,89.267)
        (255.311,85.783)
        (257.767,82.674)
        (260.000,80.000)
 
\path(260,80)   (263.382,76.435)
        (266.169,73.694)
        (270.000,70.000)
 
\put(-100,10){\makebox(0,0)[lb]{\raisebox{0pt}[0pt][0pt]
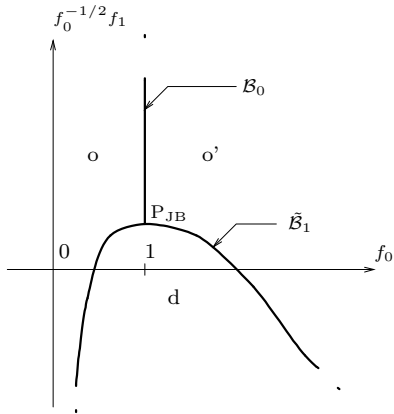
{\shortstack[l]{\footnotesize {\bf Figure(16)} Phase diagram the $\Gamma_{2}$ 
Potts model in the neighborhood of the }}}} 
\put(-10,-13){\makebox(0,0)[lb]{\raisebox{0pt}[0pt][0pt]
{\shortstack[l]{\footnotesize  JB integrable point. The three phases o, o', d 
are separated by ${\cal B}_{0}$  }}}}
\put(-10,-36){\makebox(0,0)[lb]{\raisebox{0pt}[0pt][0pt]
{\shortstack[l]{\footnotesize and $\tilde{{\cal B}}_{1}$, which are 
invariant curves under rotation of all the faces. }}}}
\put(45,165){\makebox(0,0)[lb]{\raisebox{0pt}[0pt][0pt]
{\shortstack[l]{\footnotesize 0}}}}
\put(320,165){\makebox(0,0)[lb]{\raisebox{0pt}[0pt][0pt]
{\shortstack[l]{${\scriptstyle f_{0}}$}}}}
\put(40,370){\makebox(0,0)[lb]{\raisebox{0pt}[0pt][0pt]
{\shortstack[l]{\footnotesize ${\scriptstyle f_{0}^{-1/2}f_{1}}$}}}}
\put(120,165){\makebox(0,0)[lb]{\raisebox{0pt}[0pt][0pt]
{\shortstack[l]{\footnotesize 1}}}}
\put(140,125){\makebox(0,0)[lb]{\raisebox{0pt}[0pt][0pt]
{\shortstack[l]{\footnotesize d}}}}
\put(170,250){\makebox(0,0)[lb]{\raisebox{0pt}[0pt][0pt]
{\shortstack[l]{\footnotesize o'}}}}
\put(70,250){\makebox(0,0)[lb]{\raisebox{0pt}[0pt][0pt]
{\shortstack[l]{\footnotesize o}}}}
\put(245,190){\makebox(0,0)[lb]{\raisebox{0pt}[0pt][0pt]
{\shortstack[l]{${\scriptstyle \tilde{\cal B}_{1}}$}}}}
\put(205,310){\makebox(0,0)[lb]{\raisebox{0pt}[0pt][0pt]
{\shortstack[l]{${\scriptstyle {\cal B}_{0}}$}}}}
\put(125,200){\makebox(0,0)[lb]{\raisebox{0pt}[0pt][0pt]
{\shortstack[l]{${\scriptstyle {\rm P}_{\rm JB}}$}}}}
\end{picture}
\end{center} 

The typical phase diagram is shown in fig.(16). 

Finally we want to remark that the above scenario relies upon the 
assumption that there is no other phase transition lines that 
bifurcate from the JB integrable line at some $Q\geq4$. Recall that 
such phenomena occurs in the Ashkin-Teller model where the self 
dual line (where the underlying staggered vertex model becomes 
homogeneous) bifurcates into two phase transition lines. 

\section{The $\Gamma_{2}$ Model}
 
While the models for $Q>4$ are generally expected to be non critical, 
the phase diagrams for $Q<4$ should exhibit several kinds of second 
order phase transitions. We shall here discuss in some details the 
case of $\Gamma_2$. We restrict to the non staggered case (which would 
be the two lines $x^2=1$ in the $\Gamma_1$ case) and to the  geometry  
of cylinder, ie. periodicity in  time direction, with free boundary
condition on the top and bottom rows. This ensures quantum group 
symmetry, which will turn out to be quite a useful ingredient.
To start, we discuss the related  one dimensional quantum spin chain.

\subsection{The quantum spin chain}
 
\hspace{5 mm}
   
 The hamiltonian can always be written as a general linear combination 
of the projectors as follows;   
\begin{equation}
H=\sum_{i=1}^{2l-1}(\sin\omega-\cos\omega)(Q-1)P_{0}(i,i+1)-\cos
\omega(Q-2)P_{1}(i,i+1)     
\label{eq:h}
\end{equation}
where $P_{j}(i,i+1)$'s are projectors that  project onto the 
irreducible spin-$j$ representation from the tensor product of the 
spin-1 states at sites $i$ and $i+1$, and the spin chain has free 
boundary conditions.  We have chosen  the coefficients of the projectors 
for later convenience. The parameter $\omega$ takes values in $[0,2\pi]$ 
and $q$ is restricted to the case $|q|=1$, we define as before 
$q={\rm e}^{{\rm i}\gamma}\;,\gamma\in{\bf R}$ and introduce the 
parameter $\delta=\pi/\gamma$. Owing to the existence of the 
unitary transformation
\begin{equation}
     UP_{j}(q)U^{-1}=P_{j}(-q^{-1})\;\;\;;j=0,1,2  
\end{equation}
where 
\begin{eqnarray*}
U&=&{\bf 1}+\frac{1}{2}S^{z}\otimes S^{z}-\frac{1}{2}S^{z^{2}}
\otimes S^{z^{2}}-\frac{1}{4}S^{+^{2}}
\otimes S^{-^{2}}-\frac{1}{4}S^{-^{2}}\otimes S^{+^{2}}\\
&&\mbox{}+\beta S^{+}S^{z}\otimes (1-S^{z})S^{-}
+\beta^{-1}S^{-}S^{z}\otimes(S^{z}-1)S^{+}+\alpha S^{+}(1+S^{+})
\otimes S^{z}S^{-}\\
&&\mbox{}+\alpha^{-1}S^{-}(1+S^{z})\otimes S^{z}S^{+}
\;\;\;\hspace{20mm};\alpha,\beta\in{\bf C}
\end{eqnarray*}
and
\[S^{+}=\left( \begin{array}{ccc}
  0&\sqrt{2}&0\\
  0&0&\sqrt{2}\\
  0&0&0   \end{array}  \right)\;,\;\;  
  S^{-}=\left( \begin{array}{ccc}
  0&0&0\\
  \sqrt{2}&0&0\\
  0&\sqrt{2}&0   \end{array}  \right)\;,\;\;  
  S^{z}=\left( \begin{array}{ccc}
  1&0&0\\
  0&0&0\\
  0&0&-1   \end{array}  \right)\]  
are the usual su(2) generators, it suffices to consider $\gamma$ 
in the domain $[0,\pi/2]$. The hamiltonian is in general not hermitian 
with
\begin{equation}
H_{i}^{\dagger}(q) = H_{i}(q^{-1})\;,
\end{equation}
however, through relabeling of the spin sites, $H_{i}(q^{-1})$ can be 
made equivalent to $H_{i}(q)$ and therefore the energy eigenvalues are 
real or occur in conjugate pairs. The hamiltonian can be written in 
terms of the more familiar su(2) spin operators as,
\begin{eqnarray}
H&=&\sum_{i=1}^{2l-1}\{-2\cos\omega\cos2\gamma+\cos\omega-\sin\omega
+\cos\omega\sigma_{i}+\sin\omega\sigma_{i}^{2}+\sin^{2}
\gamma[(\sin\omega-\cos\omega)(\sigma_{i}^{z}-\sigma_{i}^{z^{2}})
 \nonumber   \\
 & &\mbox{}-2\cos\omega(S_{i}^{z^{2}}+S_{i+1}^{z^{2}})]
 +\frac{\textstyle {\rm i}\sin2\gamma}{\textstyle 2}(\sin\omega
+\cos\omega)\sigma_{i}^{z}(S_{i}^{z}-S_{i+1}^{z}) 
+\frac{\textstyle {\rm i}\sin\gamma}{\textstyle 2}(\sin\omega
+\cos\omega)  
\nonumber  \\
 & &[\sigma^{\bot}_{i}(S_{i}^{z}-S_{i+1}^{z})
+(S_{i}^{z}-S_{i+1}^{z})\sigma^{\bot}_{i}]+(\sin\omega-\cos\omega)
(\cos\gamma-1)(\sigma^{\bot}_{i}\sigma^{z}_{i}+
\sigma^{z}_{i}\sigma^{\bot}_{i})\}   \nonumber \\
& &\mbox{}+{\rm i}\cos\omega\sin2\gamma(S_{1}^{z}-S_{2l}^{z})  
\label{eq:sh}
\end{eqnarray}
where
\[ \begin{array}{lllll}
  \sigma_{i}&=&\vec{S}_{i}\cdot\vec{S}_{i+1}&=&
\sigma^{\bot}_{i}+\sigma_{i}^{z}\;, \\
    \sigma_{i}^{z}&=&S_{i}^{z}S_{i+1}^{z}&\;.&  \end{array} \]
 Written in this manner, this hamiltonian can 
therefore be regarded as a special case of the spin-1 XXZ chain 
where the boundary terms ensure ${\rm U}_{q}{\rm su(2)}$ symmetry.
 
The hamiltonian reduces to that of the bilinear biquadratic spin 
chain with su(2) symmetry at $\gamma=0$ (or $q=1$). It has the 
simple expression
\begin{equation}
       H=\sum_{i=1}^{2l-1}[\cos\omega\sigma_{i}+\sin\omega
\sigma_{i}^{2}]\;.
\end{equation}
This model has a nontrivial phase diagram. We summarize in fig.(17)
and below certain features of the phase diagram\cite{abp}:

\begin{center} 
\input{f17}
\end{center} 
 
The phase diagram is essentially divided into four regions
$\omega\in(\frac{\pi}{2},\:\frac{5\pi}{4}),\;(\frac{5\pi}{4},
\:\frac{3\pi}{2})
,\;(\frac{3\pi}{2},\:\frac{\pi}{4})$ and $(\frac{\pi}{4},
\:\frac{\pi}{2})$ 
depending on the ground state of the spin chain. For 
$\frac{\pi}{2}<\omega<\frac{5\pi}{4}$, the ground state is 
ferromagnetic. The model is integrable at $\omega=\frac{3\pi}{4}$ 
and has (formally) central charge $c=-\infty$. At the boundaries, 
$\omega=\frac{\pi}{2}$ and $\frac{5\pi}{4}$ the symmetry is augmented 
from su(2) to su(3). More specifically, at $\frac{\pi}{2}$, neighboring 
spins assume the representation $(\bar{3},\:3)$, the hamiltonian
being proportional to $3P_{0}$ is related to the spin-$\frac{1}{2}$ 
Heisenberg antiferromagnetic spin chain via the Temperley Lieb algebra. 
At $\frac{5\pi}{4}$ neighboring spins belong to (3,3) representation, 
the spin chain is the permutation model studied by Sutherland 
{\em et al}\cite{lai} and is found to have central charge equal to 2. 
In the interval $\frac{5\pi}{4}<\omega<\frac{3\pi}{2}$, the spin chain 
is found using a semi classical approach to have vanishing magnetization 
but nonzero tensorial order parameter, the ground state therefore 
exhibits a "nematic order" . At $\frac{3\pi}{2}$ the spin chain has 
hamiltonian $-P_{0}$ which differs from that at $\frac{\pi}{2}$ by a
sign, it is again in the representation $(\bar{3},\:3)$. The ground 
state is found to have massive excitation. The interval 
$\frac{3\pi}{2}<\omega<\frac{\pi}{4}$, where we have identified the 
point $\omega=0$ and $2\pi$, has antiferromagnetic ground state and 
contains the Takhtajan-Babujian model\cite{tb} at $\omega=\frac{7\pi}{4}$, 
this point is solvable with $c=\frac{3}{2}$. In addition, the exact 
valence bond ground state\cite{aklt} can be constructed at
 $\omega=\tan^{-1}\frac{1}{3}$, and the spin chain is shown to 
have massive excitation. The vicinity of this point, which includes 
the Heisenberg antiferromagnetic model at $\omega=0$, belongs to an 
antiferromagnetic fluid phase or disorder flat phase\cite{den} where 
there is long range antiferromagnetic spin order and position disorder. 
For $\frac{\pi}{4}<\omega<\frac{\pi}{2}$, the ground state is 
dimerized, and at $\omega=\frac{\pi}{4}$, where phase transition 
occurs, the model is integrable and has su(3) symmetry.

\subsection{The Integrable Lines}
 
\hspace{5mm}
 
The $\Gamma_{2}$ model has mainly been studied along the integrable 
lines $(\!~\ref{eq:FZline}\,)$,$(\!~\ref{eq:IKline}\,)$,
$(\!~\ref{eq:TLline}\,)$, they are given, in terms of parameters of 
the spin chain, as
\begin{eqnarray*}       
{\rm(FZ)}\;\;\;\;\;\;\tan\omega&=&-1\;,\\
{\rm(IK)}\;\;\;\;\;\;\tan\omega&=&\frac{1}{Q-3}\;,\\
{\rm(TL)}\;\;\;\;\;\;\cos\omega&=&0\;.
\end{eqnarray*}
  The first two are related respectively to the ${\rm A}^{(1)}_{1}$ 
and ${\rm A}^{(2)}_{2}$ solutions to the Yang-Baxter equation\cite{jim1}, 
they have been studied first by Fateev-Zamolodchikov\cite{fz}  and 
Izergin-Korepin\cite{ik}. The one labelled by TL has hamiltonian 
proportional to the spin-0 projector which is known to satisfy the 
Temperley-Lieb algebra\cite{saa}. We also want to point out that the FZ 
and TL lines are respectively the $k=2$ element of the family of
integrable models denoted as JB and TL in previous sections. Each of 
these equations gives rise to two lines in the $\omega-\gamma$ phase 
diagram where the hamiltonians differ by an overall sign. In the limit 
$\gamma=0$, they reduce to the integrable points of the su(2) invariant 
bilinear biquadratic spin chain. In this section, we shall examine the 
phase diagram beginning with these integrable lines, they will serve
as benchmarks for the understanding of the critical properties of the 
general phase diagram. In fig.(17), we summarized the features of the 
phase diagram. 
        
\subsubsection{The TL case}

 It is governed by the hamiltonian 
\begin{equation}
H=\epsilon\sum_{i=1}^{2l-1}(Q-1)P_{0}(i,i+1)       
\label{eq:h1}
\end{equation}
where 
\[\epsilon=\left\{\begin{array}{ll}
 1&\mbox{if $\omega=\pi/2$}\;,\\
-1&\mbox{if $\omega=3\pi/2$}\;.
\end{array}
\right. \]
The projectors $(Q-1)P_{0}$ satisfies the Temperley Lieb algebra 
$(\!~\ref{eq:TL}\,)$ with 
\[e_{i}=(Q-1)P_{0}\]
and \[e_{i}^{2}=(Q-1)e_{i}\;.\]
The corresponding vertex model has transfer matrix that satisfies the 
Yang-Baxter equation, the model is therefore integrable. In principle,  
energy eigenvalues and hence the critical properties can be deduced 
from the Bethe anatz solution. However, we shall instead  employ all
that is known about the ${\rm U}_{q}{\rm su(2)}$ invariant 
spin-$\frac{1}{2}$ chain\cite{sal}, which also has hamiltonian given by 
sum of Temperley Lieb generators, to understand this integrable case. 
Since the two hamiltonians are related to the same algebra, they  share 
the same set of  eigenvalues. On the other hand, as the representations 
are different, we do not expect the degeneracy to be identical. Also the 
same eigenvalue may appear in different spin sectors in the two models. 
To overcome these difficulties, we compare numerically their eigenvalues. 
It is worth pointing out that the hamiltonian has a hidden 
${\rm U}_{q}{\rm sl(3)}$ symmetry\cite{bmn} where neighboring spin 
sites are in the $(3,\bar{3})$ or $(\bar{3},3)$ representation of  the 
quantum group, and $P_{0}$ can be regarded as the operator that projects 
the above representation onto the trivial representation.
 
The spin-$\frac{1}{2}$ ${\rm U}_{q}{\rm su(2)}$ invariant spin chain with 
free boundary condition has hamiltonian
\begin{equation}
H=-\sum_{i=1}^{2l-1}\sqrt{Q^{'}}P_{0}(i,i+1)   
\label{eq:h2}
\end{equation}
it is the extreme anisotropic limit of the self dual six vertex model 
$(\!~\ref{eq:6vertex}\,)$, and
the projector $\sqrt{Q^{'}}P_{0}$ satisfies the Temperley Lieb algebra 
with \[e_{i}=\sqrt{Q^{'}}P_{0}(i,i+1)\;.\]
The spin chain is critical for $\sqrt{Q^{'}}\in[0,2]$ (or 
$\delta^{'}\in [2,\infty]$), and the central charge depends on 
$\delta^{'}$ as\cite{dnd}
\begin{equation}
c=1-\frac{6}{\delta^{'}(\delta^{'}-1)}\;.  \label{eq:cc}
\end{equation}  
The ground state energy of spin-$j$ sector $\varepsilon_{j}^{1}$ scales 
as\cite{cdy}
\begin{equation}
\frac{l(\varepsilon_{0}^{1}-
\varepsilon_{j}^{1})}{\xi\pi}\stackrel{l\rightarrow\infty}{=}
h_{j}    \label{eq:weight1}
\end{equation}                                     
where
\begin{equation}
h_{j}=\frac{j[j(\delta^{'}-1)-1]}{\delta^{'}}
                      \label{eq:cweight2}
\end{equation}             
and the sound velocity
\[\xi=\frac{\delta^{'}}{2}\sin\frac{\pi}{\delta^{'}}\]
is  obtained from the Bethe anatz solution\cite{abb}. For $q$ a root of 
unity, the central charge belongs to the minimal series\cite{bpz} with   
\begin{equation}
c=1-\frac{6}{m(m+1)}
\end{equation}  
and $m=\delta^{'}-1$. The conformal weight is given by
\begin{equation}
h_{r,s}=\frac{[(m+1)r-ms]^{2}-1}
{4m(m+1)}
\end{equation}
and therefore 
\begin{equation}
h_{j}=h_{1,1+2j}\;.
\end{equation}
For $Q^{'}>4$, the spin chain is noncritical and has a massive
excitation\cite{bax}. 
  
The spin-$\frac{1}{2}$ chain also has a ferromagnetic counter 
part\cite{bax}, whose has hamiltonian differs from $(\!~\ref{eq:h2}\,)$ 
by an overall sign. The negated hamiltonian can in fact be obtained 
from $(\!~\ref{eq:h2}\,)$ by  rewriting the coefficient\cite{sal}
\[\sqrt{Q^{'}}=-2\cos(\pi(1-\frac{1}{\delta^{'}}))\] and extending the 
domain of $\delta^{'}$ to [1,2] so that $(1-\frac{1}{\delta^{'}})^{-1}
\in[2,\infty]$ or $-\sqrt{Q^{'}}\in[0,2]$. At $\sqrt{Q^{'}}=0$, it can 
be shown that the eigenvalues are symmetric about zero, the hamiltonian 
is therefore equivalent to its ferromagnetic counter part. We can 
therefore regard the ferromagnetic counter part as an extension of 
$(\!~\ref{eq:h2}\,)$ where the domain of $\sqrt{Q^{'}}$ is enlarged to 
include [-2,0] or $\delta^{'}\in[1,2]$ as well. More importantly, it 
is found that for $\delta^{'}\in[1,2]$ the central charge and conformal 
weights are correctly given by $(\!~\ref{eq:cc}\,)$ and 
$(\!~\ref{eq:weight1}\,)$, $(\!~\ref{eq:cweight2}\,)$. In other words, 
the above results apply to the hamiltonian $(\!~\ref{eq:h2}\,)$  with  
$\delta^{'}\in[1,\infty]$. 
 
At $\sqrt{Q^{'}}=-2$, the spin chain is noncritical with $c=-\infty$, 
and for $\sqrt{Q^{'}}\leq-2$, the ground state has massive excitation.
 
Comparing the hamiltonians $(\!~\ref{eq:h1}\,)$ and  $(\!~\ref{eq:h2}\,)$, 
we see that the two spin chain have the same set of eigenvalues when
\begin{equation}
\sqrt{Q^{'}}=\epsilon(1-Q)\;,  \label{eq:m1}
\end{equation}
since the Temperley Lieb algebra realized by them have the same 
$\sqrt{"Q"}$ parameter. Numerical studies of the eigenvalues for the 
spin-1 chain of finite size ( $2l<10$ ) reveals that the spin-1 energy 
spectrum contains many crossing of eigenvalues due to the additional 
${\rm U}_{q}{\rm sl(3)}$ symmetry. In particular we find,  
\begin{equation}
\varepsilon^{1}_{2j-1}=\varepsilon^{1}_{2j}\;\;\;\mbox{ for } j\geq1
\label{eq:e12}
\end{equation}             
always hold. By comparing the ground state energies for various spin 
sectors of the spin-$\frac{1}{2}$ and spin-1 chain, we find, when 
$Q^{'}$ and $Q$ are related by $(\!~\ref{eq:m1}\,)$,
\begin{equation}
\begin{array}{lllllllll}
\epsilon&=&-1&:&\varepsilon_{2j}^{1}&=&\varepsilon_{j}^{'1}\;\;\;;j
\geq0\;\;\;&\mbox{for}&\;Q\in[0,4]\\  \\
\epsilon&=&1&:&\varepsilon_{2j}^{1}&=&\varepsilon_{j}^{'1}\;\;\;;j
\geq0\;\;\;&\mbox{for}&\;Q\in[0,2]
\end{array}  \label{eq:qmap}
\end{equation}
where eigenvalues with prime belong to the spin-$\frac{1}{2}$ chain.
For $\epsilon=-1$, this identification implies that the spin-1 chain 
is critical for $Q\in[0,3]$, the central charge is given by 
$(\!~\ref{eq:cc}\,)$ with $\delta^{'}$ related to $\delta$ by 
$(\!~\ref{eq:m1}\,)$ or \[2\cos\frac{\pi}{\delta^{'}}=2\cos\frac{2\pi}
{\delta}+1\;,\] and it increases from -7 to 1 as $Q$ varies from 0 to 3. 
Moreover, the ground state of the spin-$j$ sectors scales as  
\begin{equation}   
\frac{l(\varepsilon_{0}^{1}-\varepsilon_{2j}^{1})}{\xi\pi}\stackrel
{N\rightarrow\infty}{=}\frac{j[j(\delta^{'}-1)-1]}{\delta^{'}}
                      \label{eq:ws}
\end{equation}
with the sound velocity given as before. Using $(\!~\ref{eq:e12}\,)$, 
the scaling behavior of the odd spin sectors can also be deduced. 
Since $Q\in[3,4]$ is mapped to the noncritical region $\sqrt{Q^{'}}
\in[2,3]$ of the spin-$\frac{1}{2}$ chain, the spin-1 chain in this 
interval is therefore massive. 
 
For $\epsilon=1$, the identification is valid for $Q\in[0,2]$ and the 
spin chain is critical with central charge given by $(\!~\ref{eq:cc}\,)$ 
but in this case $\delta^{'}$ and $\delta$ are related by 
\[2\cos\frac{\pi}{\delta^{'}}=-1-2\cos\frac{2\pi}{\delta}\;,\] hence 
the central charge varies from 0 to -7. For $Q\in[2,4]$ the mapping
$(\!~\ref{eq:qmap}\,)$ does not hold anymore, and it is not clear how 
to use  the spin-$\frac{1}{2}$ chain  to deduce the critical properties 
of the spin-1 model.  
\par

\subsubsection{The FZ case}
 
The FZ integrable spin chain has hamiltonian
\begin{equation}
H=\epsilon\sum_{i=1}^{2l-1}[2(Q-1)P_{0}(i,i+1)+
     (Q-2)P_{1}(i,i+1)]\;;\;\;\;\epsilon=\pm1\;,
\end{equation}
it is the extreme anisotropic limit of the vertex model given in 
$(\!~\ref{eq:u}\,)$ for $k=2$. For the $\epsilon=-1$ regime, the model 
is critical with central charge given by\cite{mtn}
\begin{equation}
c=\frac{3}{2}-\frac{12}{\delta(\delta-2)}\;;\;\;\;\delta\in[2,\infty)\;,
      \label{eq:FZc}
\end{equation}
while the lowest eigenvalue of each spin-$j$ sector we found numerically 
to scale as
\begin{equation}
\frac{l(\varepsilon_{j}^{1}-\varepsilon_{0}^{1})}{\xi\pi}=
\frac{j((\delta-2)j-2)}{2\delta}+\frac{1}{2}\delta_
{j,\mbox{\footnotesize odd }}  
\label{eq:FZw}
\end{equation}
where
\[\xi=\frac{\pi\sin2\delta}{2\delta}\]
denotes the sound velocity. For $q$ a root of unity, $\delta$ becomes 
rational, the central charge belongs to the superconformal 
series\cite{fqs} where the conformal weight reads
\begin{equation}
h_{p,q}=\frac{(p\delta-q(\delta-2))^{2}-4}{8\delta(\delta-2)}
+\frac{1}{32}[1-(-1)^{p-q}]\;.
\end{equation}
Substituting $p=1$ and $q=2j+1$ into the above,
\begin{equation}
h_{1,2j+1}=\frac{j((\delta-2)j-2)}{2\delta}\;,
\end{equation}
we recover $(\!~\ref{eq:FZw}\,)$ except for the additional term 
$\frac{1}{2}\delta_{j,\mbox{\footnotesize odd }}$.
The spin chain is therefore related to the Neveu-Schwarz sector of the
minimal superconformal series for $q$ a root of unity. Furthermore, 
only the lowest eigenvalues of the even spin sectors are related simply 
to the primary states $|h_{1,i+2j}>$, while for the odd spin sectors they 
are related to $G_{-\frac{1}{2}}|h_{1,i+2j}>$ where $G_{-\frac{1}{2}}$ 
is the fermionic raising generator of the global superconformal group 
OSP$(2|1)$. This extra factor accounts for the term $\frac{1}{2}
\delta_{j,\mbox{\footnotesize odd }}$ in $(\!~\ref{eq:FZw}\,)$. Another 
interesting phenomena occurs at the point $\gamma=\frac{\pi}{4}$. It has 
been  noted that the numerical estimate of the central charge is exactly 
zero and does not suffers from finite size correction. This point is 
related in fact to the $N=2$ supersymmetric series\cite{sal1}.
 
The $\epsilon=1$ regime of the FZ line has drastically different behavior 
from its $\epsilon=-1$ counterpart\cite{acr}. It has been studied for 
toroidal boundary conditions where 
\[\begin{array}{lllll}
S^{\pm}_{2l+1}&=&{\rm e}^{\pm{\rm i}\phi}S^{\pm}_{1}\;,\;\;\;
S^{z}_{2l+1}&=&S^{z}_{1}\;.
\end{array}\]
The "effective" central charge depends also on $\phi$ as
\begin{equation}
c=1-\frac{3\phi^{2}}{2\pi\gamma}\;;\;\;\;\gamma\in[0,\frac{\pi}{2}]\;.
\end{equation}
It is however well known that with appropriate value for $\phi$, this 
formula gives the central charge for the free boundary spin 
chain\cite{pqs,agr}. Indeed, putting $\phi=2\pi-2\gamma$, we get
\begin{equation}
c=1-\frac{6(\delta-1)^{2}}{\delta}\;;\;\;\;\delta\in[2,\infty]\;.
\label{eq:FZcp}
\end{equation}
This expression is the same as that for the spin-$\frac{1}{2}$ 
${\rm U}_{q}{\rm su(2)}$ invariant spin chain in the domain
$\delta^{'}\in[1,2]$ as can be seen by replacing $\delta^{'}$ by
$\frac{\delta^{'}}{\delta{'}-1}$ in $(\!~\ref{eq:cc}\,)$.  We have also
verified numerically for the $\epsilon=1$ FZ line that the ground state 
energy scales according to
\begin{equation}
\frac{l(\varepsilon_{j}^{1}-\varepsilon_{0}^{1})}{\xi\pi}=
\frac{j(j-\delta+1)}{\delta}  
\label{eq:FZwp}
\end{equation}
where
\[\xi=\frac{\pi\sin2\delta}{2\pi-2\delta}\;.\]
The above formula is in fact equal to $(\!~\ref{eq:cweight2}\,)$ after 
the replacement 
\[\delta^{'}\longrightarrow\frac{\delta^{'}}{\delta{'}-1}\;.\]
This regime is therefore in the same universality class as the 
spin-$\frac{1}{2}$ chain $(\!~\ref{eq:h2}\,)$ in the
interval $\delta^{'}\in[1,2]$.
 
\subsubsection{The IK case}

The IK integrable spin chain has hamiltonian given by
\begin{equation}
H=\frac{\epsilon}{\sqrt{1+(Q-3)^{2}}}\sum_{i=1}^{2l-1}[(4-Q)(Q-1)
P_{0}(i,i+1)+(3-Q)(Q-2)P_{1}]
\end{equation}
where $\epsilon=1$ corresponds to 
\[\omega=\tan^{-1}(\frac{1}{Q-3})\in[0,\pi]\]
and $\epsilon=-1$ to
\[\omega=\tan^{-1}(\frac{1}{Q-3})+\pi\;.\]
The model coincides with the FZ chain at $\gamma=\frac{\pi}{4}$ and 
the TL chain at $\gamma=\frac{\pi}{6}$. Exact Bethe anatz solution has 
been worked out for the model with toroidal boundary condition\cite{wbn}. 
It was found that the critical behaviors are classified according to 
the following regimes;
\begin{equation}
\begin{array}{lllrllll}
\mbox{regime I}&\epsilon&=&1\;\;\;&c&=&1-\frac{3\phi^{2}}{2\pi\gamma}
\;\;\;\;&;\gamma\in(0,\frac{\pi}{2})\\
\mbox{regime II}&\epsilon&=&-1\;\;\;&c&=&\frac{3}{2}-\frac{3\phi^{2}}
{\pi(\pi-2\gamma)}\;\;\;\;&;\gamma\in(\frac{\pi}{6},\frac{\pi}{2})\\
\mbox{regime III}&\epsilon&=&-1\;\;\;&c&=&\left\{\begin{array}{ll}
    2-\frac{3\phi^{2}}{2\pi\gamma}&\;\;\phi\leq 2\gamma\\
    -1+\frac{3(\phi-\pi)^{2}}{\pi(\pi-2\gamma)}&\;\;\phi\geq 2\gamma
 \end{array}\right.  \;\;\;\;&;\gamma\in(0,\frac{\pi}{6});.
\end{array}
\end{equation}
As in the previous case, these results can be used to obtain the 
central charge for the free boundary spin chain. We verify numerically 
that in regime II the central charge is given by $\phi=2\gamma$ where 
the above formula becomes
\begin{equation}
c=\frac{3}{2}-\frac{12}{\delta(\delta-2)}\;\;\;;\delta\in(2,6)\;.
\end{equation}
This expression is identical to $(\!~\ref{eq:FZc}\,)$ of the $\epsilon=1$ 
FZ line. Moreover, numerical check of the energy eigenvalues shows that 
the ground state of each spin-$j$ sector scales as $(\!~\ref{eq:FZw}\,)$ 
but with the sound velocity given in this case, following \cite{wbn} by
\[\xi=\frac{2\pi\sin2\gamma\cos3\gamma}{(\pi-6\gamma)
\sqrt{Q+Q(Q-3)^{2}}}\;.\]
We therefore conclude that regime II is in the same universality class 
as the $\epsilon=-1$ FZ line for $\gamma\in(\frac{\pi}{6},\frac{\pi}{2})$. 
However beyond $\gamma=\frac{\pi}{6}$ ie. regime III, the IK model has 
different critical behavior. Our numerical checks of the conformal weight 
proved inconclusive due to poor finite size convergence. On the other hand, 
it is known that the spin chain at $\gamma=0$ is related to the 
permutation model studied by Sutherland {\em et al}  where $c=2$ and to 
TL model at $\gamma=\frac{\pi}{6}$ where $c=1$.
It is therefore likely that regime III has
\begin{equation}
c=2-\frac{6}{\delta}
\end{equation} 
which is obtained from above with the substitution $\phi=2\gamma$.
As for regime I, numerical check suggests again that the critical 
properties are again different in the two regimes
\[\begin{array}{lllll}
\mbox{regime I$^{'}$}&\epsilon&=&1&\;\;\;\gamma\in(0,\frac{\pi}{6})\\     
\mbox{regime I$^{''}$}&\epsilon&=&1&\;\;\;\gamma\in(\frac{\pi}{6},
\frac{\pi}{2})\;.\end{array}\]     
In regime I$^{''}$, the model is found to be in the same universality 
class as the $\epsilon=-1$ FZ line  where the central charge has 
expression $(\!~\ref{eq:FZcp}\,)$ which is obtained from the above by 
taking $\phi=2\pi-2\gamma$, and the scaling behavior of the energy 
eigenvalues is given as in $(\!~\ref{eq:FZwp}\,)$ with sound velocity 
\[\xi=-\frac{2\pi\sin2\gamma\cos3\gamma}{3(\pi-2\gamma)
\sqrt{Q+Q(Q-3)^{2}}}\;.\]
for regime I$^{'}$, finite size convergence is poor and classification 
of the regime is uncertain.
 
It is intriguing to find that the IK and FZ line in both $\epsilon=\pm1$ 
have the same critical properties for $\gamma\in(\frac{\pi}{6},
\frac{\pi}{2})$. 
To elucidate this we performed further numerical study for models 
"in between". The sound velocities are not known then, and the difference  
of ground state energies $\varepsilon^{1}_{j}-\varepsilon^{1}_{0}$ is 
more difficult to use to deduce the critical behavior ( such as 
$(\!~\ref{eq:FZw}\,)$ ) of the model. One can still  study  quantities
that do not depend on the sound velocity and compare them with  the FZ
and IK integrable cases. For given $\gamma$ in the shaded regions a and b 
in fig.(17), we found that such scaled quantities approach those  common 
to the two integrable lines as $l$ increases. Further, the ordering 
of levels with respect to $j$ is the same as that of the integrable lines. 
This behavior suggets that the shaded regions are massless phases, with 
the same universality class as the integrable lines. Other indications 
come from the fact that on the integrable lines, the ground state energies 
of certain spin sectors coincide at special values of $\gamma$ such as
\begin{equation}
\begin{array}{lcrcl}
\epsilon&=&-1\hspace{10mm}\delta\varepsilon^{1}_{1}(\frac{2\pi}{5})
&=&\delta\varepsilon^{1}_{3}(\frac{2\pi}{5})\;,\\ 
&&\delta\varepsilon^{1}_{1}(\frac{\pi}{4})&=&\delta
\varepsilon^{1}_{2}(\frac{\pi}{4})\;,
\end{array}
\end{equation}
and
\begin{equation}
\begin{array}{lcrcl}
\epsilon&=&1\hspace{10mm}
\delta\varepsilon^{1}_{1}(\frac{\pi}{5})&=&\delta
\varepsilon^{1}_{3}(\frac{\pi}{5})\;,\\ 
&&\delta\varepsilon^{1}_{1}(\frac{\pi}{4})&=&\delta
\varepsilon^{1}_{2}(\frac{\pi}{4})\;,\\ 
&&\delta\varepsilon^{1}_{0}(\frac{\pi}{4})&=&\delta
\varepsilon^{1}_{3}(\frac{\pi}{4})\;, 
\end{array}
\end{equation}
which can be seen from $(\!~\ref{eq:FZw}\,)$ and 
$(\!~\ref{eq:cweight2}\,)$ respectively. Numerical study shows that 
such crossings still hold in the shaded regions. Finally at
$\gamma=\frac{\pi}{4}$ in region a, moving away from the integrable 
lines along the $\omega$ direction, the central charge is found to be 
exactly zero without finite size correction as in the integrable cases. 
A study of the operator contents of the continuum theories on the FZ 
line suggests that  the operator of the $N=1$ supersymmetric series 
that correspond to perturbing spin chain from the FZ line in the 
$\omega$ direction and which respect the quantum group symmetry of the 
spin chain has conformal weight $h_{3,1}=(\delta+2)/(2\delta-4)$
ie is irrelevant for $\gamma> \pi/6$ (region a).  
As for region b, it is likely that   the operator of the minimal series 
that drives the perturbation from the integrable line as again weight 
$h_{3,1}=2\delta-1$, ie is irrelevant for $\gamma\in[0,\pi/2]$.

\subsection{$P_{2}$ projector and the $q$-deformed valence bond states}
 
\hspace{5mm} 

   Besides the integrable lines, the hamiltonian obtained by summing 
projectors $P_{2}(i,i+1)$ deserves further examination. Recall that 
for $q=1$, the ground state of vanishing energy could  be exactly 
constructed using Valence Bond States \cite{aklt}. It turns out that 
the construction generalizes to arbitrary $q$. We shall again refer to 
the corresponding  state as VBS. Notice that for $q={\rm e}^{i\gamma}$, 
 such a state needs not always be  the ground state. We shall in fact 
observe other  eigenenergies  crossing  0  as $\gamma$ deviates from
zero. 
 
The hamiltonian considered lies along the line
\begin{equation}
\tan\omega=\frac{1}{Q-1}\;\;\;;\;\omega\in[0,\pi]  \label{eq:p2}
\end{equation}
with
\begin{equation}
H=\sum_{i=1}^{N-1}(Q-1)(Q-2)P_{2}(i,i+1)\;,  \label{eq:p2ham}
\end{equation}
where we have dropped an overall positive coefficient and a constant 
term. This restricts to the  ($q=1$) su(2) invariant model as a special 
case. Let us  now  extend the valence bond state method, used so far  
for the su(2) model, to the above hamiltonian. As a first step, we regard 
the spin-1 state $\psi_{\alpha\beta}$ as being formed by the $q$-
symmetric product of two spin-$\frac{1}{2}$ states defined as 
\begin{equation}
\psi_{\alpha\beta}=g_{\alpha\beta}\phi_{\alpha}\otimes\phi_{\beta}+
g_{\beta\alpha}\phi_{\beta}\otimes\phi_{\alpha}\;\;\;;\;\alpha,\beta=1,2
\end{equation}
where $g_{\alpha\beta}$ is the matrix element of 
\[g=\left(\begin{array}{cc}
\frac{1}{\sqrt{2}}&\frac{\sqrt{2}q^{\frac{1}{2}}}{q+q^{-1}}\\
\frac{\sqrt{2}q^{-\frac{1}{2}}}{q+q^{-1}}&\frac{1}{\sqrt{2}}
\end{array}\right)\;,\]
and $\phi_{\alpha}$ denotes the orthonormal basis of the spin-
$\frac{1}{2}$ states. The spin-1 state $\psi_{\alpha\beta}$ is by 
construction symmetric in the two indices $\alpha,\beta$ and is related 
to the orthonormal basis $|\pm>,|0>$ as 
\[\psi_{11}=\sqrt{2}|+>,\psi_{22}=\sqrt{2}|->,\psi_{12}=\psi_{21}=
\sqrt{\frac{2}{q+q^{-1}}}|0>\;.\]
Note that in our notation, the two spin-$\frac{1}{2}$ states are 
labelled by 1,2, and the three spin-1 states are labelled by $\pm$, 0.
Consider two neighboring spin-1 states $\psi_{\alpha\alpha_{1}}$ and
$\psi_{\beta_{1}\beta}$, we construct a tensor product using two of the
spin-$\frac{1}{2}$ states (one from each spin-1 state) such that the 
total spin can only be 0 or 1, this is achieved with the help of the tensor
\begin{equation}
\epsilon=\left(\begin{array}{cc}
0&q^{-1}\\
-q&0
\end{array}\right)
\end{equation}
and the tensor product is defined as
\begin{equation}
\Omega_{\alpha\beta}=\sum_{\alpha_{1},\beta_{1}}\psi_{\alpha\alpha_{1}}
\epsilon_{\alpha_{1}\beta_{1}}\psi_{\beta_{1}\beta}\;.     \label{eq:tpd}
\end{equation}
In the $q=1$  case, $\epsilon$ reduces to the usual antisymmetric tensor 
with $\epsilon_{12}=1$. One can check that the resulting element indeed 
belongs to the ${\rm U}_{q}{\rm su(2)}$ spin 1 representation by 
expressing the above as
\begin{eqnarray}
\Omega_{11}&=&\frac{-2}{\sqrt{q+q^{-1}}}(q^{2}
+q^{-2})^{\frac{1}{2}}|1,1>\;,   \nonumber \\
\Omega_{22}&=&\frac{-2}{\sqrt{q+q^{-1}}}(q^{2}
+q^{-2})^{\frac{1}{2}}|1,-1>\;,\\
q\Omega_{12}+q^{-1}\Omega_{21}&=&-2(q^{2}
+q^{-2})^{\frac{1}{2}}|1,0> \nonumber\\
\mbox{and }\hspace{10mm}\Omega_{12}-\Omega_{21}&=&2(q^{2}+1
+q^{-2})^{\frac{1}{2}}|0,0>   \nonumber\;
\end{eqnarray}
where $\{|1,\pm>,\;|1,0>\}$ and $\{|0,0>\}$ are respectively the 
orthonormal basis of the irreducible spin-1 and spin-0 representations, 
which are constructed out of two copies of spin-1 (four copies of 
spin $1/2$) orthonormal bases from sites $i$ and $i+1$ as follows
\begin{equation}
\begin{array}{llll}
\mbox{spin-0}\;:\;\;&|0,0>&=&(q^{2}+1+q^{-2})^{-\frac{1}{2}}
(q^{-1}|+->-|00>+q|-+>)\\             \\
        &|1,1>&=&(q^{2}+q^{-2})^{-\frac{1}{2}}(q|0+>-q^{-1}|+0>)\\
\mbox{spin-1}\;:\;\;&|1,0>&=&(q^{2}+q^{-2})^{-\frac{1}{2}}(|-+>
+(q-q^{-1})|00>)-|+->)\\
        &|1,-1>&=&(q^{2}+q^{-2})^{-\frac{1}{2}}(q|-0>-q^{-1}|0->)\;.
\end{array}
\end{equation}
These formulae show the important fact that the tensor product 
$(\!~\ref{eq:tpd}\,)$ satisfies
\begin{equation}
P_{2}(i,i+1)\Omega_{\alpha\beta}=0\;.
\end{equation}
Such a construction can be extended to the whole spin chain by 
tensoring neighboring spin-1 states with $\epsilon$ giving 
\begin{equation}
\Omega_{\alpha\beta}^{(N)}=\sum_{\begin{array}{c}
{\scriptstyle \alpha_{i}}\\
{\scriptstyle i\in[2,N]}
\end{array}}
\sum_{\begin{array}{c}
{\scriptstyle \beta_{j}}\\
{\scriptstyle j\in[1,N-1]}
\end{array}}
\psi_{\alpha\beta_{1}}\epsilon_{\beta_{1}\alpha_{2}}
\psi_{\alpha_{2}\beta_{2}}\cdots\epsilon_{\beta_{N-1}\alpha_{N}}
\psi_{\alpha_{N}\beta}
\end{equation}
the VBS state. It satisfies 
\begin{equation}
H\Omega^{(N)}_{\alpha\beta}=0
\end{equation}
which can be checked by considering the action of individual 
$P_{2}(i,i+1)$.
 
For the chain with free boundary conditions, the indices $\alpha,\beta$ 
give rise to four states which, when expressed as linear combination of 
strings of $|\pm\!>,|0\!>$, have the characteristic that a nonzero 
state $|+\!>$ ($|-\!>$) must be followed by a $|0\!>$ or $|-\!>$ 
($|+\!>$). Thus $|+\!>$ and $|-\!>$ appear alternately in the VBS and 
there can be any number of $|0\!>$ between the $|+\!>$ and $|-\!>$. The 
four states are distinguished by the following
\begin{description}
\item[$\Omega^{N}_{11}$] The first nonzero states is $|+\!>$ and the 
number of $|+\!>$ states exceeds that of the $|-\!>$ states by 1.
\item[$\Omega^{N}_{12}$] The string has equal number of $|+\!>$ and 
$|-\!>$ states or all $|0\!>$ states. 
\item[$\Omega^{N}_{21}$] Same as in the $\Omega_{12}$ case. 
\item[$\Omega^{N}_{22}$] Same as in the $\Omega_{11}$ case with 
$|+\!>$ replaced by $|-\!>$. 
\end{description}
From the lattice gas point of view (where $|0\!>$ is regarded as 
a vacancy), the VBS exhibits a perfect antiferromagnetic order and 
positional disorder.

Since $\Omega^{N}_{11}$ ($\Omega^{N}_{22}$) contains an extra $|+>$ 
($|->$), we have,
\begin{eqnarray}
S^{z}\Omega^{N}_{11}&=&\Omega^{N}_{11}\;,   \nonumber \\ 
S^{z}\Omega^{N}_{22}&=&-\Omega^{N}_{22}\;,\\ 
\mbox{and}\hspace{10mm}S^{z}\Omega^{N}_{12(21)}&=&0\;.  \nonumber
\end{eqnarray}  
As in the $q=1$ case, it can be proved that the four states belongs to 
the spin-0 and spin-1 irreducible representations, namely,
\begin{eqnarray*}
\Omega^{N}_{11}&\propto&|1,1>\;,\\ 
\Omega^{N}_{22}&\propto&|1,-1>\;,\\ 
q\Omega^{N}_{12}+q^{-1}\Omega^{N}_{21}&\propto&|1,0>\\ 
\mbox{and}\hspace{10mm}\;\;\;\Omega^{N}_{12}-\Omega^{N}_{21}
&\propto&|0,0>\;.
\end{eqnarray*}
The norm can also be computed. We define scalar products  by treating  
$q$  formally as a real parameter so that the  conjugate of the raising 
operator is the lowering operator, and vice versa. As an example the 
state \[q|+\!>+q^{-1}|-\!>\] has norm $q^{2}+q^{-2}$ instead of 2. With 
this convention, eigenstates with different eigenvalues continue to be 
orthogonal for complex $q$. But we loose positivity and definiteness 
in general. The computation is now done using graphical means which 
generalize the method of \cite{aklt}. The contraction
$(\Omega_{\gamma\delta}^{(N)},\Omega_{\alpha\beta}^{(N)})$ is represented 
graphically as two parallel series of horizontal links and dots 
(see fig.(18))
\begin{center} 
\setlength{\unitlength}{0.008in}
\begin{picture}(313,95)(0,-10)
\put(204,60){\circle*{4}}
\put(174,60){\circle*{4}}
\put(139,60){\circle*{4}}
\put(167,60){\circle*{4}}
\put(56,60){\circle*{4}}
\put(131,60){\circle*{4}}
\put(102,60){\circle*{4}}
\put(65,60){\circle*{4}}
\put(95,60){\circle*{4}}
\put(276,60){\circle*{4}}
\put(248,60){\circle*{4}}
\put(286,60){\circle*{4}}
\put(286,30){\circle*{4}}
\put(248,30){\circle*{4}}
\put(276,30){\circle*{4}}
\put(95,30){\circle*{4}}
\put(65,30){\circle*{4}}
\put(102,30){\circle*{4}}
\put(131,30){\circle*{4}}
\put(56,30){\circle*{4}}
\put(167,30){\circle*{4}}
\put(139,30){\circle*{4}}
\put(174,30){\circle*{4}}
\put(204,30){\circle*{4}}
\thicklines
\path(174,60)(204,60)
\path(139,60)(167,60)
\path(102,60)(131,60)
\path(65,60)(95,60)
\path(248,60)(276,60)
\path(248,30)(276,30)
\path(65,30)(95,30)
\path(102,30)(131,30)
\path(139,30)(167,30)
\path(174,30)(204,30)
\put(211,56){\makebox(0,0)[lb]{\raisebox{0pt}[0pt][0pt]
{\shortstack[l]{$\cdots$}}}}
\put(211,27){\makebox(0,0)[lb]{\raisebox{0pt}[0pt][0pt]
{\shortstack[l]{$\cdots$}}}}
\put(-5,56){\makebox(0,0)[lb]{\raisebox{0pt}[0pt][0pt]
{\shortstack[l]{${\scriptstyle \Omega_{\alpha\beta}}$}}}}
\put(-5,27){\makebox(0,0)[lb]{\raisebox{0pt}[0pt][0pt]
{\shortstack[l]{${\scriptstyle \Omega_{\gamma\delta}}$}}}}
\put(40,56){\makebox(0,0)[lb]{\raisebox{0pt}[0pt][0pt]
{\shortstack[l]{${\scriptstyle \alpha}$}}}}
\put(295,56){\makebox(0,0)[lb]{\raisebox{0pt}[0pt][0pt]
{\shortstack[l]{${\scriptstyle \beta}$}}}}
\put(295,27){\makebox(0,0)[lb]{\raisebox{0pt}[0pt][0pt]
{\shortstack[l]{${\scriptstyle \delta}$}}}}
\put(40,27){\makebox(0,0)[lb]{\raisebox{0pt}[0pt][0pt]
{\shortstack[l]{${\scriptstyle \gamma}$}}}}
\put(-40,0){\makebox(0,0)[lb]{\raisebox{0pt}[0pt][0pt]
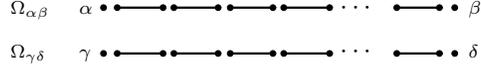
{\shortstack[l]{\footnotesize {\bf Figure(18)} Graphical representation of 
the valence bond states}}}}
\end{picture}
\end{center} 
where each pair of closely spaced dots represents the two spin-
$\frac{1}{2}$ states at each spin-1 site and the horizontal links 
represent the presence of the valence bond, ie. the $\epsilon$ tensor. 
Each pair of dots has contraction only with that directly below (or above) 
it, which gives the contraction of the spin-1 states from 
$\Omega^{(N)}_{\alpha\beta}$ and $\Omega^{(N)}_{\gamma\delta}$ at the 
same sites. We first examine the one particle norm
\begin{equation}
(\psi_{\gamma\delta},\psi_{\alpha\beta})=K_{\alpha\beta}
(\delta_{\alpha\gamma}\delta_{\beta\delta}+\delta_{\alpha\delta}
\delta_{\beta\gamma})   \label{eq:1n}
\end{equation}
where
\[K_{\alpha\beta}=g_{\alpha\beta}^{2}+g_{\beta\alpha}^{2}\;.\]
The rhs of $(\!~\ref{eq:1n}\,)$ can be represented graphically as
(figs.(19a)(19b))
\begin{center} 
\setlength{\unitlength}{0.008in}
\begin{picture}(165,106)(0,-10)
\put(132,45){\circle*{4}}
\put(116,45){\circle*{4}}
\put(50,45){\circle*{4}}
\put(34,45){\circle*{4}}
\put(34,78){\circle*{4}}
\put(50,78){\circle*{4}}
\put(116,78){\circle*{4}}
\put(132,78){\circle*{4}}
\thicklines
\path(132,78)(116,45)
\path(116,78)(132,45)
\path(50,78)(50,45)
\path(34,78)(34,45)
\put(-20,-10){\makebox(0,0)[lb]{\raisebox{0pt}[0pt][0pt]
{\shortstack[l]{\footnotesize Contraction of two spin-1 states}}}}
\put(104,15){\makebox(0,0)[lb]{\raisebox{0pt}[0pt][0pt]
{\shortstack[l]{\footnotesize {\bf Figure(19b)}}}}}
\put(5,15){\makebox(0,0)[lb]{\raisebox{0pt}[0pt][0pt]
{\shortstack[l]{\footnotesize {\bf Figure(19a)}}}}}
\put(137,33){\makebox(0,0)[lb]{\raisebox{0pt}[0pt][0pt]
{\shortstack[l]{${\scriptstyle \delta}$}}}}
\put(108,33){\makebox(0,0)[lb]{\raisebox{0pt}[0pt][0pt]
{\shortstack[l]{${\scriptstyle \gamma}$}}}}
\put(108,82){\makebox(0,0)[lb]{\raisebox{0pt}[0pt][0pt]
{\shortstack[l]{${\scriptstyle \alpha}$}}}}
\put(137,82){\makebox(0,0)[lb]{\raisebox{0pt}[0pt][0pt]
{\shortstack[l]{${\scriptstyle \beta}$}}}}
\put(25,37){\makebox(0,0)[lb]{\raisebox{0pt}[0pt][0pt]
{\shortstack[l]{${\scriptstyle \gamma}$}}}}
\put(54,37){\makebox(0,0)[lb]{\raisebox{0pt}[0pt][0pt]
{\shortstack[l]{${\scriptstyle \delta}$}}}}
\put(54,82){\makebox(0,0)[lb]{\raisebox{0pt}[0pt][0pt]
{\shortstack[l]{${\scriptstyle \beta}$}}}}
\put(25,82){\makebox(0,0)[lb]{\raisebox{0pt}[0pt][0pt]
{\shortstack[l]{${\scriptstyle \alpha}$}}}}
\end{picture}
\end{center} 
and we shall refer to these two geometrical objects as the {\bf 
parallel} and {\bf crossed} vertical links respectively.
It is now clear that 
$(\Omega^{(N)}_{\gamma\delta},\Omega^{(N)}_{\alpha\beta})$ gives 
$2^{N}$ possible graphs which are obtained by replacing each of the one 
particle contraction by fig.(19a) or fig.(19b). A typical graph for 
$N=4$ looks like
\begin{center} 
\setlength{\unitlength}{0.008in}
\begin{picture}(168,99)(0,-10)
\put(149,66){\circle*{4}}
\put(108,66){\circle*{4}}
\put(141,66){\circle*{4}}
\put(16,66){\circle*{4}}
\put(100,66){\circle*{4}}
\put(67,66){\circle*{4}}
\put(26,66){\circle*{4}}
\put(59,66){\circle*{4}}
\put(59,33){\circle*{4}}
\put(26,33){\circle*{4}}
\put(67,33){\circle*{4}}
\put(100,33){\circle*{4}}
\put(16,33){\circle*{4}}
\put(141,33){\circle*{4}}
\put(108,33){\circle*{4}}
\put(149,33){\circle*{4}}
\thicklines
\path(108,66)(141,66)
\path(67,66)(100,66)
\path(26,66)(59,66)
\path(26,33)(59,33)
\path(67,33)(100,33)
\path(108,33)(141,33)
\path(16,66)(16,33)
\path(26,66)(26,33)
\path(59,66)(67,33)
\path(67,66)(59,33)
\path(100,66)(100,33)
\path(108,66)(108,33)
\path(141,66)(149,33)
\path(149,66)(141,33)
\put(0,75){\makebox(0,0)[lb]{\raisebox{0pt}[0pt][0pt]
{\shortstack[l]{${\scriptstyle \alpha}$}}}}
\put(162,75){\makebox(0,0)[lb]{\raisebox{0pt}[0pt][0pt]
{\shortstack[l]{${\scriptstyle \beta}$}}}}
\put(162,15){\makebox(0,0)[lb]{\raisebox{0pt}[0pt][0pt]
{\shortstack[l]{${\scriptstyle \delta}$}}}}
\put(0,15){\makebox(0,0)[lb]{\raisebox{0pt}[0pt][0pt]
{\shortstack[l]{${\scriptstyle \gamma}$}}}}
\put(25,75){\makebox(0,0)[lb]{\raisebox{0pt}[0pt][0pt]
{\shortstack[l]{${\scriptstyle \beta_{1}}$}}}}
\put(54,75){\makebox(0,0)[lb]{\raisebox{0pt}[0pt][0pt]
{\shortstack[l]{${\scriptstyle \bar{\beta}_{1}}$}}}}
\put(112,15){\makebox(0,0)[lb]{\raisebox{0pt}[0pt][0pt]
{\shortstack[l]{${\scriptstyle \bar{\beta}}$}}}}
\put(137,15){\makebox(0,0)[lb]{\raisebox{0pt}[0pt][0pt]
{\shortstack[l]{${\scriptstyle \beta}$}}}}
\put(137,75){\makebox(0,0)[lb]{\raisebox{0pt}[0pt][0pt]
{\shortstack[l]{${\scriptstyle \beta}$}}}}
\put(112,75){\makebox(0,0)[lb]{\raisebox{0pt}[0pt][0pt]
{\shortstack[l]{${\scriptstyle \bar{\beta}}$}}}}
\put(25,15){\makebox(0,0)[lb]{\raisebox{0pt}[0pt][0pt]
{\shortstack[l]{${\scriptstyle \beta_{1}}$}}}}
\put(66,15){\makebox(0,0)[lb]{\raisebox{0pt}[0pt][0pt]
{\shortstack[l]{${\scriptstyle \bar{\beta}_{1}}$}}}}
\put(54,15){\makebox(0,0)[lb]{\raisebox{0pt}[0pt][0pt]
{\shortstack[l]{${\scriptstyle \bar{\beta}_{1}}$}}}}
\put(70,75){\makebox(0,0)[lb]{\raisebox{0pt}[0pt][0pt]
{\shortstack[l]{${\scriptstyle \bar{\beta}_{1}}$}}}}
\put(95,75){\makebox(0,0)[lb]{\raisebox{0pt}[0pt][0pt]
{\shortstack[l]{${\scriptstyle \beta_{1}}$}}}}
\put(95,15){\makebox(0,0)[lb]{\raisebox{0pt}[0pt][0pt]
{\shortstack[l]{${\scriptstyle \beta_{1}}$}}}}
\put(-20,-10){\makebox(0,0)[lb]{\raisebox{0pt}[0pt][0pt]
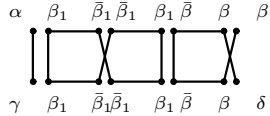
{\shortstack[l]{\footnotesize {\bf Figure(20)} Typical graph for 
${\scriptstyle N=4}$}}}}
\end{picture}
\end{center} 
which carries a weight
\[\sum_{\beta_{1}}\delta_{\alpha\beta}K_{\alpha\beta_{1}}
(\epsilon^{2}_{\beta_{1}\bar{\beta}_{1}}\epsilon^{2}_{\bar{\beta}_{1}
\beta_{1}})K_{\bar{\beta}_{1}\beta_{1}}(\epsilon^{2}_{\bar{\beta}\beta}
\delta_{\beta\delta})\]
and we have introduced the notation
\[\bar{\beta}=\left\{\begin{array}{ll}
1&\mbox{if}\;\; \beta=2\;,\\ 
2&\mbox{if}\;\; \beta=1\;.
\end{array}\right.\]
Our task now is to sum up the weights of the $2^{N}$ graphs. Unlike 
the $q=1$ model where the sum can be performed using combinatoric 
arguments only, the $q$ dependence of $K_{\alpha\beta}$ and $\epsilon$ 
complicates this approach and the sum has to be done with the help 
of recursion relations.
 
We shall define the collection of horizontal or vertical links which 
are connected together as a {\bf circuit}, thus each graph is made 
up of disconnected circuits. Notice that the circuits come in two 
different forms, which are distinguished by the type of vertical links 
at the right and left most ends. In the above example, there are three 
disconnected circuits; The one in the middle has its rightmost vertical 
link formed by one of the parallel vertical links fig.(19a), while the 
circuit on the right has its rightmost vertical link given by fig.(19b) 
and left most link given by fig.(19a). We shall refer to them as 
circuits of type A and B respectively, note that in our definition, 
circuit of type B is characterized by the vertical links at its two ends, 
while circuit of type A by its rightmost end only. We also introduce the 
notion of length for these circuits, namely the length of a circuit is 
equal to half of the number of valence bond it covers. Therefore the B 
and A circuits in the example have length 2 and 1 respectively. Having 
established the notations, we are in a position to characterize the
graphs. Any graph belong to one of the following types:
\begin{eqnarray*}
\delta_{\alpha\gamma}\delta_{\beta\delta}A^{{\rm o}(N)}_{\alpha\beta}
&:&\mbox{graphs whose rightmost circuit is an odd length type A 
circuit,}\\
\delta_{\alpha\gamma}\delta_{\beta\delta}A^{{\rm e}(N)}_{\alpha\beta}
&:&\mbox{graphs whose rightmost circuit is an even length type A 
circuit,}\\
\delta_{\alpha\gamma}\delta_{\beta\delta}B^{(N)}_{\alpha\beta}
&:&\mbox{graphs whose rightmost circuit is type B,}\\
C^{(N)}_{\alpha\beta\gamma\delta}&:&\mbox{graph which does not 
contain any parallel vertical link given in fig.(21).}
\end{eqnarray*}
In the above definitions, $N$ denotes the size of the spin chain, e 
and o denote the parity of the rightmost circuit. It is not difficult 
to see that the above four cases exhaust all the possible types of graph 
and are mutually exclusive. Among the $2^{N}$ graphs, only one of them 
is of type $C^{(N)}_{\alpha\beta\gamma\delta}$, it is made up two 
disconnected circuits running from one end of the graph to the other as 
given by the last figure in fig.(21). Since the first three types of 
graphs always come with the factor $\delta_{\alpha\gamma}
\delta_{\beta\delta}$, we explicitly separate the factor from the rest 
of the contribution. Graphically these four types of graphs have the 
following features (fig.(21))
\begin{center} 
\setlength{\unitlength}{0.008in}
\begin{picture}(374,187)(0,-10)
\put(341,120){\circle*{4}}
\put(331,120){\circle*{4}}
\put(331,148){\circle*{4}}
\put(341,148){\circle*{4}}
\put(181,148){\circle*{4}}
\put(171,148){\circle*{4}}
\put(171,120){\circle*{4}}
\put(181,120){\circle*{4}}
\put(244,120){\circle*{4}}
\put(231,120){\circle*{4}}
\put(231,148){\circle*{4}}
\put(242,148){\circle*{4}}
\put(39,148){\circle*{4}}
\put(29,148){\circle*{4}}
\put(29,120){\circle*{4}}
\put(39,120){\circle*{4}}
\put(100,120){\circle*{4}}
\put(89,120){\circle*{4}}
\put(89,148){\circle*{4}}
\put(100,148){\circle*{4}}
\put(80,25){\circle*{4}}
\put(155,25){\circle*{4}}
\put(127,25){\circle*{4}}
\put(91,25){\circle*{4}}
\put(116,25){\circle*{4}}
\put(116,54){\circle*{4}}
\put(91,54){\circle*{4}}
\put(127,54){\circle*{4}}
\put(155,54){\circle*{4}}
\put(80,54){\circle*{4}}
\put(240,54){\circle*{4}}
\put(205,54){\circle*{4}}
\put(233,54){\circle*{4}}
\put(233,25){\circle*{4}}
\put(205,25){\circle*{4}}
\put(240,25){\circle*{4}}
\path(341,148)(341,120)
\path(331,148)(331,120)
\path(327,141)(309,124)
\path(313,144)(292,124)
\path(299,144)(284,131)
\path(281,120)(331,120)
\path(281,148)(331,148)
\path(228,141)(209,124)
\path(213,144)(192,124)
\path(199,144)(185,131)
\path(181,120)(231,120)
\path(181,148)(231,148)
\path(167,137)(153,124)
\path(163,144)(146,127)
\path(157,144)(146,134)
\path(142,148)(171,148)
\path(142,120)(171,120)
\path(171,148)(171,120)
\path(181,148)(181,120)
\path(242,148)(231,120)
\path(231,148)(242,120)
\path(39,95)(39,113)
\path(41.000,105.000)(39.000,113.000)(37.000,105.000)
\path(85,141)(67,124)
\path(71,144)(50,124)
\path(57,144)(43,131)
\path(39,120)(89,120)
\path(39,148)(89,148)
\path(25,137)(10,124)
\path(21,144)(4,127)
\path(14,144)(4,134)
\path(0,148)(29,148)
\path(0,120)(29,120)
\path(29,148)(29,120)
\path(39,148)(39,120)
\path(100,148)(89,120)
\path(89,148)(100,120)
\path(89,95)(89,113)
\path(91.000,105.000)(89.000,113.000)(87.000,105.000)
\path(181,95)(181,113)
\path(183.000,105.000)(181.000,113.000)(179.000,105.000)
\path(231,95)(231,113)
\path(233.000,105.000)(231.000,113.000)(229.000,105.000)
\path(205,25)(198,46)
\path(205,54)(198,33)
\path(155,25)(158,33)
\path(155,54)(158,46)
\path(80,54)(91,25)
\path(91,54)(80,25)
\path(127,54)(116,25)
\path(116,54)(127,25)
\path(127,25)(155,25)
\path(91,25)(116,25)
\path(91,54)(116,54)
\path(127,54)(155,54)
\path(205,54)(233,54)
\path(205,25)(233,25)
\path(233,54)(240,25)
\path(240,54)(233,25)
\put(281,159){\makebox(0,0)[lb]{\raisebox{0pt}[0pt][0pt]
{\shortstack[l]{${\scriptstyle B_{\alpha\beta}^{(N)}}$}}}}
\put(352,116){\makebox(0,0)[lb]{\raisebox{0pt}[0pt][0pt]
{\shortstack[l]{${\scriptstyle \delta}$}}}}
\put(352,144){\makebox(0,0)[lb]{\raisebox{0pt}[0pt][0pt]
{\shortstack[l]{${\scriptstyle \beta}$}}}}
\put(153,159){\makebox(0,0)[lb]{\raisebox{0pt}[0pt][0pt]
{\shortstack[l]{${\scriptstyle A_{\alpha\beta}^{{\rm e} (N)}}$}}}}
\put(187,98){\makebox(0,0)[lb]{\raisebox{0pt}[0pt][0pt]
{\shortstack[l]{\scriptsize even}}}}
\put(253,116){\makebox(0,0)[lb]{\raisebox{0pt}[0pt][0pt]
{\shortstack[l]{${\scriptstyle \delta}$}}}}
\put(253,144){\makebox(0,0)[lb]{\raisebox{0pt}[0pt][0pt]
{\shortstack[l]{${\scriptstyle \beta}$}}}}
\put(47,98){\makebox(0,0)[lb]{\raisebox{0pt}[0pt][0pt]
{\shortstack[l]{\scriptsize odd}}}}
\put(110,116){\makebox(0,0)[lb]{\raisebox{0pt}[0pt][0pt]
{\shortstack[l]{${\scriptstyle \delta}$}}}}
\put(110,144){\makebox(0,0)[lb]{\raisebox{0pt}[0pt][0pt]
{\shortstack[l]{${\scriptstyle \beta}$}}}}
\put(14,159){\makebox(0,0)[lb]{\raisebox{0pt}[0pt][0pt]
{\shortstack[l]{${\scriptstyle A_{\alpha\beta}^{{\rm o} (N)}}$}}}}
\put(56,-10){\makebox(0,0)[lb]{\raisebox{0pt}[0pt][0pt]
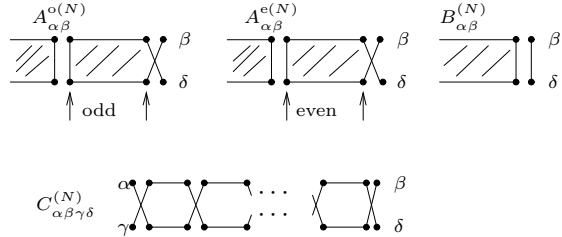
{\shortstack[l]{\footnotesize {\bf Figure(21)} The four types of graphs}}}}
\put(17,36){\makebox(0,0)[lb]{\raisebox{0pt}[0pt][0pt]
{\shortstack[l]{${\scriptstyle C_{\alpha\beta\gamma\delta}^{(N)}}$}}}}
\put(162,29){\makebox(0,0)[lb]{\raisebox{0pt}[0pt][0pt]
{\shortstack[l]{$\cdots$}}}}
\put(162,43){\makebox(0,0)[lb]{\raisebox{0pt}[0pt][0pt]
{\shortstack[l]{$\cdots$}}}}
\put(70,22){\makebox(0,0)[lb]{\raisebox{0pt}[0pt][0pt]
{\shortstack[l]{${\scriptstyle \gamma}$}}}}
\put(70,50){\makebox(0,0)[lb]{\raisebox{0pt}[0pt][0pt]
{\shortstack[l]{${\scriptstyle \alpha}$}}}}
\put(251,50){\makebox(0,0)[lb]{\raisebox{0pt}[0pt][0pt]
{\shortstack[l]{${\scriptstyle \beta}$}}}}
\put(251,22){\makebox(0,0)[lb]{\raisebox{0pt}[0pt][0pt]
{\shortstack[l]{${\scriptstyle \delta}$}}}}
\end{picture}
\end{center} 
The weight $A^{(N)},B^{(N)},C^{(N)}$ of $N$ sites can be related to that 
of $N+1$ sites as 
\begin{equation}
\begin{array}{lll}
A^{{\rm o}(N+1)}_{\alpha\beta}&=&A^{{\rm e}(N)}_{\alpha\bar{\beta}}
\epsilon^{2}_{\bar{\beta}\beta}+B^{(N)}_{\alpha\bar{\beta}}
\epsilon^{2}_{\bar{\beta}\beta}\;,\\
A^{{\rm e}(N+1)}_{\alpha\beta}&=&A^{{\rm o}(N)}_{\alpha\bar{\beta}}
\epsilon^{2}_{\bar{\beta}\beta}\;,\\
B^{(N+1)}_{\alpha\bar{\beta}}&=&\sum_{\beta_{1}}(A^{{\rm e}(N)}_
{\alpha\beta_{1}}\epsilon^{2}_{\beta_{1}\bar{\beta}_{1}}K_{
\bar{\beta}_{1}\beta}+A^{{\rm e}(N)}_{\alpha\beta_{1}}\epsilon^{2}_
{\beta_{1}\bar{\beta}_{1}}K_{\bar{\beta}_{1}\beta}
+B^{(N)}_{\alpha\beta_{1}}\epsilon^{2}_{\beta_{1}\bar{\beta}_{1}}
K_{\bar{\beta}_{1}\beta})\\
& &\mbox{}+\epsilon^{2}_{\alpha\bar{\alpha}}K_{\bar{\alpha}\beta}
\delta_{N,\mbox{\footnotesize odd}}+K_{\alpha\beta}
\delta_{N,\mbox{\footnotesize even}}\;,        
\end{array}   \label{eq:rc}
\end{equation}
which correspond to the various ways of appending an additional site 
to the $N$-site graphs (see fig.(22))
\begin{center} 
\setlength{\unitlength}{0.008in}
\begin{picture}(768,250)(0,60)
\put(276,131){\circle*{4}}
\put(250,131){\circle*{4}}
\put(241,131){\circle*{4}}
\put(241,105){\circle*{4}}
\put(250,105){\circle*{4}}
\put(276,105){\circle*{4}}
\put(215,131){\circle*{4}}
\put(215,105){\circle*{4}}
\put(182,105){\circle*{4}}
\put(208,105){\circle*{4}}
\put(208,131){\circle*{4}}
\put(182,131){\circle*{4}}
\put(172,131){\circle*{4}}
\put(172,105){\circle*{4}}
\put(81,287){\circle*{4}}
\put(81,261){\circle*{4}}
\put(34,261){\circle*{4}}
\put(25,261){\circle*{4}}
\put(25,287){\circle*{4}}
\put(34,287){\circle*{4}}
\put(125,287){\circle*{4}}
\put(91,287){\circle*{4}}
\put(116,287){\circle*{4}}
\put(116,261){\circle*{4}}
\put(91,261){\circle*{4}}
\put(125,261){\circle*{4}}
\put(182,261){\circle*{4}}
\put(182,287){\circle*{4}}
\put(125,183){\circle*{4}}
\put(91,183){\circle*{4}}
\put(117,183){\circle*{4}}
\put(117,209){\circle*{4}}
\put(91,209){\circle*{4}}
\put(125,209){\circle*{4}}
\put(36,209){\circle*{4}}
\put(26,209){\circle*{4}}
\put(26,183){\circle*{4}}
\put(36,183){\circle*{4}}
\put(81,183){\circle*{4}}
\put(81,209){\circle*{4}}
\put(81,131){\circle*{4}}
\put(81,105){\circle*{4}}
\put(36,105){\circle*{4}}
\put(26,105){\circle*{4}}
\put(26,131){\circle*{4}}
\put(36,131){\circle*{4}}
\put(125,131){\circle*{4}}
\put(91,131){\circle*{4}}
\put(117,131){\circle*{4}}
\put(117,105){\circle*{4}}
\put(91,105){\circle*{4}}
\put(125,105){\circle*{4}}
\put(192,261){\circle*{4}}
\put(218,261){\circle*{4}}
\put(218,287){\circle*{4}}
\put(192,287){\circle*{4}}
\put(224,287){\circle*{4}}
\put(224,261){\circle*{4}}
\put(402,105){\circle*{4}}
\put(402,131){\circle*{4}}
\put(408,131){\circle*{4}}
\put(408,105){\circle*{4}}
\put(376,131){\circle*{4}}
\put(376,105){\circle*{4}}
\put(343,105){\circle*{4}}
\put(369,105){\circle*{4}}
\put(369,131){\circle*{4}}
\put(343,131){\circle*{4}}
\path(250,131)(276,131)
\path(250,105)(276,105)
\path(241,131)(250,105)
\path(250,131)(241,105)
\path(276,131)(279,125)
\path(276,105)(279,111)
\path(208,131)(215,105)
\path(215,131)(208,105)
\path(182,131)(182,105)
\path(182,105)(208,105)
\path(182,131)(208,131)
\path(169,121)(156,108)
\path(166,127)(150,111)
\path(160,127)(150,118)
\path(146,131)(172,131)
\path(146,105)(172,105)
\path(172,131)(172,105)
\path(81,287)(91,261)
\path(91,287)(81,261)
\path(36,287)(36,261)
\path(26,287)(26,261)
\path(0,261)(26,261)
\path(0,287)(26,287)
\path(13,284)(3,274)
\path(19,284)(3,268)
\path(23,277)(10,264)
\path(36,287)(81,287)
\path(36,261)(81,261)
\path(52,284)(39,270)
\path(65,284)(45,264)
\path(78,280)(62,264)
\path(91,287)(117,287)
\path(91,261)(117,261)
\path(117,287)(125,261)
\path(125,287)(117,261)
\path(36,238)(36,254)
\path(38.000,246.000)(36.000,254.000)(34.000,246.000)
\path(81,238)(81,254)
\path(83.000,246.000)(81.000,254.000)(79.000,246.000)
\path(182,287)(182,261)
\path(156,261)(182,261)
\path(156,287)(182,287)
\path(169,284)(160,274)
\path(176,284)(160,268)
\path(179,277)(166,264)
\path(81,161)(81,177)
\path(83.000,169.000)(81.000,177.000)(79.000,169.000)
\path(36,161)(36,177)
\path(38.000,169.000)(36.000,177.000)(34.000,169.000)
\path(125,209)(117,183)
\path(117,209)(125,183)
\path(91,183)(117,183)
\path(91,209)(117,209)
\path(78,203)(62,186)
\path(65,206)(45,186)
\path(52,206)(39,193)
\path(36,183)(81,183)
\path(36,209)(81,209)
\path(23,199)(10,186)
\path(19,206)(3,189)
\path(13,206)(3,196)
\path(0,209)(26,209)
\path(0,183)(26,183)
\path(26,209)(26,183)
\path(36,209)(36,183)
\path(91,209)(81,183)
\path(81,209)(91,183)
\path(81,131)(91,105)
\path(91,131)(81,105)
\path(36,131)(36,105)
\path(26,131)(26,105)
\path(0,105)(26,105)
\path(0,131)(26,131)
\path(13,127)(3,118)
\path(19,127)(3,111)
\path(23,121)(10,108)
\path(36,131)(81,131)
\path(36,105)(81,105)
\path(52,127)(39,115)
\path(65,127)(45,108)
\path(78,125)(62,108)
\path(91,131)(117,131)
\path(91,105)(117,105)
\path(117,131)(125,105)
\path(125,131)(117,105)
\path(36,82)(36,99)
\path(38.000,91.000)(36.000,99.000)(34.000,91.000)
\path(81,82)(81,99)
\path(83.000,91.000)(81.000,99.000)(79.000,91.000)
\path(192,287)(192,261)
\path(192,261)(218,261)
\path(192,287)(218,287)
\path(218,287)(224,261)
\path(224,287)(218,261)
\path(402,131)(402,105)
\path(408,131)(408,105)
\path(402,131)(376,131)
\path(402,105)(376,105)
\path(376,131)(369,105)
\path(369,131)(376,105)
\path(343,105)(369,105)
\path(343,131)(369,131)
\path(343,131)(337,109)
\path(343,105)(337,123)
\put(417,131){\makebox(0,0)[lb]{\raisebox{0pt}[0pt][0pt]
{\shortstack[l]{${\scriptstyle \beta}$}}}}
\put(417,105){\makebox(0,0)[lb]{\raisebox{0pt}[0pt][0pt]
{\shortstack[l]{${\scriptstyle \delta}$}}}}
\put(129,131){\makebox(0,0)[lb]{\raisebox{0pt}[0pt][0pt]
{\shortstack[l]{${\scriptstyle \beta}$}}}}
\put(129,105){\makebox(0,0)[lb]{\raisebox{0pt}[0pt][0pt]
{\shortstack[l]{${\scriptstyle \delta}$}}}}
\put(132,209){\makebox(0,0)[lb]{\raisebox{0pt}[0pt][0pt]
{\shortstack[l]{${\scriptstyle \beta}$}}}}
\put(132,183){\makebox(0,0)[lb]{\raisebox{0pt}[0pt][0pt]
{\shortstack[l]{${\scriptstyle \delta}$}}}}
\put(132,287){\makebox(0,0)[lb]{\raisebox{0pt}[0pt][0pt]
{\shortstack[l]{${\scriptstyle \beta}$}}}}
\put(132,261){\makebox(0,0)[lb]{\raisebox{0pt}[0pt][0pt]
{\shortstack[l]{${\scriptstyle \delta}$}}}}
\put(230,287){\makebox(0,0)[lb]{\raisebox{0pt}[0pt][0pt]
{\shortstack[l]{${\scriptstyle \beta}$}}}}
\put(230,261){\makebox(0,0)[lb]{\raisebox{0pt}[0pt][0pt]
{\shortstack[l]{${\scriptstyle \delta}$}}}}
\put(220,131){\makebox(0,0)[lb]{\raisebox{0pt}[0pt][0pt]
{\shortstack[l]{${\scriptstyle \beta}$}}}}
\put(220,105){\makebox(0,0)[lb]{\raisebox{0pt}[0pt][0pt]
{\shortstack[l]{${\scriptstyle \delta}$}}}}
\put(17,295){\makebox(0,0)[lb]{\raisebox{0pt}[0pt][0pt]
{\shortstack[l]{${\scriptstyle A_{\alpha\beta}^{{\rm o} (N+1)}}$}}}}
\put(283,121){\makebox(0,0)[lb]{\raisebox{0pt}[0pt][0pt]
{\shortstack[l]{$\cdots$}}}}
\put(283,108){\makebox(0,0)[lb]{\raisebox{0pt}[0pt][0pt]
{\shortstack[l]{$\cdots$}}}}
\put(13,219){\makebox(0,0)[lb]{\raisebox{0pt}[0pt][0pt]
{\shortstack[l]{${\scriptstyle A_{\alpha\beta}^{{\rm e} (N+1)}}$}}}}
\put(0,55){\makebox(0,0)[lb]{\raisebox{0pt}[0pt][0pt]
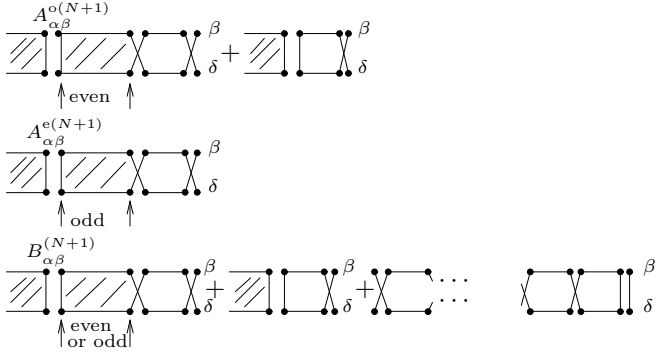
{\shortstack[l]{\footnotesize {\bf Figure(22)} Graphical representation of the 
recursion relations $(\!~\ref{eq:rc}\,)$.}}}}
\put(40,82){\makebox(0,0)[lb]{\raisebox{0pt}[0pt][0pt]
{\shortstack[l]{\scriptsize or odd}}}}
\put(42,92){\makebox(0,0)[lb]{\raisebox{0pt}[0pt][0pt]
{\shortstack[l]{\scriptsize even}}}}
\put(130,115){\makebox(0,0)[lb]{\raisebox{0pt}[0pt][0pt]
{\shortstack[l]{+}}}}
\put(41,242){\makebox(0,0)[lb]{\raisebox{0pt}[0pt][0pt]
{\shortstack[l]{\scriptsize even}}}}
\put(140,270){\makebox(0,0)[lb]{\raisebox{0pt}[0pt][0pt]
{\shortstack[l]{+}}}}
\put(42,161){\makebox(0,0)[lb]{\raisebox{0pt}[0pt][0pt]
{\shortstack[l]{\scriptsize odd}}}}
\put(13,141){\makebox(0,0)[lb]{\raisebox{0pt}[0pt][0pt]
{\shortstack[l]{${\scriptstyle B_{\alpha\beta}^{(N+1)}}$}}}}
\put(228,115){\makebox(0,0)[lb]{\raisebox{0pt}[0pt][0pt]
{\shortstack[l]{+}}}}
\end{picture}
\end{center} 
The weight of $C^{(N)}_{\alpha\beta\gamma\delta}$ can be calculated 
directly as
\begin{equation}
C^{(N)}_{\alpha\beta\gamma\delta}=\left\{\begin{array}{ll}
(K_{\alpha\gamma})^{N-1}K_{\beta\delta}\epsilon_{\alpha\beta}
\epsilon_{\gamma\delta}\;\;\;&;N\in\mbox{even}\;,\\
(K_{\alpha\gamma})^{N}\delta_{\alpha\delta}\delta_{\gamma\beta}
\;\;\;&;N\in\mbox{odd}\;.
\end{array}\right.   \label{eq:cform}
\end{equation}
 
The set of recursion relations $(\!~\ref{eq:rc}\,)$ can be solved 
easily when they are iterated once to relate graphs whose length are of 
the same parity. This gives, for \mbox{\boldmath $N$} {\bf odd},
\begin{equation}
\begin{array}{lll}
A^{{\rm o}(N+2)}_{\alpha\beta}&=&A^{{\rm o}(N)}_{\alpha\beta}
+\sum_{\beta_{1}}(A^{{\rm o}(N)}_{\alpha\beta_{1}}\tilde{K}_
{\beta_{1}\beta}+A^{{\rm e}(N)}_{\alpha\beta_{1}}\tilde{K}_
{\beta_{1}\beta}+B^{(N)}_{\alpha\beta_{1}}\tilde{K}_
{\beta_{1}\beta})+\tilde{K}_{\alpha\beta}\;,\\
A^{{\rm e}(N+2)}_{\alpha\beta}&=&A^{{\rm e}(N)}_{\alpha\beta}+
B^{(N)}_{\alpha\beta}\;,\\
B^{(N+2)}_{\alpha\beta}&=&\sum_{\beta_{1}}(A^{{\rm o}(N)}_
{\alpha\beta_{1}}K_{\beta_{1}\beta}+A^{{\rm e}(N)}_{\alpha\beta_{1}}
K_{\beta_{1}\beta}+B^{(N)}_{\alpha\beta_{1}}\tilde{K}_{\beta_{1}\beta}
+\sum_{\alpha_{1}}(A^{{\rm o}(N)}_{\alpha\alpha_{1}}\tilde{K}_
{\alpha_{1}\beta_{1}}K_{\beta_{1}\beta}\\
& &\mbox{}+A^{{\rm e}(N)}_{\alpha\alpha_{1}}\tilde{K}_
{\alpha_{1}\beta_{1}}K_{\beta_{1}\beta}+B^{(N)}_{\alpha\alpha_{1}}
\tilde{K}_{\alpha_{1}\beta_{1}}K_{\beta_{1}\beta})+\tilde{K}_
{\alpha\beta_{1}}K_{\beta_{1}\beta})+K_{\alpha\beta}\;.
\end{array}
\end{equation}                                                             
where
\[\tilde{K}_{\alpha\beta}=\epsilon^{2}_{\alpha\bar{\alpha}}
K_{\alpha\beta} \epsilon^{2}_{\bar{\beta}\beta}\;.\]
The above may be written more compactly as the matrix equation 
\begin{equation}
{\bf G}^{(N+2)}+{\bf 1}=({\bf G}^{(N)}+{\bf 1})({\bf 1}
+{\bf \tilde{K}})({\bf 1}+{\bf K})       \label{eq:G}                     
\end{equation}
where
\[{\bf G}^{(N)}={\bf A}^{{\rm o}(N)}+{\bf A}^{{\rm e}(N)}
+{\bf B}^{(N)}\]
is a $2\times2$ matrix whose indices are labelled by $\alpha$ 
and $\beta$, and from the definition, it includes $2^{N}-1$ graphs, 
the missing one being $C^{(N)}_{\alpha\beta\gamma\delta}$.
 
For \mbox{\boldmath $N$} {\bf even}, similar calculation gives
\begin{equation}
{\bf \tilde{G}}^{(N+2)}+{\bf 1}=({\bf \tilde{G}}^{(N)}+{\bf 1})
({\bf 1}+{\bf K})({\bf 1}+\tilde{{\bf K}})  \label{eq:GP}             
\end{equation}
where
\[\tilde{G}^{(N)}_{\alpha\beta}=G^{(N)}_{\alpha\bar{\beta}}
\epsilon^{2}_{\bar{\beta}\beta}\;.\]
Equations $(\!~\ref{eq:G}\,)$, $(\!~\ref{eq:GP}\,)$ lead to the results
\begin{equation}
\begin{array}{llll}
{\bf G}^{(N)}&=&({\bf 1}+{\bf K})[({\bf 1}+{\bf \tilde{K}})({\bf 1}
+{\bf K})]^{\frac{N-1}{2}}-{\bf 1}\;\;\;&\;N\in\mbox{odd}\;,\\            
\tilde{{\bf G}}^{(N)}&=&[({\bf 1}+{\bf K})({\bf 1}+{\bf 
\tilde{K}})]^{\frac{N}{2}}-{\bf 1}\;\;\;&\;N\in\mbox{even}\;.            
\end{array}
\end{equation}
Taking $q=1$, the rhs of the above formulae reduce to 
\begin{equation}
\frac{3^{N}-1}{2}\left(\begin{array}{cc}
1&1\\
1&1
\end{array}\right)\;,
\end{equation}
which is precisely the result derived in \cite{aklt}. For arbitrary $q$, 
these formulae can be further simplified by noting that
\begin{equation}
\begin{array}{rcl}
({\bf 1}+\tilde{\bf K})({\bf 1}+{\bf K})&=&\frac{4}{(q+q^{-1})^{2}}
({\bf 1}+(q^{2}+q^{-2})(q+q^{-1})^{2}{\bf P})\\
({\bf 1}+{\bf K})({\bf 1}+\tilde{\bf K})&=&\frac{4}{(q+q^{-1})^{2}}
({\bf 1}+(q^{2}+q^{-2})(q+q^{-1})^{2}{\bf P}^{'})
\end{array}
\end{equation}
where
\[{\bf P}=\frac{1}{q+q^{-1}}\left(\begin{array}{cc}
q^{-1}&q^{-2}\\
q^{2}&q
\end{array}\right)\] 
and
\[{\bf P}^{'}=\frac{1}{q+q^{-1}}\left(\begin{array}{cc}
q&q^{-2}\\
q^{2}&q^{-1}
\end{array}\right)\]
satisfy the property
\begin{equation}
{\bf P}^{(')2}={\bf P}^{(')}\;.
\end{equation}
This relation, when combine with $(\!~\ref{eq:cform}\,)$, gives the 
result of the contraction
\begin{eqnarray}
(\Omega^{(N)}_{\gamma\delta},\Omega^{(N)}_{\alpha\beta})
(\frac{q+q^{-1}}{2})^{N}&=&
\frac{1}{q+q^{-1}}\left(\begin{array}{cc}
\Lambda^{N}-(-1)^{N}&q^{-1}\Lambda^{N}+(-1)^{N}q^{-3}\\
q\Lambda^{N}+(-1)^{N}q^{3}&\Lambda^{N}-(-1)^{N}
\end{array}\right)_{\alpha\beta}
\delta_{\alpha\gamma}\delta_{\beta\delta}   \label{eq:norm}  \\ 
&&\mbox{}-(-1)^{N}\left(\begin{array}{cc}
0&1\\
1&0\end{array}\right)_{\alpha\beta}\delta_{\alpha\delta}
\delta_{\beta\gamma} 
\hspace{20mm}     \nonumber
\end{eqnarray}
where 
\[\Lambda=q^{2}+1+q^{-2}\;.\]
Again the $q=1$ limit of this formula recovers the result of \cite{aklt}.

With this expression for the norm, the spin-spin correlation functions 
defined as
\[<S^{\mu}_{i}S^{\nu}_{j}>_{\rm VBS}=(\Omega^{(N)}_{\gamma\delta},
S^{\mu}_{i}S^{\nu}_{j}\Omega^{(N)}_{\alpha\beta})/
(\Omega^{(N)}_{\gamma\delta},\Omega^{(N)}_{\alpha\beta})  
\;\;\;;\mu,\nu\in{\pm,z}\]
in the VBS states can be computed by breaking down the numerator into
\[\begin{array}{l}
(\Omega^{(i-1)}_{\gamma\delta_{i-1}},\Omega^{(i-1)}_{\alpha\beta_{i-1}})
\epsilon_{\delta_{i-1}\gamma_{i}}\epsilon_{\beta_{i-1}\alpha_{i}}
(\psi_{\gamma_{i}\delta_{i}}, S^{\mu}_{i}\psi_{\alpha_{i}\beta_{i}})
\epsilon_{\delta_{i}\gamma_{i+1}}\epsilon_{\beta_{i}\alpha_{i+1}}
(\Omega^{(j-i-1)}_{\gamma_{i+1}\delta_{j-1}},
\Omega^{(j-i-1)}_{\alpha_{i+1}\beta_{j-1}})  
\epsilon_{\delta_{j-1}\gamma_{j}}\epsilon_{\beta_{j-1}\alpha_{j}}\\
(\psi_{\gamma_{j}\delta_{j}}, S^{\nu}_{j}\psi_{\alpha_{j}\beta_{j}})
\epsilon_{\delta_{j}\gamma_{j+1}}\epsilon_{\beta_{j}\alpha_{j+1}}
(\Omega^{(N-j)}_{\gamma_{j+1}\delta},\Omega^{(N-j)}_{\alpha_{j+1}\beta})
\end{array}\]
and applying $(\!~\ref{eq:norm}\,)$. We shall display only the result 
for the special case $\alpha=\gamma$ and $\beta=\delta$, where the 
nonvanishing correlation functions are
\begin{equation}
\begin{array}{lll}
<S_{i}^{+}S_{j}^{-}>_{11}&=&
(-1)^{j-1}(q+q^{-1})[a\Lambda^{N-j+i-1}-(-1)^{i}q^{-1}b
\Lambda^{N-j}-(-1)^{N-j}qb\Lambda^{i-1}\\
&&\mbox{}+(-1)^{N-j+i}c]/(\Lambda^{N}-(-1)^{N})\\   \\
<S_{i}^{+}S_{j}^{-}>_{12}&=&
(-1)^{j-1}(q+q^{-1})[a\Lambda^{N-j+i-1}-(-1)^{i}q^{-1}b
\Lambda^{N-j}+(-1)^{N-j}q^{-1}b\Lambda^{i-1}\\
&&\mbox{}-(-1)^{N-j+i}q^{-2}c]/
(\Lambda^{N}+q^{-2}(-1)^{N})\;,\\              \\
<S_{i}^{+}S_{j}^{-}>_{21}&=&
(-1)^{j-1}(q+q^{-1})[a\Lambda^{N-j+i-1}+(-1)^{i}qb
\Lambda^{N-j}-(-1)^{N-j}qb\Lambda^{i-1}\\
&&\mbox{}-(-1)^{N-j+i}q^{2}c]/
(\Lambda^{N}+q^{2}(-1)^{N})\;,\\                 \\
<S_{i}^{+}S_{j}^{-}>_{22}&=&
(-1)^{j-1}(q+q^{-1})[a\Lambda^{N-j+i-1}+(-1)^{i}qb\Lambda^{N-j}
+(-1)^{N-j}q^{-1}b\Lambda^{i-1}\\
&&\mbox{}+(-1)^{N-j+i}c]/
(\Lambda^{N}-(-1)^{N})\;,\\                   \\
<S_{i}^{-}S_{j}^{+}>_{\alpha\beta}&=&<S_{i}^{+}S_{j}^{-}>
_{\alpha\beta}\;,\\ \\
<S_{i}^{z}S^{z}_{j}>_{\alpha\alpha}&=&
(-1)^{j-i}(q+q^{-1})^{2}\left(\Lambda^{N-j+i-1}-(-1)^{N}
\Lambda^{j-i}\right)/(\Lambda^{N}-(-1)^{N})\;,\\           \\
<S_{i}^{z}S^{z}_{j}>_{12}&=&(-1)^{j-i}(q+q^{-1})^{2}\left
(q^{-2}\Lambda^{N-j+i-1}+(-1)^{N}\Lambda^{j-i}\right)/
(\Lambda^{N}+q^{-2}(-1)^{N})\;,\\                \\
<S_{i}^{z}S^{z}_{j}>_{21}&=&(-1)^{j-i}(q+q^{-1})^{2}\left
(q^{2}\Lambda^{N-j+i-1}+(-1)^{N}\Lambda^{j-i}\right)/
(\Lambda^{N}+q^{2}(-1)^{N})
\end{array}  
\end{equation}
where
\[\begin{array}{lll}
a&=&q^{3}+2+q^{-3}\\
b&=&q^{2}-q+q^{-1}-q^{-2}\\
c&=&q+q^{-1}-2\;.
\end{array}\]
Before interpreting these formulae, we first examine the role of the 
VBS in the spectrum of the hamiltonian. For $q$ real, one can extend 
the proof of the $q=1$ case and show that the eigenvalues are always 
nonnegative, and VBS are the only ground states. In the infinite $N$ 
limit, the ground state is unique with massive excitation. Hence, the 
model is noncritical with spin-spin correlation functions in the VBS 
given by
\begin{equation}
\begin{array}{rll}
<S_{i}^{+}S_{j}^{-}>=<S_{i}^{-}S_{j}^{+}>&\stackrel{N\rightarrow
\infty}{=}&(-\Lambda)^{-j+i}(q^{3}+2+q^{-3})/(q^{2}+1+q^{-2})\;,\\
<S_{i}^{z}S_{j}^{z}>&\stackrel{N\rightarrow\infty}{=}&
(-\Lambda)^{-j+i}(q^{2}+2+q^{-2})/(q^{2}+1+q^{-2})\;.
\end{array}
\end{equation}
The correlation length is therefore $1/\ln(q^{2}+1+q^{-2})$ and notice 
that the nonisotropy of $S^{\pm}_{i}$ and $S^{z}_{i}$ in the hamiltonian 
$(\!~\ref{eq:p2ham}\,)$ due to the quantum group symmetry is manifested 
in the above. Only when $q\rightarrow1$ where su(2) symmetry is present 
will isotropy in the spin components be restored.
 
Recall that the $q=1$ model belongs to the more general antiferromagnetic 
fluid phase or disorder flat phase (DOF)\cite{den}. It can likewise be 
demonstrated that for real $q$ the VBS ground states have the type of long 
range order and disorder associated with DOF phase. The various correlation 
functions introduced in \cite{den} that distinguish the DOF phase 
can be calculated. We list, in particular, the density-density 
correlation function
\begin{equation}
<(S_{i}^{z})^{2}(S_{j}^{z})^{2}>_{\rm VBS}\stackrel{N\rightarrow 
\infty}{=}<(S_{i}^{z})^{2}>_{\rm VBS}<(S_{j}^{z})^{2}>_{\rm VBS}
\stackrel{N\rightarrow \infty}{=}\frac{4}{(q^{2}+1+q^{-2})^{2}}    
\end{equation}
which confirms that spin positions are completely uncorrelated, and the
correlation function which exhibits antiferromagnetic ordering,
\begin{equation}
G_{s}(j-i)=<S_{i}^{z}{\rm e}^{(S_{i}^{z}+\cdots+S_{j}^{z})}
S_{j}^{z}>_{\rm VBS}\stackrel{N\rightarrow \infty}{=}\frac{4}{(q^{2}
+1+q^{-2})^{2}}\;.    
\end{equation}
The lack of distance dependence shows that AF spin order is perfect.
 
For $q={\rm e}^{i\gamma}\;,\gamma\in{\bf R}$, the configuration space 
belongs to $({\bf C}^{3})^{N}$ and the reasoning that led to the proof 
of  massive excitations for real $q$ no longer holds.  
The eigenvalues can in fact be negative.

Numerical check reveals that for finite $N$ the VBS continue to be the 
only ground state for $\gamma\in[0,\frac{\pi}{6})$ and 
$\gamma\in(\frac{2\pi}{5}, \frac{\pi}{2}]$. In the first domain, 
$q^{2}+1+q^{-2}>1$, so we expect the properties for real $q$ to be still 
qualitatively valid, with  massive excitations and a kind of  DOF phase. 
The second domain has $|q^{2}+1+q^{-2}|<1$ so the behavior is now 
possibly  different from the real $q$ model, in particular, it is not
sure whether  excitations are still  massive. For 
$\gamma\in(\frac{\pi}{6},\frac{2\pi}{5})$, there are negative 
eigenenergies so we certainly  expect  different  properties. 

That another eigenenergy crosses the value zero at  
$\gamma=\frac{\pi}{6}$ can be shown  using ${\rm U}_{q}{\rm su(2)}$ 
symmetry. Indeed the projector $(Q-1)(Q-2)P_{2}$ when restricted to
type II representations satisfies the Temperley Lieb algebra 
with\cite{saa}
\begin{equation}
\begin{array}{rll}
2P_{2}&=&e_{i}      \\
  \mbox{ and }\hspace{5mm}e_{i}^{2}&=&2e_{i}\;.
\end{array}
\end{equation}
The same algebra in the spin-$\frac{1}{2}$ representation given by
\begin{equation} 
e_{i}=2P_{0}
\end{equation}     
with $2P_{0}$ acts on ${\bf C}^{2}\otimes{\bf C}^{2}$ has $"q"$ 
parameter of the quantum group given by $"q"=1$ or ($\sqrt{"Q"}=2$). The 
type II spectrum of $2P_{2}$ at $\gamma=\frac{\pi}{6}$ and the entire 
spectrum of $2P_{0}$ at $"q"=1$ share the same set of eigenvalues. 
Moreover the $q$-dimensions (defined as $(2j+1)_{q}$) of the spin sectors 
from the two representations which share the same eigenvalues are
equal. In particular, the zero eigenvalue, which occurs in the highest 
spin sector ($j=\frac{N}{2}$) of the spin-$\frac{1}{2}$ representation, 
has $q$-dimension
\[(2j+1)_{1}=(N+1)_{1}=N+1\;,\]
while in the spin-1 representation, the contribution to the $q$-dimension 
of the VBS states, which have $j=0,1$, amounts to
\[(2\cdot0+1)_{q}+(2\cdot1+1)_{q}=1+q^{2}+1+q^{-2}=3<N+1\;\;\;;\;
\mbox{ for }N>2\]
implying that new zero eigenvalues must emerge for spin chains with $N>2$. 
As an example, at $N=3$, we find a new zero eigenstate with $j=2$, the
$q$-dimension of which 
\[(2\cdot2+1)_{q}=1\] 
adds to the above giving the total contribution 
\[4(=N+1)\;.\]
 
The crossing of eigenvalues  $\frac{2\pi}{5}$ (and also at 
$\frac{\pi}{5}$) can also be explained. The  $j=2$ spin representations 
are then type I representations and $(Q-1)(Q-2)P_{2}$ vanishes when 
restricted to type II representations. Thus all type II eigenvalues vanish. 

Despite the fact that quantum group symmetry implies additional zero 
eigenstates have to emerge at $\frac{\pi}{6}$ and $\frac{2\pi}{5}$, 
we only have numerical support that this
does not happen outside the domain $[\frac{\pi}{6},\frac{2\pi}{5}]$.

We did not get definite numerical evidence for possible critical 
properties in the domain $[\frac{\pi}{6},\frac{2\pi}{5}]$. Notice 
however the special value 
$\gamma=\frac{\pi}{3}$ where the $P_{2}$ projector line 
$(\!~\ref{eq:p2}\,)$ meets the TL line, so there we have criticality
 with $c=-2$.

\subsection{The $\Gamma_{2}$ Potts model}
 
\hspace{5 mm}
    We have discussed the phase diagram of the quantum spin chain because 
it is the simplest and has most immediate applications. It is not always 
easy to discuss the relation of this study with the two dimensional Potts 
model. Clearly the above hamiltonians, although considered so far as 
acting on spins, can be rewritten as Potts hamiltonians using the 
appropriate representation of the projectors discussed earlier. It is 
reasonable to hope that the physics of these hamiltonians is the same as 
the one of a two dimennsional strongly anisotropic Potts model whose 
elementary transfer matrix reads $1+\epsilon H$. This correspondence is 
enough to apply to the quantum spin chain duality arguments deduced for a 
two dimensional (not necessarily isotropic) Potts model.  However when 
couplings in two directions take comparable values, it is not clear whether 
the physics will or not be qualitatively different. This is especially true 
in our case where there are both ferromagnetic and antiferromagnetic 
interactions. In the integrable cases however, one can usually connect the 
physics for different isotropies by changing the spectral parameter, and 
exact solutions usually show that properties are the same provided this 
parameter runs in a certain range. Let us  write the isotropic interactions 
associated with the three integrable lines discussed earlier ( they will 
be recovered in the hamiltonian limit  $u=0$, $\epsilon=-1$. The case for 
$\epsilon=1$ can similarly be studied. ) 
 
The Boltzmann weight associated to the fundamental block
${\cal G}_{2}$ has  physical expression given by
\begin{equation}
W_{abcd}=\exp {\cal E}_{abcd}
\end{equation}
where the interaction energy
\[{\cal E}={\cal K}_{0}(Q-1)P_{0}+{\cal K}_{1}(Q-1)P_{1}\;.\]
Written in terms of the four sites $a,\cdots,d$, the energy reads
\begin{eqnarray}
{\cal E}_{abcd}&=&({\cal K}_{0}-4{\cal K}_{1})Q^{-2}
 -({\cal K}_{0}-2{\cal K}_{1})Q^{-1}(\delta_{ab}+\delta_{cd})
+{\cal K}_{1}Q^{-1}(\delta_{ac}+\delta_{bd}+\delta_{ad}+\delta_{bc})\\ 
& & \mbox{}+({\cal K}_{0}-{\cal K}_{1})\delta_{ab}\delta_{cd}
 -{\cal K}_{1}(\delta_{abc}+\delta_{abd}+\delta_{bcd}+\delta_{acd})
+{\cal K}_{1}Q\delta_{abcd}\;, \nonumber
\end{eqnarray}
which shows that the various interactions; nearest neighbors, next to 
nearest neighbors etc. can either be ferromagnetic or antiferromagnetic 
depending on the values of the coupling constants ${\cal K}_{0}$ and 
${\cal K}_{1}$. The dual model  involves similar expressions. 

The TL integrable line is given by
\[Q^{-1/2}f_{1}=0\]
which translates into
\begin{equation}
\begin{array}{rcl}
{\cal K}_{0}&=&\ln 2/(Q-1)\;,\\
\mbox{and }\hspace{5mm}{\cal K}_{1}&=&0\;.
\end{array}
\end{equation}
using $(\!~\ref{eq:coupling}\,)$ and $f_{0}=1$. In this case the energy 
expression becomes a product of  $Q^{-1}-\delta_{ab}$ and 
$Q^{-1}-\delta_{cd}$ with ${\cal K}_{0}$ being the overall coefficient. 
The ferro- and antiferro-magnetic nature of the interactions therefore 
depend only on the sign of ${\cal K}_{0}$, which  flips at $Q=1$. For 
$Q>1$ the model is characterized by antiferromagnetic nearest neighbors 
interaction and ferromagnetic four sites interaction 
$\delta_{ab}\delta_{cd}$, and the converse  for $Q<1$. It should be 
noted that the point $Q=1$ corresponds to $Q'=0$ of the $\Gamma_{1}$
( or standard ) Potts model via the Temperley Lieb algebra. For the
$\Gamma_{1}$ Potts model on the TL integrable line 
$(\!~\ref{eq:TLintegrable}\,)$, $Q'=0$ is 
precisely the point that divides the ferro- and anti-ferromagnetic
regimes\cite{sal}.  

The FZ line is given by
\begin{equation}
Q^{-1/2}f_{1}=\frac{1}{\sqrt{Q}+1}\;,
\end{equation}
which is equivalent to 
\begin{equation}
\begin{array}{rcl}
{\cal K}_{0}&=&\frac{\textstyle \ln(Q+\sqrt{Q}-1)}{\textstyle Q-1}\;,\\
{\cal K}_{1}&=&\frac{\textstyle \ln(Q+\sqrt{Q}-1)-\ln(\sqrt{Q}+1)}
{\textstyle Q-2}
\end{array}
\end{equation}
where the coupling constants at $Q=1$ and $Q=2$ are defined by continuity.
It is easy to see that both ${\cal K}_{i}$'s are nonnegative functions 
of $Q$. However ${\cal K}_{0}$ becomes complex for $Q<(3-\sqrt{5})/2$ 
( or $\gamma>2\pi/5$ ) and the Potts model beyond that point is not 
physical. In the domain $Q>(3-\sqrt{5})/2$, the magnetic nature of the 
 interaction terms in the energy expression remain unchanged being always 
ferro- or antiferro-magnetic  as the coefficients ${\cal K}_{1}$, 
${\cal K}_{0}-{\cal K}_{1}$ and ${\cal K}_{0}-2{\cal K}_{1}$ 
are always positive. 

The same analysis can be performed on the IK integrable line, which is 
given by
\begin{equation}
Q^{-1/2}f_{1}=-1+\sqrt{4-Q}
\end{equation}
or equivalently
\begin{equation}
\begin{array}{rcl}
{\cal K}_{0}&=&\frac{\textstyle \ln(1+(Q-1)\sqrt{4-Q})}
{\textstyle Q-1}\;,\\
{\cal K}_{1}&=&\frac{\textstyle \ln(3-Q+(Q-2)\sqrt{4-Q})}
{\textstyle Q-2}\;.
\end{array}
\end{equation}
The coupling constants are real for $Q\in(0.77,3.80)$ approximately. 
As $Q$ increases from 0.77 in this domain, the majority of the 
interaction terms which have coefficient proportional to ${\cal K}_{1}$ 
changes from ferromagnetic to antiferromagnetic or vice versa at $Q=3$. 
In the phase diagram this is the point that divides the various regimes 
of the integrable line. Since ${\cal K}_{1}>0$ as in the FZ case, the 
magnetic natures of the majority of the interactions of the IK integrable 
model for $Q<3$ are the same as that of the FZ integrable model. 
The exception being the interactions $\delta_{ab}\delta_{cd}$ and 
$\delta_{ab}+\delta_{cd}$, whose respective coefficients 
${\cal K}_{0}-{\cal K}_{1}$ and ${\cal K}_{0}-2{\cal K}_{1}$ change 
from negative to positive at $Q\simeq 0.8$ and $Q\simeq 1.9$. This 
similarity in the physical behaviors of the interactions supports the 
conclusion reached in  the spin chain study that the two lines are in 
the same universality for $Q<3$.

\section{Conclusion}
\hspace{5mm}
 
$\Gamma_k$ Potts models provide a rather different kind of physical 
models associated with  spin-$k/2$  representations of 
${\rm U}_{q}{\rm su}(2)$, where the higher symmetry constraints are 
encoded in a pattern of complicated interactions on a plaquette. Besides 
their "academic" interest we hope they can provide new insight on the 
physics of related solutions of Yang Baxter equation, or universality 
classes. For instance   the standard ${\rm su(2)}$ symmetric quantum 
spin chains are related to  $Q=4$ states Potts models with a mixture of
ferromagnetic and antiferromagnetic interactions. The splitting between 
integer and half integer spin is very naturally observed in this picture. 
A translation invariant quantum spin chain is the anisotropic limit of 
a four state Potts model based on a homogeneous vertex model. For
half integer spin  ($k$ odd) this Potts model turns out to be necessarily 
self dual. One therefore expect it, by standard arguments, to be at a 
critical point. On the other hand for integer spin ($k$ even) the Potts 
model is not self dual, and generically is expected to be in some non 
critical state. This is quite close to the Haldane conjecture. 

The technlogy of quantum groups, Temperley Lieb algebras and graphical
representations is known under other names in the condensed matter 
literature\cite{affleck}. In particular it was remarked in \cite{affleck} 
that the standard $Q$ state Potts model can be related to a quantum  
spin chain  with ${\rm su(n)}$ symmetry, with the fundamental 
representation on a sub lattice and its conjugate on the other, and 
$n=\sqrt{Q}$. More generally one can speculate that systems with quantum
group symmetries provide proper analytic continuations of models with 
ordinary symmetries when the rank of the algebra or the size of the 
representation assume "intermediate" values. For instance the spin 1 
${\rm U}_{q}{\rm su}(2)$ model, or equivalently the $\Gamma_2$ Potts 
model, can be related to a quantum spin chain with ${\rm su(n)}$ symmetry, 
once again $n=\sqrt{Q}$, but with the adjoint representation on every site. 
This is because $(3)_q=n^2-1$. As shown in section 5 of this paper, the 
phase diagram is rich. In particular several critical lines and massless 
phases are met in the continuation from $n=2$ ($q=1$) to $n=0$ ($q=i$), 
which is of interest for the quantum Hall effect \cite{affleck}
. As the representation gets more complex we expect this continuation to 
"go through" a more and more complicated phase diagram. 

Finally we remark that the quantum group symmetric models are  also 
a particular example of anisotropic quantum spin chains. From that point 
of view, the last paragraph in section 5 represents an extension of the 
valence bond method to  a particular anisotropic situation. 

\bigskip
{\bf Acknowledgments}: I. Affleck, B.Nienhuis and N.Read are thanked 
for useful discussions.

\pagebreak

\appendix
\section{Boltzmann Factor of $\Gamma_{3}$ Potts model}
\hspace{5mm}
The Boltzmann weight of the $\Gamma_{3}$ Potts model contains 
interactions between all the six sites $a,\dots, f$, the explicit form 
is given by
\begin{eqnarray*}
\lefteqn{W(u)_{aebcfd}=(Q-1)^{-2}\{\delta_{ef}-(Q-1)\delta_{ef}
(\delta_{bd}+\delta_{ac})-(\delta_{efa}+\delta_{efb}+\delta_{efc}
+\delta_{efd})+\delta_{abef}+\delta_{adef}}\\
&&\mbox{}+\delta_{bcef}+\delta_{cdef}+Q(\delta_{bdef}+\delta_{acef})
+(Q-1)(\delta_{bd}\delta_{aef}+\delta_{bd}\delta_{cef}+\delta_{ac}
\delta_{bef}+\delta_{ac}\delta_{def})\\
&&\mbox{}-Q(\delta_{abdef}+\delta_{acdef}+\delta_{abcef}
+\delta_{bcdef})-Q(Q-1)(\delta_{ac}\delta_{bdef}+\delta_{bd}
\delta_{acef})\\
&&\mbox{}+(Q-1)^{2}\delta_{ac}\delta_{bd}\delta_{ef}+Q^{2}
\delta_{abcdef}\}\\
&&\mbox{}+Q^{-1/2}h(Q-1)^{-2}\{1-(\delta_{cf}+\delta_{fd}+\delta_{be}
+\delta_{ae})-(Q-1)(\delta_{ab}+\delta_{cd})+Q(\delta_{abe}
+\delta_{cdf})\\
&&\mbox{}+\delta_{ae}\delta_{cf}+\delta_{ae}\delta_{df}+\delta_{be}
\delta_{cf}+\delta_{be}\delta_{df}+(Q-1)(\delta_{be}\delta_{cd}
+\delta_{ae}\delta_{cd}+\delta_{ab}\delta_{cf}+\delta_{ab}\delta_{df})\\
&&\mbox{}+(Q-1)^{2}\delta_{ab}\delta_{cd}-Q(\delta_{ae}\delta_{cfd}
+\delta_{be}\delta_{cfd}+\delta_{cf}\delta_{abe}+\delta_{df}\delta_{abe})
-Q(Q-1)(\delta_{ab}\delta_{cdf}+\delta_{cd}\delta_{abe})\\
&&\mbox{}+Q^{2}\delta_{abe}\delta_{cdf}\}\\
&&\mbox{}+f(Q-1)^{-4}\{2-3Q+(3Q-2)(\delta_{be}+\delta_{cf}
+\delta_{df}+\delta_{ae})+(Q-1)(\delta_{bf}+\delta_{de}+\delta_{ce}
+\delta_{af})\\
&&\mbox{}+\delta_{ef}+(Q-1)^{2}(\delta_{bd}+\delta_{cd}+\delta_{bc}
+\delta_{ad}+\delta_{ab}+\delta_{ac})-Q(\delta_{bef}+\delta_{def}
+\delta_{aef}+\delta_{cef})\\
&&\mbox{}-(Q-1)^{2}(\delta_{cde}+\delta_{abf})-Q(Q-1)(\delta_{bde}
+\delta_{bdf}+\delta_{bcf}+\delta_{ade}+\delta_{bce}+\delta_{adf}
+\delta_{acf}+\delta_{ace})  \\
&&\mbox{}+(1-Q-Q^{2})(\delta_{cfd}+\delta_{abe})-(Q-1)^{3}
(\delta_{bcd}+\delta_{abd}+\delta_{abc}+\delta_{acd})+(2-3Q)
(\delta_{be}\delta_{cf}\\
&&\mbox{}+\delta_{be}\delta_{df}+\delta_{ae}\delta_{df}+\delta_{ae}
\delta_{cf})-(Q-1)(\delta_{de}\delta_{cf}+\delta_{ae}\delta_{bf}
+\delta_{df}\delta_{ec}+\delta_{be}\delta_{af})\\
&&\mbox{}-(Q-1)^{2}(\delta_{bd}\delta_{cf}+\delta_{be}\delta_{cd}
+\delta_{bc}\delta_{df}+\delta_{ae}\delta_{cd}+\delta_{bc}\delta_{ae}
+\delta_{ad}\delta_{be}+\delta_{ab}\delta_{df}+\delta_{ad}\delta_{cf}
+\delta_{ac}\delta_{be}\\
&&\mbox{}+\delta_{ac}\delta_{df}+\delta_{ab}\delta_{fc}
+\delta_{bd}\delta_{ae})+Q^{2}(\delta_{bdef}+\delta_{acef}
+\delta_{bcef}+\delta_{cdef}+\delta_{abef}+\delta_{adef})\\
&&\mbox{}+Q(Q-1)^{2}(\delta_{bcde}+\delta_{acde}
+\delta_{abdf}+\delta_{abcf})+Q^{2}(Q-1)(\delta_{abce}+\delta_{abde}
+\delta_{bcdf}+\delta_{acdf})\\
&&\mbox{}+(Q^{2}+Q-1)(\delta_{cdf}\delta_{be}
+\delta_{cdf}\delta_{ae}+\delta_{abe}\delta_{df}+\delta_{abe}
\delta_{cf})+Q(Q-1)(\delta_{bde}\delta_{cf}+\delta_{bdf}\delta_{ae}\\
&&\mbox{}+\delta_{acf}\delta_{be}+\delta_{ace}\delta_{df}
+\delta_{ade}\delta_{cf}+\delta_{bcf}\delta_{ae}+\delta_{bce}
\delta_{df}+\delta_{adf}\delta_{be})+(Q-1)^{2}(\delta_{abe}\delta_{cd}
+\delta_{cdf}\delta_{ab})\\
&&\mbox{}+\delta_{abcd}+(Q-1)^{3}(\delta_{abc}\delta_{df}
+\delta_{acd}\delta_{be}+\delta_{dcb}\delta_{ae}+\delta_{abd}
\delta_{fc})\\
&&\mbox{}-Q^{3}(\delta_{bcdef}+\delta_{abdef}+\delta_{abcef}
+\delta_{acdef})-Q^{2}(Q-1)^{2}(\delta_{abcde}+\delta_{abcdf})
-Q(2Q-1)\delta_{abe}\delta_{cdf}\\
&&\mbox{}+(Q-1)^{2}(\delta_{bd}\delta_{fc}\delta_{ae}
+\delta_{bc}\delta_{fd}\delta_{ae}+\delta_{ad}\delta_{be}
\delta_{fc}+\delta_{ac}\delta_{be}\delta_{fd})+Q^{4}\delta_{abcdef}\\
&&\mbox{}-Q^{2}(Q-1)(\delta_{abde}\delta_{cf}+\delta_{bcdf}\delta_{ae}
+\delta_{abce}\delta_{df}+\delta_{acdf}\delta_{be})\}\\
&&\mbox{}+Q^{-1/2}g(Q-1)^{-4}\{Q^{2}+(1-2Q)\delta_{ef}-Q(Q-1)^{2}
(\delta_{ac}+\delta_{bd})-Q^{2}(\delta_{ae}+\delta_{be}+\delta_{cf}
+\delta_{df})\\
&&\mbox{}-Q(Q-1)(\delta_{ed}+\delta_{ec}+\delta_{bf}+\delta_{af})
+(Q^{2}+Q-1)(\delta_{efd}+\delta_{efb}+\delta_{efc}+\delta_{efa})
+Q^{2}(\delta_{cfd}+\delta_{abe})\\
&&\mbox{}+Q(Q-1)(\delta_{bfc}+\delta_{bec}+\delta_{adf}+\delta_{ade})
+Q^{2}(Q-1)(\delta_{bed}+\delta_{bfd}+\delta_{acf}+\delta_{ace})\\
&&\mbox{}+(Q-1)^{2}(\delta_{afb}+\delta_{cde}+\delta_{ac}\delta_{ef}
+\delta_{bd}\delta_{ef})+Q(Q-1)^{2}(\delta_{bd}\delta_{fc}
+\delta_{bd}\delta_{ae}+\delta_{be}\delta_{ac}+\delta_{ac}\delta_{df})\\
&&\mbox{}+Q^{2}(\delta_{be}\delta_{fc}+\delta_{ae}\delta_{fd}
+\delta_{ae}\delta_{fc}+\delta_{be}\delta_{df})+(Q-1)^{3}
(\delta_{af}\delta_{bd}+\delta_{ec}\delta_{bd}+\delta_{ac}\delta_{bf}
+\delta_{ac}\delta_{de})\\
&&\mbox{}+Q(Q-1)(\delta_{ec}\delta_{fd}+\delta_{af}\delta_{be}
+\delta_{ae}\delta_{bf}+\delta_{cf}\delta_{ed})+(Q-1)^{2}(\delta_{bf}
\delta_{ec}+\delta_{af}\delta_{ed})\\&&
\mbox{}+(Q-1)^{4}\delta_{ac}\delta_{bd}+Q(1-Q-Q^{2})(\delta_{acef}
+\delta_{bdef})+Q(2-3Q)(\delta_{cdef}+\delta_{abef}
+\delta_{bcef}+\delta_{adef})\\
&&\mbox{}-Q^{2}(Q-1)(\delta_{acdf}+\delta_{bcdf}+\delta_{abde}
+\delta_{abce})-Q(Q-1)^{2}(\delta_{bcde}+\delta_{acde}+\delta_{abcf}
+\delta_{abdf})\\
&&\mbox{}-Q^{2}(\delta_{be}\delta_{cfd}+\delta_{ae}\delta_{cfd}
+\delta_{abe}\delta_{cf}+\delta_{abe}\delta_{df})
-Q^{2}(Q-1)(\delta_{bde}\delta_{fc}+\delta_{acf}\delta_{be}
+\delta_{ace}\delta_{fd}+\delta_{bdf}\delta_{ae})\\
&&\mbox{}-Q(Q-1)^{2}(\delta_{ace}\delta_{bf}+\delta_{bdf}\delta_{ec}
+\delta_{efc}\delta_{bd}+\delta_{bde}\delta_{af}+\delta_{bef}\delta_{ac}
+\delta_{efd}\delta_{ac}+\delta_{afc}\delta_{ed}+\delta_{aef}\delta_{bd})\\
&&\mbox{}-Q(Q-1)(\delta_{bec}\delta_{fd}+\delta_{adf}\delta_{be}
+\delta_{aed}\delta_{fc}+\delta_{bcf}\delta_{ae})
+Q^{2}\delta_{abe}\delta_{cdf}+Q^{2}(2-3Q)\delta_{abcdef}\\
&&\mbox{}-Q(Q-1)^{3}(\delta_{bde}\delta_{ac}+\delta_{bdf}\delta_{ac}
+\delta_{acf}\delta_{bd}+\delta_{ace}\delta_{bd})+Q^{2}(3Q-2)
(\delta_{abcef}+\delta_{abdef}+\delta_{bcdef}\\
&&\mbox{}+\delta_{acdef})+Q^{2}(Q-1)^{2}(\delta_{abcdf}+\delta_{abcde}
+\delta_{bdef}\delta_{ac}+\delta_{acef}\delta_{bd}+\delta_{afc}
\delta_{bde}+\delta_{ace}\delta_{bdf})\\
&&\mbox{}-Q(Q-1)^{2}(\delta_{ac}\delta_{be}\delta_{df}
+\delta_{bd}\delta_{ae}\delta_{cf})+Q^{2}(Q-1)(\delta_{acdf}
\delta_{be}+\delta_{bcdf}\delta_{ae}+\delta_{abde}\delta_{cf}
+\delta_{abce}\delta_{df})\}
\end{eqnarray*}

\section{Loop Model Formulation of the $\Gamma_{k}$ model}
\hspace{5mm}
We present in this appendix a brief review of the known 
loop model formulations of the $\Gamma_{k}$ model. 

\subsection{}
First we recall the standard graphical representation of the 
Temperley-Lieb algebra\cite{kfm,mar} with the representation space 
being a set of strands. The generator $e_{i}$ acts on two
neighboring strands and produces the following configurations
\begin{center} 
\setlength{\unitlength}{0.01in}
\begin{picture}(163,49)(0,-10)
\thicklines
\path(120,0)(137,16)(120,34)
\path(163,0)(146,16)(163,34)
\path(30,26)(47,26)
\path(30,8)(47,8)
\thinlines
\path(65,16)(103,16)
\path(95.000,14.000)(103.000,16.000)(95.000,18.000)
\put(0,16){\makebox(0,0)[lb]{\raisebox{0pt}[0pt][0pt]{\shortstack[l]
{$e_{i}$}}}}
\end{picture}

\end{center} 
while the identity leaves the strands unaltered
\begin{center} 
\setlength{\unitlength}{0.01in}
\begin{picture}(141,55)(0,-10)
\thicklines
\path(37,28)(53,28)
\path(37,12)(53,12)
\path(109,40)(125,24)(141,40)
\path(109,0)(125,16)(141,0)
\thinlines
\path(69,20)(105,20)
\path(97.000,18.000)(105.000,20.000)(97.000,22.000)
\put(0,15){\makebox(0,0)[lb]{\raisebox{0pt}[0pt][0pt]{\shortstack[l]
{{\bf 1}}}}}
\end{picture}

\end{center} 
With this definition, the algebraic relations $(\!~\ref{eq:TL}\,)$ are 
represented as
\begin{center} 
\setlength{\unitlength}{0.01in}
\begin{picture}(226,170)(0,-10)
\thicklines
\path(179,112)(218,112)
\path(218,124)(202,140)(218,155)
\path(179,124)(195,140)(179,155)
\path(62,155)(47,140)(58,124)
	(43,108)(27,108)
\path(62,155)(78,140)(66,124)
	(82,108)(97,108)
\path(101,124)(86,140)(101,155)
\path(23,124)(39,140)(23,155)
\thinlines
\path(125,132)(160,132)
\path(152.000,130.000)(160.000,132.000)(152.000,134.000)
\thicklines
\path(31,23)(47,38)(31,54)
\path(70,23)(55,38)(70,54)
\path(70,23)(86,38)(70,54)
\path(109,23)(93,38)(109,54)
\path(226,23)(210,38)(226,54)
\path(187,23)(202,38)(187,54)
\thinlines
\path(125,38)(160,38)
\path(152.000,36.000)(160.000,38.000)(152.000,40.000)
\put(0,60){\makebox(0,0)[lb]{\raisebox{0pt}[0pt][0pt]
{\shortstack[l]{and}}}}
\put(160,35){\makebox(0,0)[lb]{\raisebox{0pt}[0pt][0pt]
{\shortstack[l]{$\sqrt{Q}$}}}}
\put(44,0){\makebox(0,0)[lb]{\raisebox{0pt}[0pt][0pt]
{\shortstack[l]{$e_{i}^{2}=\sqrt{Q}e_{i}$}}}}
\put(44,84){\makebox(0,0)[lb]{\raisebox{0pt}[0pt][0pt]
{\shortstack[l]{$e_{i}e_{i+1}e_{i}=e_{i}$}}}}
\end{picture}
\end{center} 
which are easily seen to be satisfied in this representation provided 
every loop is given a weight $\sqrt{Q}$. Such a reformulation was 
rediscovered many times, in particular in \cite{affleck} using valence 
bond language.

 This gives a geometrical reformulation of the six-vertex model (known 
also as loop model formulation) where the vertex
$x_{1}{\bf 1}+e_{2i-1}$ or ${\bf 1}+x_{2}e_{2i}$ are replaced by the 
graphical combinations
\begin{center} 
\setlength{\unitlength}{0.01in}
\begin{picture}(312,56)(0,-10)
\thicklines
\path(193,41)(210,24)(226,41)
\path(193,0)(210,16)(226,0)
\path(312,3)(295,20)(312,36)
\path(271,3)(287,20)(271,36)
\path(30,0)(46,16)(63,0)
\path(30,41)(46,24)(63,41)
\path(91,3)(107,20)(91,36)
\path(132,3)(115,20)(132,36)
\put(230,16){\makebox(0,0)[lb]{\raisebox{0pt}[0pt][0pt]
{\shortstack[l]{+}}}}
\put(245,16){\makebox(0,0)[lb]{\raisebox{0pt}[0pt][0pt]
{\shortstack[l]{$x_{2}$}}}}
\put(74,16){\makebox(0,0)[lb]{\raisebox{0pt}[0pt][0pt]
{\shortstack[l]{+}}}}
\put(0,16){\makebox(0,0)[lb]{\raisebox{0pt}[0pt][0pt]
{\shortstack[l]{$x_{1}$}}}}
\put(157,16){\makebox(0,0)[lb]{\raisebox{0pt}[0pt][0pt]
{\shortstack[l]{or}}}}
\end{picture}
\end{center} 
and the lattice is  accordingly covered by a collections of closed 
loops. From the Poots model point of view, these are the surrounding 
polygons of clusters    high temperature expansion.

Recall that there are some subtleties about the models correspondence 
due to boundary conditions. 

Using the fusion procedure, a loop model formulation can be given to the
$\Gamma_{k}$ vertex model \cite{kauff,mar,sal1}.
 Graphically, this is done by replacing each of the
six-vertex in $(\!~\ref{eq:Xu}\,)$ by one of the above configurations. 
Each spin-$\frac{k}{2}$ vertex therefore acts on $2k$ strands, the 
symmetrizer $S_{k}$ that acts on $k$ strands is represented by
\begin{center} 
\setlength{\unitlength}{0.01in}
\begin{picture}(108,58)(0,-10)
\thicklines
\path(35,38)(104,38)
\path(35,34)(104,34)
\path(35,30)(104,30)
\path(35,8)(104,8)
\path(35,4)(104,4)
\path(52,43)(52,0)
\put(65,15){\makebox(0,0)[lb]{\raisebox{0pt}[0pt][0pt]
{\shortstack[l]{$\vdots$}}}}
\put(0,17){\makebox(0,0)[lb]{\raisebox{0pt}[0pt][0pt]
{\shortstack[l]{$S_{k}$}}}}
\put(-40,-15){\makebox(0,0)[lb]{\raisebox{0pt}[0pt][0pt]
{\shortstack[l]{\footnotesize {\bf Figure(B0)} The composite $k$-strand
that denotes the symmetrizer $S_{k}$}}}}
\end{picture}
\end{center} 
which is a composite object given by $(\!~\ref{eq:sym}\,)$. As an example 
for $k=2$
\[S_{2}={\bf 1}-\frac{1}{(2)_{q}}e\]
has graphical representation 
\begin{center} 
\setlength{\unitlength}{0.01in}
\begin{picture}(256,62)(0,-10)
\thicklines
\path(45,29)(83,29)
\path(45,19)(83,19)
\path(65,33)(65,14)
\path(256,5)(238,23)(256,43)
\path(209,5)(229,23)(209,43)
\path(120,0)(140,19)(158,0)
\path(120,47)(140,29)(158,47)
\put(0,19){\makebox(0,0)[lb]{\raisebox{0pt}[0pt][0pt]
{\shortstack[l]{$S_{2}$}}}}
\put(155,19){\makebox(0,0)[lb]{\raisebox{0pt}[0pt][0pt]
{\shortstack[l]{$- 1/(2)_{q}$}}}}
\put(102,19){\makebox(0,0)[lb]{\raisebox{0pt}[0pt][0pt]
{\shortstack[l]{=}}}}
\end{picture}
\end{center} 
The internal vertices $r_{k}(u_{1}),\cdots,r_{k}(u_{k^{2}})$ that are 
inserted between the four copies of $S_{k}$ represented above produce 
in general $2^{k^{2}}$  configurations. However, configurations that
have any two strands that originate from the same symmetrizer joined 
together have vanishing weight, since this implies the presence of the 
factor 
\[P_{1}P_{0}=0\]
where $P_{1}$ comes from the symmetrizer and $P_{0}\propto e$ from the 
fusing of the two strands. Hence for $k=2$, the nonvanishing 
configurations are those shown in fig.(B1). 
\begin{center} 
\setlength{\unitlength}{0.008in}
\begin{picture}(320,145)(0,-10)
\path(116,61)(145,89)(116,119)
\path(121,72)(130,61)
\path(121,107)(130,119)
\path(121,58)(153,89)(121,122)
\path(190,58)(159,89)(190,122)
\path(190,107)(179,119)
\path(190,72)(179,61)
\path(195,61)(165,89)(195,119)
\path(3,130)(32,101)(60,130)
\path(14,125)(3,114)
\path(50,125)(60,114)
\path(0,125)(32,93)(63,125)
\path(0,55)(32,87)(63,55)
\path(50,55)(60,66)
\path(14,55)(3,66)
\path(3,51)(32,79)(60,51)
\path(248,122)(272,98)(297,122)
\path(243,69)(268,93)(243,119)
\path(297,66)(272,89)(248,66)
\path(300,119)(276,93)(300,69)
\path(291,122)(300,112)
\path(243,76)(254,66)
\path(291,66)(300,76)
\path(243,112)(254,122)
\put(24,15){\makebox(0,0)[lb]{\raisebox{0pt}[0pt][0pt]
{\shortstack[l]{\footnotesize {\bf Figure(B1)} The nonvanishing strand 
configurations of the }}}}
\put(59,0){\makebox(0,0)[lb]{\raisebox{0pt}[0pt][0pt]
{\shortstack[l]{\footnotesize loop model formulation of the 
${\scriptstyle \Gamma_{2}}$ model}}}}
\put(32,37){\makebox(0,0)[b]{\raisebox{0pt}[0pt][0pt]
{\shortstack{${\scriptstyle 1}$}}}}
\put(155,37){\makebox(0,0)[b]{\raisebox{0pt}[0pt][0pt]
{\shortstack{${\scriptstyle f_{0}}$}}}}
\put(272,41){\makebox(0,0)[b]{\raisebox{0pt}[0pt][0pt]
{\shortstack{${\scriptstyle f_{1}}$}}}}
\end{picture}
\end{center} 
In general, the $\Gamma_{k}$ vertex is replaced by $k+1$ nonvanishing 
strands configurations. (For $k=3$, see fig.(B2)) These strands are again 
the surrounding polygons of the fused Potts models introduced earlier. In 
this fused loop model, a closed composite $k$-strand obtained by fusing 
the individual strands in fig.(B0) into loops carries a weight $(k+1)_{q}$.

  In such a formulation the numbers of degrees of freedom at each vertex 
is vastly reduced, but one has instead to deal with nonlocal quantities.

\begin{center} 
\setlength{\unitlength}{0.005in}
\begin{picture}(628,185)(0,-10)
\path(18,25)(58,65)(98,25)
\path(13,30)(58,75)(103,30)
\path(8,35)(58,85)(108,35)
\path(28,25)(8,45)
\path(88,25)(108,45)
\path(88,155)(108,135)
\path(28,155)(8,135)
\path(8,145)(58,95)(108,145)
\path(13,150)(58,105)(103,150)
\path(18,155)(58,115)(98,155)
\path(148,125)(188,85)(148,45)
\path(153,130)(198,85)(153,40)
\path(158,135)(208,85)(158,35)
\path(148,115)(168,135)
\path(148,55)(168,35)
\path(278,55)(258,35)
\path(278,115)(258,135)
\path(268,135)(218,85)(268,35)
\path(273,130)(228,85)(273,40)
\path(278,125)(238,85)(278,45)
\path(623,165)(563,105)(503,165)
\path(618,170)(563,115)(508,170)
\path(628,160)(568,100)(628,40)
\path(623,35)(563,95)(503,35)
\path(618,30)(563,85)(508,30)
\path(498,160)(558,100)(498,40)
\path(518,170)(498,150)
\path(518,30)(498,50)
\path(628,50)(608,30)
\path(628,150)(608,170)
\path(333,160)(313,140)
\path(433,160)(453,140)
\path(453,50)(433,30)
\path(313,50)(333,30)
\path(323,30)(383,90)(443,30)
\path(453,150)(398,95)(453,40)
\path(448,155)(388,95)(448,35)
\path(323,160)(383,100)(443,160)
\path(313,150)(368,95)(313,40)
\path(318,155)(378,95)(318,35)
\put(20,-20){\makebox(0,0)[lb]{\raisebox{0pt}[0pt][0pt]
{\shortstack[l]
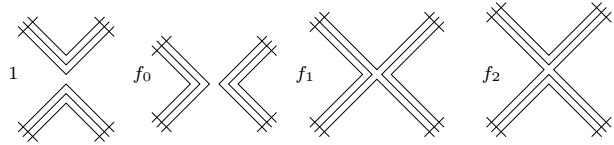
{\footnotesize {\bf Figure(B2)} Nonvanishing strand configurations for the 
$\Gamma_{3}$ model}}}}
\put(138,90){\makebox(0,0)[b]{\raisebox{0pt}[0pt][0pt]{\shortstack
{${\scriptstyle f_{0}}$}}}}
\put(3,90){\makebox(0,0)[b]{\raisebox{0pt}[0pt][0pt]{\shortstack
{${\scriptstyle 1}$}}}}
\put(308,90){\makebox(0,0)[b]{\raisebox{0pt}[0pt][0pt]
{\shortstack{${\scriptstyle f_{1}}$}}}}
\put(503,90){\makebox(0,0)[b]{\raisebox{0pt}[0pt][0pt]
{\shortstack{${\scriptstyle f_{2}}$}}}}
\end{picture}
\end{center} 
 
\subsection{}
For $k=2$, there exist other loop model formulations. The first is due 
to the observation that\cite{wda} 
\begin{equation}
\begin{array}{lll}
b_{i}&=&q^{-2}-(q^{2}+q^{-2})P_{1}(i,i+1)+q(q^{3}-q^{-3})P_{0}(i,i+1)\;,\\
e_{i}&=&(q^{2}+1+q^{-2})P_{0}(i,i+1)
\end{array}
\end{equation}
satisfy the Birman Wenzl Murakami (BWM) algebra\cite{bwm}. The latter
 contains the Temperley-Lieb algebra generated by $e_{i}$ with 
$\sqrt{"Q"}=q^{2}+1+q^{-2}$ as a subalgebra, and $b_{i}$ satisfies the 
braid group relation
\begin{equation}
b_{i}b_{i\pm1}b_{i}=b_{i\pm1}b_{i}b_{i\pm1}\;. 
\end{equation}
They have graphical representations defined by the action on two 
neighboring strands;
\begin{center} 
\setlength{\unitlength}{0.008in}
\begin{picture}(161,167)(0,-10)
\thicklines
\path(124,152)(141,136)(157,152)
\path(124,111)(141,128)(157,111)
\path(124,91)(157,58)
\path(157,91)(145,78)
\path(124,58)(137,70)
\path(120,0)(137,16)(120,33)
\path(161,0)(145,16)(161,33)
\path(42,12)(58,12)
\path(42,20)(58,20)
\path(42,78)(58,78)
\path(42,70)(58,70)
\path(42,128)(58,128)
\path(42,136)(58,136)
\thinlines
\path(71,74)(108,74)
\path(100.000,72.000)(108.000,74.000)(100.000,76.000)
\path(66,132)(104,132)
\path(96.000,130.000)(104.000,132.000)(96.000,134.000)
\path(66,16)(104,16)
\path(96.000,14.000)(104.000,16.000)(96.000,18.000)
\put(0,128){\makebox(0,0)[lb]{\raisebox{0pt}[0pt][0pt]
{\shortstack[l]{${\bf 1} :$}}}}
\put(0,70){\makebox(0,0)[lb]{\raisebox{0pt}[0pt][0pt]
{\shortstack[l]{$b_{i} :$}}}}
\put(0,12){\makebox(0,0)[lb]{\raisebox{0pt}[0pt][0pt]
{\shortstack[l]{$e_{i} :$}}}}
\end{picture}
\end{center} 
Besides the Temperley Lieb and braid group relations, these generators 
satisfy some other algebraic relations which can be represented 
graphically. Most of these relations are then straightforwardly expressed 
by regular isotopy of the diagrams. The others are :\\
1.) The first Reidemester move\\ 
\begin{center} 
\setlength{\unitlength}{0.01in}
\begin{picture}(120,47)(0,-10)
\thicklines
\path(0,32)(32,0)
\path(32,32)(20,20)
\path(0,0)(12,12)
\path(32,0)(48,16)(32,32)
\path(104,0)(120,16)(104,32)
\put(57,12){\makebox(0,0)[lb]{\raisebox{0pt}[0pt][0pt]
{\shortstack[l]{$= q^{4}$}}}}
\end{picture}

\end{center} 
produces a factor $q^{4}$.\\
2.) The relation $b_{i}-b^{-1}_{i}=(q^{-2}-q^{2})({\bf 1}-e_{1})$ holds, 
which  can be represented graphically as 
\begin{center} 
\setlength{\unitlength}{0.01in}
\begin{picture}(311,55)(0,-10)
\thicklines
\path(191,40)(207,24)(223,40)
\path(191,0)(207,16)(223,0)
\path(258,4)(274,19)(258,35)
\path(298,4)(282,19)(298,35)
\path(95,35)(64,4)
\path(95,4)(83,16)
\path(64,35)(75,24)
\path(0,4)(11,16)
\path(32,35)(19,24)
\path(0,35)(32,4)
\put(238,16){\makebox(0,0)[lb]{\raisebox{0pt}[0pt][0pt]
{\shortstack[l]{-}}}}
\put(43,16){\makebox(0,0)[lb]{\raisebox{0pt}[0pt][0pt]
{\shortstack[l]{-}}}}
\put(125,16){\makebox(0,0)[lb]{\raisebox{0pt}[0pt][0pt]
{\shortstack[l]{$q^{-2}-q^{2}$}}}}
\put(103,16){\makebox(0,0)[lb]{\raisebox{0pt}[0pt][0pt]
{\shortstack[l]{=}}}}
\put(183,16){\makebox(0,0)[lb]{\raisebox{0pt}[0pt][0pt]
{\shortstack[l]{(}}}}
\put(306,16){\makebox(0,0)[lb]{\raisebox{0pt}[0pt][0pt]
{\shortstack[l]{)}}}}
\end{picture}
\end{center}    
3.) A loop carries a weight $(3)_{q}$.
\begin{center} 
\setlength{\unitlength}{0.01in}
\begin{picture}(106,47)(0,-10)
\thicklines
\path(16,0)(0,16)(16,32)
\path(16,0)(32,16)(16,32)
\put(40,12){\makebox(0,0)[lb]{\raisebox{0pt}[0pt][0pt]
{\shortstack[l]{$= (3)_{q}$}}}}
\end{picture}

\end{center}   
In terms of these generators, the vertex given in $(\!~\ref{eq:Xu2}\,)$ 
is written as
\begin{equation}
1+q^{-2}\bar{f}_{1}+(f_{0}+q^{2}\bar{f}_{1})e_{i}-\bar{f}_{1}b_{i}
\end{equation}
where
\[\bar{f}_{1}=f_{1}/\sqrt{Q}\;.\]

\subsection{}

The two above mappings have 
the drawback that they involve dense loop coverings of the lattice. An 
elegant way of mapping the $\Gamma_{2}$ model to a "dilute"loop model is 
given in \cite{nih}. In a first step one uses the edges of the vertices 
that carry the states $|\pm>$ to form oriented loops, whose  direction is 
given by the spin arrows, while edges with the state $|0\!>$ are regarded 
as unoccupied. This gives an oriented dilute loop reformulation. The 
problem then is to find under what circumstances  one can get rid of 
the orientations. The simplest way to find a correspondence between an 
oriented and an unoriented loop model is to suppose that in the unoriented 
model loops have a fugacity. This fugacity can be obtained by a sum of 
local contributions if one gives arbitrary orientations to the loops and 
sums over all possible orientations, provided a  phase factor $e^{\pm 1}$ 
or $\tilde{e}^{\pm 1}$ has been  assigned to every turn  as follows 
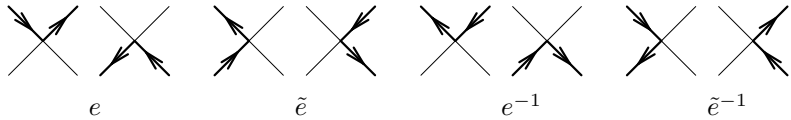
\begin{center}
\setlength{\unitlength}{0.006in}
\begin{picture}(680,110)(0,-10)
\path(30,65)(60,35)
\path(0,35)(30,65)
\thicklines
\path(50,85)(60,95)
\path(30,65)(50,85)
\path(41.515,70.858)(50.000,85.000)(35.858,76.515)
\path(0,95)(20,75)
\path(5.858,83.485)(20.000,75.000)(11.515,89.142)
\path(20,75)(30,65)
\thinlines
\path(110,65)(80,95)
\path(140,95)(110,65)
\thicklines
\path(90,45)(80,35)
\path(110,65)(90,45)
\path(98.485,59.142)(90.000,45.000)(104.142,53.485)
\path(140,35)(120,55)
\path(134.142,46.515)(120.000,55.000)(128.485,40.858)
\path(120,55)(110,65)
\path(200,55)(210,65)
\path(180,35)(200,55)
\path(191.515,40.858)(200.000,55.000)(185.858,46.515)
\path(210,65)(190,85)
\path(204.142,76.515)(190.000,85.000)(198.485,70.858)
\path(190,85)(180,95)
\thinlines
\path(240,35)(210,65)
\path(210,65)(240,95)
\thicklines
\path(300,75)(290,65)
\path(320,95)(300,75)
\path(308.485,89.142)(300.000,75.000)(314.142,83.485)
\path(290,65)(310,45)
\path(295.858,53.485)(310.000,45.000)(301.515,59.142)
\path(310,45)(320,35)
\thinlines
\path(260,95)(290,65)
\path(290,65)(260,35)
\thicklines
\path(390,65)(370,85)
\path(384.142,76.515)(370.000,85.000)(378.485,70.858)
\path(370,85)(360,95)
\path(400,75)(390,65)
\path(420,95)(400,75)
\path(408.485,89.142)(400.000,75.000)(414.142,83.485)
\thinlines
\path(390,65)(360,35)
\path(420,35)(390,65)
\thicklines
\path(470,65)(490,45)
\path(475.858,53.485)(490.000,45.000)(481.515,59.142)
\path(490,45)(500,35)
\path(460,55)(470,65)
\path(440,35)(460,55)
\path(451.515,40.858)(460.000,55.000)(445.858,46.515)
\thinlines
\path(470,65)(500,95)
\path(440,95)(470,65)
\path(600,95)(570,65)
\path(570,65)(600,35)
\thicklines
\path(540,95)(560,75)
\path(545.858,83.485)(560.000,75.000)(551.515,89.142)
\path(560,75)(570,65)
\path(550,45)(540,35)
\path(570,65)(550,45)
\path(558.485,59.142)(550.000,45.000)(564.142,53.485)
\thinlines
\path(620,35)(650,65)
\path(650,65)(620,95)
\thicklines
\path(680,35)(660,55)
\path(674.142,46.515)(660.000,55.000)(668.485,40.858)
\path(660,55)(650,65)
\path(670,85)(680,95)
\path(650,65)(670,85)
\path(661.515,70.858)(670.000,85.000)(655.858,76.515)
\put(610,0){\makebox(0,0)[lb]{\raisebox{0pt}[0pt][0pt]
{\shortstack[l]{$\tilde{e}^{-1}$}}}}
\put(430,0){\makebox(0,0)[lb]{\raisebox{0pt}[0pt][0pt]
{\shortstack[l]{${e}^{-1}$}}}}
\put(250,0){\makebox(0,0)[lb]{\raisebox{0pt}[0pt][0pt]
{\shortstack[l]{$\tilde{e}$}}}}
\put(70,0){\makebox(0,0)[lb]{\raisebox{0pt}[0pt][0pt]
{\shortstack[l]{$e$}}}}
\end{picture}
\end{center}
 For a lattice which has the geometry of a plane, a closed loop has weight
$e^{2}\tilde{e}^{2}(e^{-2}\tilde{e}^{-2})$ if the orientation is 
anticlockwise( clockwise )and so gets  fugacity (weight) $n$ equal to
$e^{2}\tilde{e}^{2}+e^{-2}\tilde{e}^{-2}$. In addition, to every edge 
of the vertex, one can assign, without altering the partition function, 
 a local phase factor  as follows
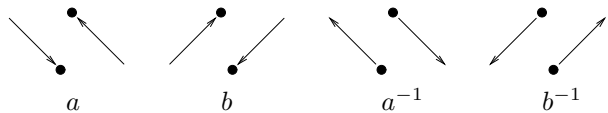
\begin{center}
\setlength{\unitlength}{0.006in}
\begin{picture}(520,120)(0,-10)
\put(45,35){\circle*{8}}
\put(55,85){\circle*{8}}
\put(195,35){\circle*{8}}
\put(185,85){\circle*{8}}
\put(335,85){\circle*{8}}
\put(325,35){\circle*{8}}
\put(475,35){\circle*{8}}
\put(465,85){\circle*{8}}
\path(0,80)(40,40)
\path(32.929,44.243)(40.000,40.000)(35.757,47.071)
\path(100,40)(60,80)
\path(67.071,75.757)(60.000,80.000)(64.243,72.929)
\path(240,80)(200,40)
\path(204.243,47.071)(200.000,40.000)(207.071,44.243)
\path(140,40)(180,80)
\path(175.757,72.929)(180.000,80.000)(172.929,75.757)
\path(372.929,44.243)(380.000,40.000)(375.757,47.071)
\path(380,40)(340,80)
\path(287.071,75.757)(280.000,80.000)(284.243,72.929)
\path(280,80)(320,40)
\path(515.757,72.929)(520.000,80.000)(512.929,75.757)
\path(520,80)(480,40)
\path(424.243,47.071)(420.000,40.000)(427.071,44.243)
\path(420,40)(460,80)
\put(465,0){\makebox(0,0)[lb]{\raisebox{0pt}[0pt][0pt]
{\shortstack[l]{$b^{-1}$}}}}
\put(325,0){\makebox(0,0)[lb]{\raisebox{0pt}[0pt][0pt]
{\shortstack[l]{$a^{-1}$}}}}
\put(185,0){\makebox(0,0)[lb]{\raisebox{0pt}[0pt][0pt]
{\shortstack[l]{$b$}}}}
\put(50,0){\makebox(0,0)[lb]{\raisebox{0pt}[0pt][0pt]
{\shortstack[l]{$a$}}}}
\end{picture}
\end{center}
where the solid dot denotes the center of the vertex.

In an unoriented model,  one has the following local loop configurations
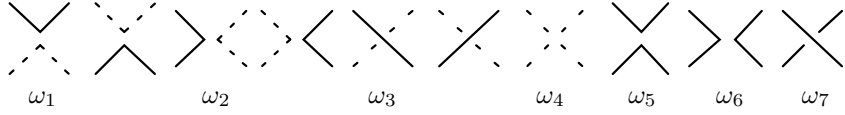
\begin{center} 
\setlength{\unitlength}{0.01in}
\begin{picture}(455,67)(0,-10)
\thicklines
\path(0,52)(16,37)(31,52)
\dashline{4.000}(0,15)(16,30)(31,15)
\path(316,15)(331,30)(345,15)
\path(316,52)(331,37)(345,52)
\path(394,48)(380,33)(394,19)
\path(356,48)(372,33)(356,19)
\dashline{4.000}(300,48)(271,19)
\dashline{4.000}(271,48)(300,19)
\path(256,48)(225,19)
\dashline{4.000}(225,48)(256,19)
\dashline{4.000}(211,48)(180,19)
\path(180,48)(211,19)
\dashline{4.000}(132,48)(147,33)(132,19)
\path(169,48)(154,33)(169,19)
\path(87,48)(102,33)(87,19)
\dashline{4.000}(124,48)(109,33)(124,19)
\dashline{4.000}(45,52)(60,37)(76,52)
\path(45,15)(60,30)(76,15)
\path(405,48)(436,19)
\path(436,48)(424,37)
\path(405,19)(416,30)
\put(276,0){\makebox(0,0)[lb]{\raisebox{0pt}[0pt][0pt]
{\shortstack[l]{$\omega_{4}$}}}}
\put(324,0){\makebox(0,0)[lb]{\raisebox{0pt}[0pt][0pt]
{\shortstack[l]{$\omega_{5}$}}}}
\put(370,0){\makebox(0,0)[lb]{\raisebox{0pt}[0pt][0pt]
{\shortstack[l]{$\omega_{6}$}}}}
\put(415,0){\makebox(0,0)[lb]{\raisebox{0pt}[0pt][0pt]
{\shortstack[l]{$\omega_{7}$}}}}
\put(10,0){\makebox(0,0)[lb]{\raisebox{0pt}[0pt][0pt]
{\shortstack[l]{$\omega_{1}$}}}}
\put(101,0){\makebox(0,0)[lb]{\raisebox{0pt}[0pt][0pt]
{\shortstack[l]{$\omega_{2}$}}}}
\put(188,0){\makebox(0,0)[lb]{\raisebox{0pt}[0pt][0pt]
{\shortstack[l]{$\omega_{3}$}}}}
\end{picture}
\end{center} 
where solid (dotted) strand denotes occupied (unoccupied) edge and
$\omega_{i}$'s, and  for the last configuration the way the two strands 
overlap has no significance. If moreover the loops have fugacity $n$ it 
is then equivalent to an oriented loop model, that is to a 19 vertex 
model, with weights which are products of $\omega_{i}$ and 
$a,b,e,\tilde{e}$
\begin{equation}
\begin{array}{lllllllll}
&&&&&&V_{00,00}&=&\omega_{4}\;,\\
&&&&V_{+0,+0}&=&V_{0-,0-}&=&\omega_{1}eab^{-1}\;,\\
&&&&V_{0+,0+}&=&V_{-0,-0}&=&\omega_{1}e^{-1}a^{-1}b\;,\\
&&&&V_{+-,00}&=&V_{00,+-}&=&\omega_{2}\tilde{e}^{-1}ab^{-1}\;,\\
&&&&V_{-+,00}&=&V_{00,-+}&=&\omega_{2}\tilde{e}a^{-1}b\;,\\
&&&&V_{++,++}&=&V_{--,--}&=&\omega_{5}+\omega_{7}\;,\\
&&&&V_{+-,-+}&=&V_{-+,+-}&=&\omega_{6}+\omega_{7}\;,\\
&&&&&&V_{+-,+-}&=&\omega_{6}\tilde{e}^{-2}a^{2}b^{-2}
+\omega_{5}e^{2}a^{2}b^{-2}\;,\\
&&&&&&V_{-+,-+}&=&\omega_{6}\tilde{e}^{2}a^{-2}b^{2}
+\omega_{5}e^{-2}a^{-2}b^{-2}\;,\\
V_{+0,0+}&=&V_{0+,+0}&=&V_{0-,-0}&=&V_{-0,0-}&=&\omega_{3}
\end{array}
\end{equation}
where $V_{ij,kl}$ denotes the vertex weight with in- and out- 
states being $ij$ and $kl$ respectively.

We thus see that the natural oriented loop model associated with 
the 19 vertex model is equivalent to an unoriented one provided the 
weights can be parametrized as above.  This gives rise to  a necessary 
condition  
\begin{equation}
\frac{\frac{V_{+0,+0}}{V_{0+,0+}} 
+\frac{V_{+-,00}}{V_{-+,00}}}
{\frac{V_{0+,0+}}{V_{+0,+0}}
+\frac{V_{-+,00}}{V_{+-,00}}}=
\frac{\left(V_{+-,-+}-V_{++,++}\right)\frac{V_{+0,+0}}{V_{0+,0+}}
+V_{+-,+-}}{\left(V_{+-,-+}-V_{++,++}\right)\frac{V_{0+,0+}}
{V_{+0,+0}}+V_{-+,-+}}\;.
\label{eq:consistency}
\end{equation}
This holds in particular for the $\Gamma_2$ vertex model (ie when 
the 19 vertex model has ${\rm U_q}{\rm su(2)}$, for which the
 correspondence between the parameters $\omega_{i},a,b,e,\tilde{e}$ 
and  $f_{0},\bar{f}_{1}$ is given by   
\begin{equation}
\begin{array}{lll}
\omega_{1}&=&(1+q^{-2}\bar{f}_{1})^{1/2}(1+q^{2}\bar{f}_{1})^{1/2}\;,\\   
\omega_{2}&=&(f_{0}+q^{-2}\bar{f}_{1})^{1/2}(f_{0}
+q^{2}\bar{f}_{1})^{1/2}\;,\\   
\omega_{3}&=&-\bar{f}_{1}\;,\\   
\omega_{4}&=&1+f_{0}+(Q-3)\bar{f}_{1}\;,\\
\omega_{5}&=&\frac{\textstyle (1+q^{2}\bar{f}_{1})(1+q^{-2}\bar{f}_{1})
(f_{0}+\bar{f}_{1})}{\textstyle f_{0}+\bar{f}_{1}
+f_{0}\bar{f}_{1}+(Q-3)\bar{f}_{1}^{2}}\;,\\  
\omega_{6}&=&\frac{\textstyle (f_{0}+q^{2}\bar{f}_{1})(f_{0}
+q^{-2}\bar{f}_{1})(1+\bar{f}_{1})}{\textstyle f_{0}+\bar{f}_{1}+
f_{0}\bar{f}_{1}+(Q-3)\bar{f}_{1}^{2}}\;,\\  
\omega_{7}&=&-\frac{\textstyle \bar{f}_{1}^{2}+(Q-3)f_{0}\bar{f}_{1}
+f_{0}\bar{f}_{1}^{2}+\bar{f}_{1}^{3}}
{\textstyle f_{0}+\bar{f}_{1}+f_{0}\bar{f}_{1}+(Q-3)\bar{f}_{1}^{2}}
\end{array}
\end{equation}
and the loop fugacity reads 
\begin{equation}
n=q^{2}\frac{(1+q^{-2}\bar{f}_{1})(f_{0}+q^{-2}\bar{f}_{1})}
{(1+q^{2}\bar{f}_{1})(f_{0}+q^{2}\bar{f}_{1})}+q^{-2}
\frac{(1+q^{2}\bar{f}_{1})(f_{0}+q^{2}\bar{f}_{1})}
{(1+q^{-2}\bar{f}_{1})(f_{0}+q^{-2}\bar{f}_{1})}\;.
\end{equation}
In this loop reformulation  there are more degrees of freedom at 
each vertex than in the first one we discussed.  Not all edges are 
occupied. The fugacity  depends on $f_0$ and $f_{1}$. 

Consider now the case 
\begin{equation}
\bar{f}_{1}=-1\;,\;\;f_{0}=0\;.
\end{equation}
The nonvanishing weights after rescaling become
\begin{equation}
\begin{array}{rllllll}
\omega_{2}&=&\omega_{3}&=&\omega_{5}&=&1\;,\\
\mbox{and}\hspace{10mm}&&\omega_{1}^{2}&=&\omega_{4}&=
&-(q-q^{-1})^{2}\;,
\end{array}
\end{equation}
the vanishing of $\omega_{5}$ and and $\omega_{7}$ implies that, if  
vertices are "expanded"  as
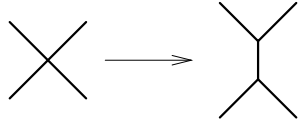
\begin{center} 
\setlength{\unitlength}{0.0125in}
\begin{picture}(120,63)(0,-10)
\thicklines
\path(32,40)(0,8)
\path(0,40)(32,8)
\path(88,48)(104,32)(120,48)
\path(88,0)(104,16)(120,0)
\path(104,32)(104,16)
\thinlines
\path(40,24)(76,24)
\path(68.000,22.000)(76.000,24.000)(68.000,26.000)
\end{picture}

\end{center} 
such that the entire lattice becomes   honeycomb , every edge can at 
most be occupied by one strand. Moreover, since the
nonvanishing weights satisfy the relations given above, the model
belongs to  a subset of the class of loop model where
loops do not intersect and the only parameters are loop and vacant 
site fugacities\cite{nih1} $\{n,\omega_{1}\}$. In this case, the two 
parameters are related as
\begin{equation}
n=2-(2-\omega_{1}^{2})^{2}\;.
\end{equation} 

The integrable lines IK, TL and FZ play distinctive roles here too, 
they arise as a result of the restriction
\[\begin{array}{rll}
\mbox{IK, TL}&:&\omega_{7}=0\\
\mbox{ FZ}&:&\mbox{ vertex weights being invariant under reversal 
of all arrows and } $n=2$.
\end{array}\]
 
For general $f_{0},\bar{f}_{1}$, the loops can be interpreted as high 
temperature expansion of an O$(n)$ model which has Boltzman weight 
\[\begin{array}{l}
\omega_{4}+\omega_{1}(\vec{s}_{i}\cdot\vec{s}_{k}+\vec{s}_{j}
\cdot\vec{s}_{l})+
\omega_{2}(\vec{s}_{i}\cdot\vec{s}_{j}+\vec{s}_{k}\cdot\vec{s}_{l}+
\omega_{3}(\vec{s}_{i}\cdot\vec{s}_{l}+\vec{s}_{j}\cdot\vec{s}_{k})\\  
\omega_{5}(\vec{s}_{i}\cdot\vec{s}_{j})(\vec{s}_{k}\cdot\vec{s}_{l})+
\omega_{6}(\vec{s}_{i}\cdot\vec{s}_{k})(\vec{s}_{j}\cdot\vec{s}_{l})+
\omega_{7}(\vec{s}_{i}\cdot\vec{s}_{l})(\vec{s}_{j}\cdot\vec{s}_{k})
\end{array}\]
where $\vec{s}_{i}$'s are $n$-component vectors situated on the edges 
of each vertex and are normalized as $\vec{s}_{i}\cdot \vec{s}_{i}=n$.
\begin{center} 
\setlength{\unitlength}{0.01in}
\begin{picture}(110,95)(0,-10)
\path(50,80)(110,20)
\path(30,60)(90,0)
\path(110,60)(50,0)
\path(90,80)(30,20)
\put(66,25){\makebox(0,0)[lb]{\raisebox{0pt}[0pt][0pt]{\shortstack[l]
{$s_{l}$}}}}
\put(66,50){\makebox(0,0)[lb]{\raisebox{0pt}[0pt][0pt]{\shortstack[l]
{$s_{k}$}}}}
\put(24,25){\makebox(0,0)[lb]{\raisebox{0pt}[0pt][0pt]{\shortstack[l]
{$s_{j}$}}}}
\put(24,50){\makebox(0,0)[lb]{\raisebox{0pt}[0pt][0pt]{\shortstack[l]
{$s_{i}$}}}}
\end{picture}
\end{center}

\end{document}